\documentclass[11pt]{article}
\usepackage[pctex32]{graphics}
\usepackage{amssymb}
\usepackage{latexsym}
\usepackage{epsfig} 
\usepackage{multirow} 
\usepackage{pstcol}
\usepackage{float}
\usepackage{epstopdf}
\usepackage{mathrsfs}
\usepackage{algorithm} 
\usepackage{algorithmic}
\usepackage{caption}
\usepackage{subfigure}

\usepackage{amsmath}
\usepackage{amssymb}
\usepackage{relsize}
\usepackage{multirow}

\usepackage{graphicx}
\usepackage{multicol} 

\usepackage{times}
\usepackage{epstopdf}

\definecolor{Red}{rgb}{1,0.,0.}

\usepackage{fancyhdr}
\usepackage{setspace}

\newcommand{\R}{{\mathbb R}}

\newcommand{\bU}{{\bf U}}
\newcommand{\bQ}{{\bf Q}}
\newcommand{\bV}{{\bf V}}

\newcommand{\bc}{{\bf c}}

\newcommand{\bu}{{\bf u}}

\newcommand{\bq}{{\bf q}}
\newcommand{\mM}{{\mathsf M}}
\newcommand{\mS}{{\mathsf S}}
\newcommand{\mK}{{\mathsf K}}

\newcommand{\mI}{{\mathsf I}}

\newcommand{\mV}{{\mathsf V}}

\newcommand{\mE}{{\mathsf E}}
\newcommand{\mB}{{\mathsf B}}

\newcommand{\mQ}{{\mathsf Q}}
\newcommand{\mO}{{\mathsf O}}

\newcommand{\mF}{{\mathsf F}}

\newcommand{\mSigma}{{\mathsf \Sigma}}

\newcommand{\bgamma}{{\mbox{\boldmath$\gamma$}}}
\newcommand{\bphi}{{\mbox{\boldmath$\phi$}}}

\graphicspath{ {./figures_spatial/} }

\usepackage[utf8]{inputenc}	
\usepackage[T1]{fontenc}	
\usepackage{xcolor}		
\usepackage[colorlinks = true,
linkcolor = purple,
urlcolor  = blue,
citecolor = cyan,
anchorcolor = black]{hyperref}	
\usepackage{booktabs} 		
\usepackage{nicefrac}		
\usepackage{microtype}		
\usepackage{lineno}		
\usepackage{float}			

\usepackage{lipsum}		

\usepackage{newfloat}
\DeclareFloatingEnvironment[name={Supplementary Figure}]{suppfigure}
\usepackage{sidecap}
\sidecaptionvpos{figure}{c}
\date{}
\usepackage{titlesec}
\titlespacing\section{0pt}{12pt plus 3pt minus 3pt}{1pt plus 1pt minus 1pt}
\titlespacing\subsection{0pt}{10pt plus 3pt minus 3pt}{1pt plus 1pt minus 1pt}
\titlespacing\subsubsection{0pt}{8pt plus 3pt minus 3pt}{1pt plus 1pt minus 1pt}

\usepackage{tikz,xcolor,hyperref}

\definecolor{lime}{HTML}{A6CE39}
\DeclareRobustCommand{\orcidicon}{
	\begin{tikzpicture}
	\draw[lime, fill=lime] (0,0) 
	circle [radius=0.16] 
	node[white] {{\fontfamily{qag}\selectfont \tiny ID}};
	\draw[white, fill=white] (-0.0625,0.095) 
	circle [radius=0.007];
	\end{tikzpicture}
	\hspace{-2mm}
}
\foreach \x in {A, ..., Z}{\expandafter\xdef\csname orcid\x\endcsname{\noexpand\href{https://orcid.org/\csname orcidauthor\x\endcsname}
		{\noexpand\orcidicon}}
}

\title{A Spatially Distributed Model of Brain Metabolism Highlights the Role of Diffusion in Brain Energy Metabolism}

\usepackage{xwatermark}
\newwatermark[firstpage,color=gray!90,angle=0,scale=0.28, xpos=0in,ypos=-5in]{*correspondence: \texttt{Daniela Calvetti - dxc57@case.edu}}

\usepackage{authblk}

\author[1]{Gideon Idumah\orcidA{}}
\author[2]{Erkki Somersalo\orcidB{}}
\author[3]{Daniela Calvetti\orcidC{}}

\affil[1, 2, 3]{Department of Mathematics, Applied Mathematics and Statistics, Case Western Reserve University, USA}

\providecommand{\keywords}[1]
{
	\small	
	\textbf{\textit{Keywords---}} #1
}

\begin{document}
\maketitle

\begin{abstract}
\noindent The different active roles of neurons and astrocytes during neuronal activation are associated with the metabolic processes necessary to supply the energy needed for their respective tasks at rest and during neuronal activation. Metabolism, in turn, relies on the delivery of metabolites and removal of toxic byproducts through diffusion processes and the cerebral blood flow. A comprehensive mathematical model of brain metabolism should account not only for the biochemical processes and the interaction of neurons and astrocytes, but also the diffusion of metabolites. In the present article, we present  a computational methodology based on a multidomain model of the brain tissue and a homogenization argument for the diffusion processes. In our spatially distributed compartment model, communication between compartments occur both through local transport fluxes, as is the case within local astrocyte-neuron complexes, and through diffusion of some substances in some of the compartments. The model assumes that diffusion takes place in the extracellular space (ECS) and in the astrocyte compartment. In the astrocyte compartment, the diffusion across the syncytium network is implemented as a function of gap junction strength. 
The diffusion process is implemented numerically by means of a finite element method (FEM) based spatial discretization, and robust stiff solvers are used to time integrate the resulting large system.  Computed experiments show the effects of ECS tortuosity, gap junction strength and spatial anisotropy in the astrocyte network on the brain energy metabolism. 
\end{abstract}

\keywords{Brain energy metabolism, Oxygen-Glucose Index, diffusion process, tortuosity, astrocyte syncytium, gap junction strength. }


	\section{Introduction}
	The brain is arguably the most important and complex organ of the human body, and the hardest to observe directly due to the protective role of the skull enclosing and the impossibility of accessing it without interfering with its regular functions. Understanding brain metabolism faces the additional challenge that most cerebral functions depend on the coordinated efforts of neurons and astrocytes, and rely on a sophisticated network of blood vessels to promptly replenish metabolites and remove waste products. Cerebral blood flow, in turn, is regulated through neurovascular coupling, a feedback system that is not yet fully understood. The large amount of energy that is needed to sustain the electrophysiological processes in the brain has prompted a lot of interest in understanding cerebral metabolism, and how it responds to different levels of neuronal activation. For the last several years, brain energy metabolism has been the topic of a large body of research, and over time mathematical models have assumed an increasingly important role. It is now acknowledged that mathematical models are necessary to put the results of experimental procedures into a proper context, as highlighted in a recent review \cite{barros2018current}.  While experimental results are vital for understanding the brain energy metabolism, the different conditions under which laboratory experiments are carried out may be the reason for the discrepancy in the findings, highlighting how it may be hard, if not impossible,  to draw definitive conclusions on the basis of laboratory data alone. Realistic mathematical models of human brain metabolism have an important role in the pharmaceutical industry also, since the efficacy of drugs in the brain depends on how they can reach the target destination and on the underlying metabolic processes \cite{vendel2019need}. Moreover, a better understanding of cerebral metabolism may shed some light on whether altered metabolism may play a role in some important pathologies, including those related to aging.
	
	It is well known that human brain requires a disproportionately  large amount of energy \cite{attwell2001energy} compared to its small size.  
	The brain does not have any reserve of fuel or oxygen, thus it requires a continuous replenishing of substances needed for energy production, mainly glucose and oxygen, and removal of byproducts like carbon dioxide and lactate. The partitioning of glucose, the main oxidative metabolite, between neurons and astrocytes has been a somewhat controversial issue, with some researchers sustaining that glucose delivered to the brain is mostly taken up by astrocytes and transformed into lactate, deemed to be the preferred oxidative fuel for neurons, and others arguing that neurons prefer and process predominantly glucose. Regardless of the roles attributed to neurons and astrocytes, the importance of astrocytes in energy metabolism is now universally acknowledged. The coordination of the metabolism in neurons and astrocytes is of crucial importance for guaranteeing the continuity of brain functions \cite{bonvento2021astrocyte}, although the details of the coupling mechanism between them is not fully understood. 
	
	Most mathematical models of brain energy metabolism are based on the well-mixed compartment assumption, a paradigm whose validity has been questioned  \cite{barros2007enquiry}. A main shortcoming of compartment models is the inability to account for the role of diffusion, a key component in many physiological processes. Diffusion in brain tissue through the extracellular space has been studied quite extensively, and the importance of the diffusion in electrolyte dynamics has been acknowledged and to some extent also addressed through mathematical models in the literature \cite{Tuttle}. In \cite{sykova2008diffusion}, the authors investigate the effect of the tortuosity of the the extracellular space as well as the size of the molecules of the different substances. The process of obtaining quantitative measures is complicated by the loss of molecules across the the blood-brain barrier or through the uptake by neuron and astrocytes. The diffusion of metabolites across groups of astrocytes in networks connected through special membrane structures known as gap junctions has been observed in cultures of astrocytes \cite{giaume1997metabolic}, thus highlighting the potentially important role of interconnected astrocytic pathways in cerebral metabolism. A recently proposed mathematical model of astrocyte syncytium has been used to study potassium buffering in connection with neuronal firing \cite{terman2019modeling}.  Still, to date, there are only very few detailed spatially distributed mathematical models of brain metabolism, in part because their large complexity and the orders of magnitude difference in typical times poses significant computational challenges. The present article is a contribution towards filling this gap.

	The systematic development towards  a spatially distributed model of brain metabolism,  initiated in \cite{CALVETTI201548}, was inspired by the bidomain models for electrolyte diffusion in myocardium. To avoid addressing the complexity of micro-geometric structure of brain tissue, the tissue is modeled as spatially distributed coexisting compartments with local interaction through transport of substrates from one compartment to another. In principle, this approach would make it possible to include complex local metabolic interactions without having the complexity become overwhelming. The cited article, where the authors presented a proof of concept prototype model of metabolism with only few metabolites tracked in each compartment, is the starting point for the present work. The main contribution of this paper is to provide a predictive spatially distributed model of brain energy metabolism that accounts for the diffusion of metabolites in extracellular space and astrocyte networks. By adhering to a multidomain setting, the metabolism is described in local terms as neuron-astrocyte complexes, coupled together by a detailed diffusion model through the ECS and astrocyte networks. The computational scheme is based on a finite element discretization of the domain.

	The reminder of the paper is organized as follows. In the Materials and Methods  section we describe the anatomical characteristics of the brain region that we consider, and we set up the multidomain framework while establishing the notation to be used in the rest of the paper. Subsection 2.1 is dedicated to diffusion in extracellular space and its mathematical formulation, while diffusion in astrocyte is discussed in subsection 2.2. In subsection 2.3 we perform a model reduction by integrating along the dimension of the orientation of the axons, and in subsection 2.4 we present the details of the finite elements discretization of the two dimensional reduced domain. Subsection 2.5 lists the cross-membrane exchanges of metabolites between compartments and the mathematical expressions of the rates at which they occur, with subsection 2.6 entirely dedicated to glutamate-glutamine cycle, which in our model is a proxy for the electrophysiological activity. In subsection 2.7 we introduce the reactions that are considered in the model and the mathematical expression of their rates. Section 3 presents the results of computed experiments related to three different protocols, designed to highlight the role of diffusion to sustain the energetic needs of neuronal activation, in addition to some conclusions and an outline of future work.
	
	\newpage
	\section{Materials and methods}
	
	We start by introducing the different components of our spatially distributed model of human brain metabolism. 
	
	\subsection{General setup} 
	
	The brain consists of a vast assemblage of densely packed cells of varying sizes, structures and functions, interspersed with a dense net of blood capillaries. For the purpose of brain energy metabolism, we will concentrate on neurons and astrocytes in the gray matter. Between cells and capillaries there is a small interstitial space called the extracellular space (ECS), that has been likened to the water phase of a foam, with the gaseous phase corresponding to cells \cite{Kuffler}.
	
\begin{figure}[tbh]
		\centerline{\includegraphics[width=0.3\textwidth]{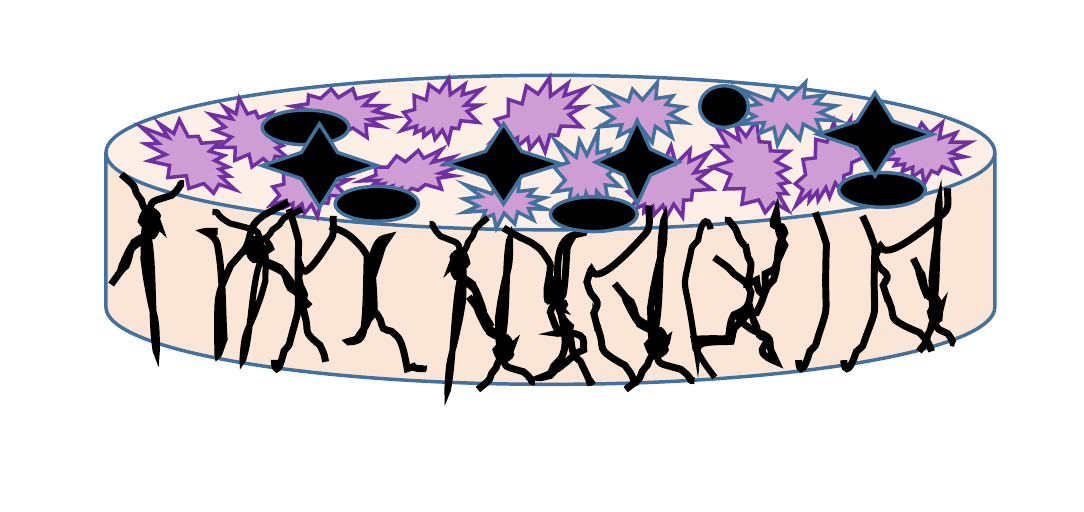}}
		\caption{\label{BrainTissue}A cartoon of the organization of neuronal axons in the region of interest.}
	\end{figure}

	We begin by considering a  three dimensional domain $W \subset \R^3$ such as gray matter, and we assume that the region of interest occupies a cylindrical subset $\Omega \subset W$,
	\[
	\Omega = B\times [0,h], \quad B\in\R^2.
	\]
	To simplify our model, we assume that in the region of the brain that we are modeling, neurons are highly organized in sieve-like fashion, with axons and dendrites predominantly perpendicular to the cortical surface, as schematically illustrated in Figure~\ref{BrainTissue}. 
	
\begin{figure}[!ht]
		\label{FourDomain} 
		\centerline{\includegraphics[width=0.3\textwidth]{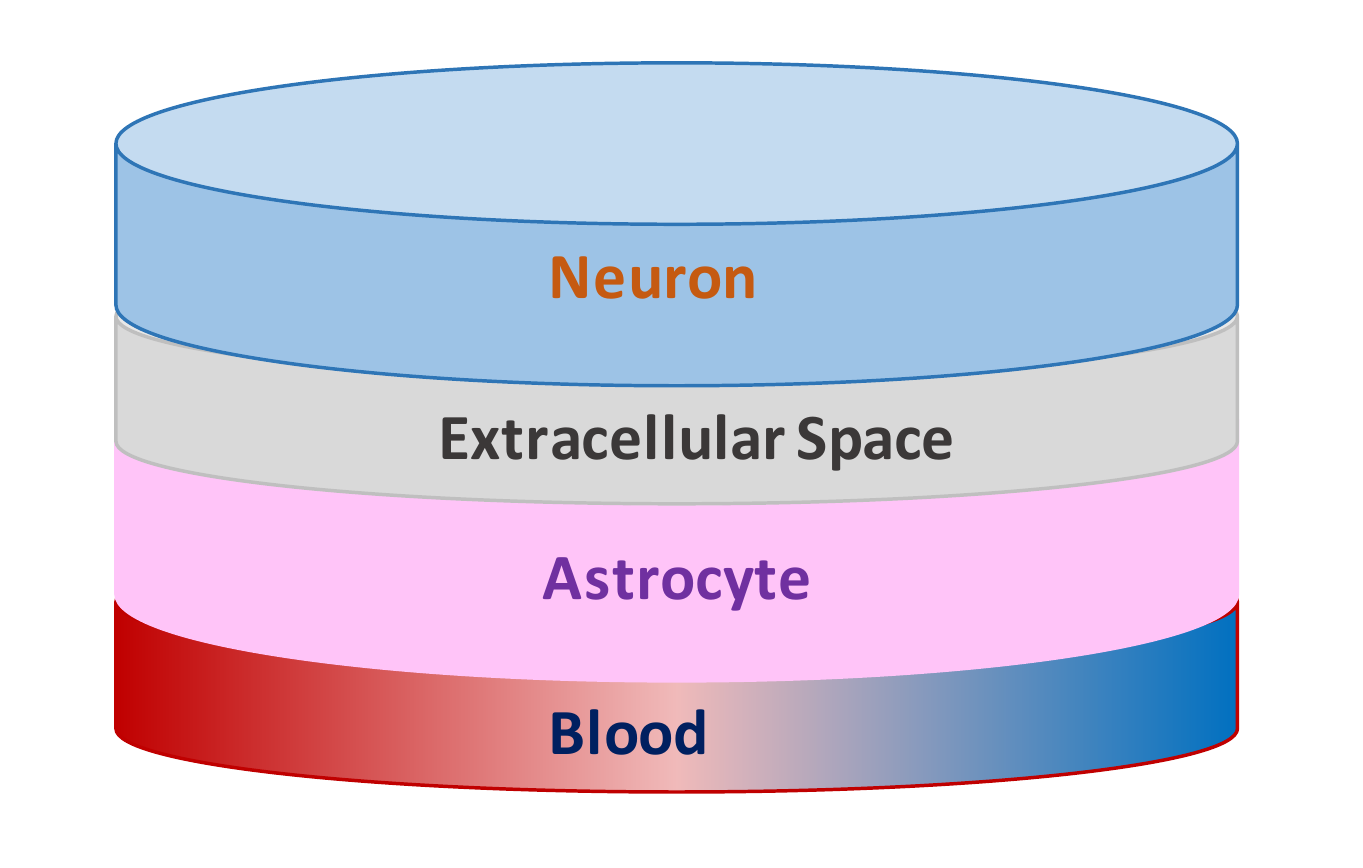}}
		\caption{\label{fig:multidomain} A rendition of the multidomain that we assume at each space location.}
	\end{figure}

	We introduce a Cartesian coordinate system with the $z$-axis perpendicular to the cortical surface. To avoid having to deal with the detailed geometric description of individual cells, following  \cite{CALVETTI201548}, we define a multi-domain structure as follows. We consider $J$ copies $\Omega^j$  of $\Omega$, and associate to each $\Omega^j$ for $1\leq j \leq J$, a positive scalar $\eta^j$, called the volume fraction of $\Omega^j$, with 
	\[
	\eta^1 + \ldots + \eta^J = 1.
	\]
	Each subdomain $\Omega^j$ represents a homogenized compartment that occupies a fraction $\eta^j$ of the total volume of the domain, and each point $x\in \Omega$ is assumed to belong simultaneously to all $J$ subdomains. 
	More formally, define the $j$th subdomain as
	\[
	\Omega^j = \left(\Omega, \eta^j \, dx\right),
	\]
	i.e., the domain $\Omega$ equipped with the Lebesgue measure weighted by $\eta^j$. For any integrable function, $f: \Omega \rightarrow \R$, define the integral over $\Omega^j$ as
	\[
	\int_{\Omega^j} f(x) ~ dx = \eta^j \int_\Omega f(x) ~ dx,
	\]
	and interpret the volume fractions as
	$$\eta^j = \frac{|\Omega^j|}{|\Omega|}$$
	with $|\cdot|$ denoting the volume of the set. We can express the multi-domain formally as a quotient space
	\[
	\overline{\Omega} = \Omega^1\times\Omega^2\times\cdots\times\Omega^J/\sim,
	\]
	where ``$\sim$'' indicates the identification of points in the sets $\Omega^j$.
	The multidomain model that we propose here consists of a coupled system of convection-reaction-diffusion equations with four subdomains,  $J=4$, corresponding to blood ($j = 1$), ECS $(j = 2)$, neurons $(j = 3)$ and astrocytes $(j = 4)$, respectively, see Figure~\ref{fig:multidomain}. The neuron and astrocyte subdomains represent the cellular compartments, and we assume that the replenishing  of metabolites and the removal of waste products occurs through the ECS.  
	
	Our model assumes that:
	\begin{itemize}
		\item[(a)] Metabolites may diffuse in domains $j=2,4$ (ECS and astrocyte).
		\item[(b)] Diffusion in domains $j=1,3$ (blood and neuron) is insignificant, thus it is neglected.
		\item[(c)] The exchange of substances across compartments is a local process, hence modeled point-wise.
		\item[(d)] Blood capillaries run rather densely through the domain $\Omega$, and the replenishing or depletion of metabolites by the blood compartment is modeled as a local process.
		\item[(e)] The synaptic cleft between pre- and postsynaptic neuron is modeled effectively as a subdomain of the ECS where the only metabolites of interest are glutamate and glutamine, and where no diffusion occurs.
		\item[(f)] In the neuron domain we do not differentiate between pre- and postsynaptic neurons, glutamatergic and GABAergic neurons.
	\end{itemize}
	
	\subsection{Diffusion in ECS}
	
	In general, diffusion of solutes in ECS plays an important role for non-synaptic cell-to-cell communication, oxygen delivery, extracellular $K^+$ and glutamate buffering during neuronal signaling, and cellular nutrient uptake  \cite{Sykova_Nicholson,SYKOVA2004453}. Diffusion in ECS is important in the local delivery of drugs, and in understanding some anomalous conditions such as  cortical spreading depolarization.
	
	Diffusion in the ECS can be modeled at the microscopic scale or macroscopic scale.  On a microscopic scale, it can be described through random walk of the molecules, using, e.g., regular and random arrays of convex polytopes to describe the ECS, whereas the process on a macroscopic scale can be described by a modified diffusion equation. In this paper we follow the latter. 
	
	The most relevant properties of ECS  with regard to diffusion at a macroscopic scale are its volume fraction $\eta$ and its tortuosity $\lambda$ \cite{2001RPPh...64..815N}. Volume fraction of ECS is defined as the ratio
	\[
	\eta = \eta_2 = V_{\rm ECS}/V_{\rm Tissue},
	\]
	where $V_{\rm ECS}$ is the volume of ECS in $\Omega$  and $V_{\rm Tissue}$ is the volume of the brain tissue in $\Omega$. In this subsection, we suppress the subindex in $\eta_2$ to keep the notation simpler.
	Tortuosity can be summarized in terms of a complex parameter  $\lambda$ which describes the average hindrance of a complex medium in comparison to an obstacle-free medium. More formally, $\lambda$ is defined as 
	\begin{equation}\label{tortuosity}
	\lambda = \sqrt{\frac{D}{D^*}}
	\end{equation}
	where $D$ is the diffusion coefficient in free medium, e.g., water or dilute gel, and $D^*$ is the effective diffusion coefficient in ECS. Tortuosity has proved to be very important in many processes in the brain, ranging from ischemia and osmotic stress to delivery of nutrients and drugs \cite{HRABE20041606}.
	
	There is a large amount of experimental data confirming that the volume fraction of  ECS under normal conditions is $\sim20 \%$, see  \cite{2001RPPh...64..815N,Sykova_Nicholson,NICHOLSON2000129}, and it can decrease to $5\%$ during ischemia due to cell swelling. In healthy brain tissues, the experimental value of tortuosity $\lambda$ is typically assessed around $1.6$, meaning that a small molecule has an effective diffusion coefficient about $2.56$ times smaller than in free solution,  and the tortuosity value increases for pathologies that involve cell swelling.

	On a microscopic scale, various studies have attempted to design computational experiments to derive diffusion parameter values using mathematical models of ECS and Monte Carlo methods. The models typically involve random walks of populations of point-mass particles in a complex geometric domain, and the parameter values are estimated from the statistical distribution of the particles. The simulations are typically time-consuming and depend on the level of details included in the geometry.
	In  \cite{Nandigam}, ECS is modeled as empty three-dimensional space between closely packed arrays of fluid membrane vesicles. These packings were generated by minimizing the configurational energy using a Monte Carlo procedure. A random walk algorithm is then used to compute the geometric tortuosities. In \cite{doi:10.1073/pnas.95.15.8975}, ECS is assumed to be a space containing a random assembly of space-filling obstacles, and the authors reported a geometric tortuosity $\lambda = 1.4-1.5$ irrespective of  the size and the shape of the ECS model. In \cite{TAO2005525} the authors employed a variety of ECS models based on an array of cubic cells containing open rectangular cavities that provided the ECS with dead-end microdomains.  Monte Carlo simulations  demonstrated that the tortuosity $\lambda$ is relatively independent of the shape of the cavities and the number of cavities per cell. The tortuosity estimated from these simulations could reach, and even substantially exceed, the experimental value of 1.6 at an ECS volume fraction $\eta = 0.2$.
	
	In \cite{JIN20081785}, the authors modeled ECS diffusion of arbitrary-size solute molecules in three dimensions for a cell array  with varying cell size, cell-cell gap dimensions, and comprising intracellular lakes, i.e., expanded regions of brain ECS at multi-cell contact points. The model predicted $D/D^* \sim 1.7$ and also  yielded predictions for the sensitivity of solute diffusion to $\eta$, cell size, solute molecule size and ECS cell-cell contact geometry. An important finding of the simulations was that solute movement in the ECS, despite its complex and crowded geometry, was generally described well by Brownian non-anomalous diffusion. Similarly, a random-walk model to simulate macromolecule diffusion in brain ECS in three dimensions was developed in  \cite{Verkman}. The input of the model included ECS volume fraction, cell size, cell-cell gap geometry, intracellular lake and molecular size of the diffusing solute. The model accurately predicted $D/D^*$ for several solute sizes.

	Another approach, outlined in \cite{HRABE20041606}, assume a macroscopically homogeneous, but not necessarily isotropic, environment composed of two phases representing the cellular obstacles and the extracellular space occupying volume fraction $\eta$ around them. The authors derive a connection between a probabilistic fine scale diffusion model and a macroscopic diffusion tensor in the mean field model, thus shedding light on the interpretation of the macroscopic parameter in terms of the fine structure.
	
	A modified version of Fick's classical equation that has been used to describe diffusion in ECS on a macroscopic scale includes parameters for volume fraction and tortousity.  The derivation of the modified equation is based on a volume averaging process over an appropriate representative elementary volume (REV) of brain tissue large enough to contain cellular and ECS elements to justify an average, yet sufficiently small to represent local values of the significant variables. We refer to \cite{2001RPPh...64..815N} for details.
	
	To keep the model description simple, in the following discussion we consider the diffusion of one substance, generalizing the formalism later to vectors of concentrations.
	Following \cite{Sykova_Nicholson}, we write the modified diffusion equation for a substance in ECS linking $\eta$ and $\lambda$  as
	\begin{equation}\label{diffusion_eqn2}
	\eta\frac{\partial u}{\partial t} = \nabla\cdot D^*\nabla u +  q , 
	\end{equation}
	where $D^* = D/\lambda^2$, $u(x, t)$ is the actual concentration of the substance in the ECS, and $D$ is the diffusion coefficient in free medium. The term $q$ is  a  source/sink term accounting for a release/uptake of the substance by the cellular domains or the blood domain through the blood-brain-barrier at the location $x$. The assumptions (c) and (d) imply that 
	\begin{equation}\label{locality}
	q = q(x,t;u(x,t)),
	\end{equation}
	that is, if $q$ depends on the concentration $u$, the dependence is local, i.e., the dependency on $u$ is restricted to the value of $u$ at the point $x$ and at time $t$.  We omit the effect of bulk flow in the ECS, considered to be insignificant over the spatial and temporal scales of interest here. The presence of the volume fraction in the diffusion model is necessary for conservation of mass. Observe that $q$ depends on the concentration, typically in a non-linear manner; the functional form of the term $q$  will be derived later.

	\subsection{Diffusion in astrocytes}
	
	Numerous recent studies in the literature have focused on the role of astrocyte syncytia in brain, by which we mean astrocytes interconnected by gap junctions.  Gap junctions are specialized intracellular junctions whereby adjacent cells are connected through protein channels \cite{terman2019modeling}. These channels connect the cytoplasms of adjacent cells, allowing the passage of molecules, ions and electrical signals without having to go through the extracellular fluid surrounding the cell.  In line with these findings, our spatially distributed model accounts for diffusion in the astrocyte domain.
	
	A detailed modeling of gap junctions would require a description of the complete network connectivity of the cells, an approach probably unrealistic because of the complexity of the network. Following the approach of \cite{Tuttle, Shapiro, Connell, MORI201594}, we describe the movement of molecules through gap junctions through the diffusion coefficient in a modified diffusion equation similar to  (\ref{diffusion_eqn2}). In this approach $D^*$ is interpreted as the driving force of the gap junction, and is therefore replaced by $D^*_a$, given as
	\begin{equation}\label{gap_junction}
	D^*_a = s \, D^*
	\end{equation}
	where $s$ describes the gap junction strength, a parameter that can be modified.
	
	\subsection{Model reduction}
	
	In this subsection we present a step by step reduction of the three dimensional model of brain metabolism introduced earlier to a computationally more feasible two dimensional one. The model reduction is motivated by the observation that in portions of the brain, for example the cerebral cortex, the axons and dendrites of the neurons are highly organized and orthogonal to the cortical surface. This implies that, at each instance, at each time, it is reasonable to expect the concentrations of metabolites to remain fairly constant along the neuronal axons, therefore suggesting that the model can be reduced by integrating along that direction. 
	
	More specifically,  we denote by $u(x,t)$ the concentration of the biochemical species of interest in a diffusive compartment with volume fraction $\eta$, and use the form of the diffusion equation given in (\ref{diffusion_eqn2}) to write
	\begin{equation}\label{diffusion}
	\eta  \frac{\partial u(x,t)}{\partial t} = \nabla\cdot D\nabla u(x,t) + q(x,t;u(x,t)),
	\end{equation}
	where $D\in\R^{3\times 3}$ is the diffusion tensor characteristic for the substance in the current compartment.
	We assume that the material in the computational domain $\Omega$ is orthotropic, that is, the diffusion tensors are block diagonal,
	\begin{equation}\label{orthotropic}
	D = \left[\begin{array}{cc} \kappa & 0 \\ 0 & k\end{array}\right],
	\end{equation}
	where $\kappa\in\R^{2\times 2}$ is the diffusion tensor in the $xy$-direction parallel to the cortical surface, and $k>0$ is the diffusion coefficient in the orthogonal $z$-direction. The boundary conditions for $\Omega$ are assigned so as to be in agreement with the following assumptions:
	\begin{itemize}
		\item[(a)] The top boundary, $z=h$, represents the surface of the brain, or pial membrane. Since there is no diffusion out of this surface, we set
		\begin{equation*}\label{bdry top}
		n\cdot D \nabla u\big|_{z=h}  = k\frac{\partial u}{\partial z}\bigg|_{z=h}= 0,  \quad \mbox{where $n = e_3$.}
		\end{equation*}
		\item[(b)] The bottom boundary, $z=0$, marks the border between gray and white matter. We assume that there is no flux between gray and white matter, and set
		\[
		n\cdot D \nabla u\big|_{z=0}  = k\frac{\partial u}{\partial z}\bigg|_{z=0}= 0, \quad \mbox{where $n = e_3$.}
		\] 
		\item[(c)] On the side boundary ${\partial B}\times[0,h]$, we put a Robin boundary condition, which is equivalent to assuming that the flux through the boundary depends linearly on the concentration gradient across the boundary, the equivalent of Fick's law for compartment models. If the outside concentrations $V$ are known, we can write
		\[
		n\cdot D\nabla u\big|_{{\partial B}\times[0,h]} =  - \lambda\big(u \big|_{{\partial B}\times[0,h]} - V\big)
		\] 
		for some $\lambda>0$.
	\end{itemize}
	
	To reduce the model, we set $x' = (x_1,x_2)$ and we integrate out the third dimension, defining
	\[
	\overline u = \frac 1h\int_0^h u(x',z) dz,\quad \overline q = \frac 1h\int_0^h q (x',z) dz.
	\]
	
	Averaging equation (\ref{diffusion})  over the vertical direction yields
	\[
	\eta \frac{\partial\overline u}{\partial t}  = \frac 1h \int_0^h \nabla\cdot D\nabla u \,dz + \overline q,
	\] 
	Observe that from the orthothrophic assumption it follows that 
	\[
	\nabla\cdot D\nabla u = \frac{\partial}{\partial z} k \frac{\partial u}{\partial z} + \nabla'\cdot \kappa \nabla' u,
	\]
	where
	\[
	\nabla' = \frac{\partial}{\partial x_1} e_1 +  \frac{\partial}{\partial x_2} e_2.
	\]
	Furthermore, letting $\kappa = \kappa(x')$, we have
	\begin{equation}
	\frac 1h \int_0^h \nabla\cdot D\nabla u \,dz = \frac 1h D \frac{\partial u}{\partial z} \bigg|_{z=0}^{z=h} + \nabla'\cdot \kappa \nabla' \overline u = \nabla'\cdot \kappa\nabla' \overline u,\phantom{XXXXXXX}
	\end{equation}
	as implied by assumptions (a) and (b).
	
	In summary, in astrocyte and extracellular space the system of three-dimensional governing equations over $\Omega$ can be reduced to a system of two-dimensional equations over $B$ of the form
	\begin{equation}\label{diffusion_eqn3}
	\eta \frac{\partial\overline u}{\partial t}  =  \nabla'\cdot \kappa\nabla' \overline u  + \overline q.
	\end{equation}
	In the neuronal domain ($j=3$), where there is no diffusion, we have:
	\[
	\eta \frac{\partial\overline u}{\partial t}  =  \overline q.
	\] 
	
	To retain the local form of the source term $q$ in the reduced model, we make an approximation
	\[
	\overline q(x',t) = \frac 1h\int_0^h q\big(x',z,t;u(x',z,t)\big) dz \approx \overline q\big(x',t;\overline u(x',t)\big),
	\]
	that is, we assume that the average source can be expressed as a function of the average concentration value at the point where $\overline q$ is evaluated.
	
	The form of the equations for the  blood domain will be discussed after we integrate the metabolism into the system. In the remainder of the paper, we will consider only the reduced model. To simplify the notation, we will simply write $u$ instead of $\overline u$, $\nabla$ instead of $\nabla'$ etc.
	
	Before describing how  the metabolism is embedded into the system, we discuss the finite element discretization of the model to obtain a predictive computational scheme. 
	
	\subsection{Finite element discretization}
	
	To discretize the diffusion equation in the spatial direction using the finite element method, assume that the domain $B$ is approximated by a polygon subdivided into conforming triangular elements. The vertices of the elements and the midpoints of the edges are the nodes $p_k$, $1\leq k\leq N$. We use second order nodal basis functions, denoted by $\psi_k.$ These are piecewise second order polynomial Lagrange basis functions with the property
	\[
	\psi_k(p_\ell) = \delta_{k\ell},\quad 1\leq k,\ell\leq N.
	\]
	Consider the diffusion equation (\ref{diffusion_eqn3}) after dimension reduction in one of the subdomains $\Omega^j$.   
	Define the inner product
	\[
	\langle u,v\rangle = \int_B u(x) v(x) dx.
	\]    
	Let $v= v(x)$ be a test function defined on $B$. The inner product of (\ref{diffusion_eqn3}) with the test function $v$ yields
	\begin{eqnarray*}
		&& \left\langle v,\eta \frac{\partial u}{\partial t}\right\rangle = \eta \frac{d}{dt}\langle v, u\rangle  \\
		&&\phantom{XXX}= \langle v,\nabla\cdot \kappa\nabla u\rangle + \langle v,q\rangle \\
		&&\phantom{XXX}= \int_{\partial B} v n\cdot\kappa\nabla u dS - \langle\nabla v,\kappa\nabla u\rangle + \langle v,q\rangle \\
		&&\phantom{XXX}=  \lambda \int_{\partial B} v (V- u) dS  - \langle\nabla v,\kappa\nabla u\rangle + \langle v,q\rangle. 
	\end{eqnarray*} 
	We write a Galerkin approximation of the concentrations and the source term as
	\[
	u(x,t) \approx \sum_{k=1}^N u(p_k,t) \psi_k(x), \quad q(x,t) \approx \sum_{k=1}^N q(p_k,t) \psi_k(x),
	\]
	and we choose the test function $v$ to be one of the basis functions, $v = \psi_\ell$. This leads us to the equation
	\begin{equation}
	\eta\sum_{k=1}^n \frac{d}{dt}u(p_k,t) \langle\psi_\ell,\psi_k\rangle =  \lambda\int_{\partial B}\psi_\ell V dS - \\ 
	\lambda \sum_{k=1}^N u(p_k,t) \int_{\partial B} \psi_\ell(x)\psi_k(x) dS  
	-\sum_{k=1}^N u(p_k,t)\langle\nabla\psi_\ell,\kappa\nabla\psi_k\rangle \\
	+\sum_{k=1}^N q(p_k,t)\langle\psi_\ell,\psi_k\rangle.
	\end{equation}
	To express the equation in matrix notation we define the mass matrix $\mM$, and the stiffness matrix $\mK$,
	\begin{eqnarray*}
		\mM_{\ell k} &=& \langle\psi_\ell,\psi_k\rangle, \\
		\mK_{\ell k} &=& \langle\nabla \psi_\ell,\kappa\nabla \psi_k\rangle,
	\end{eqnarray*}
	and the boundary mass matrix $\mB$, whose entries 
	\[
	\mB_{\ell k} =  \int_{\partial B} \psi_\ell(x)\psi_k(x) dS,
	\]
	vanish if the nodes $p_k$ and $p_\ell$ do not lie on the boundary $\partial B$. Further, we introduce the vectors
	\[
	{\bf u}(t) = \left[\begin{array}{c} u(p_1,t) \\ \vdots \\ u(p_k,t)\end{array} \right], \quad  \bq(t) = \left[\begin{array}{c} q(p_1,t) \\ \vdots \\ q(p_N,t)\end{array} \right],
	\] 
	and the boundary vector ${\bf V}$ with entries
	\[
	V_\ell =  \lambda\int_{\partial B}\psi_\ell V dS,
	\]
	that vanish for nodes outside the boundary.
	With these notations, we obtain the following system of ordinary differential equations in the diffusive compartments, astrocyte and extracellular space,
	\begin{equation}\label{single diffusion}
	\eta \,\mM \frac{d\bu}{dt} = \lambda \bV -\big(\mK + \lambda \mB\big)\bu + \mM \bq.
	\end{equation}
	In the neuron and blood compartments, where no diffusion is assumed, the governing equations reduce to 
	\begin{equation}\label{single diffusion2}
	\eta \,\mM \frac{d\bu}{dt} = \mM \bq.
	\end{equation}
	
	The next step is to vectorize these equations to include all metebolites, and to couple the equations through the fluxes of metabolites from one compartment to another, which constitutes part of the source terms $\bq$.
	
	\subsection{Vectorization}
	
	We are now ready to couple the four compartments and to include all metabolites of interest, keeping in mind that not all substances are present in all compartments. To simplify the bookkeeping, we introdue the following indexing convention: Since the discretization of each compartment is the same, the number $N$ of nodes in the finite element discretization is the same for every substance in every compartment.  We introduce the labeling of the vectors,
	\[
	\bu^\ell_k \in\R^N,
	\]
	where
	\begin{eqnarray*}
		\ell &=& \mbox{index of the compartment, $1\leq \ell\leq 4$,}\\
		k &=& \mbox{index of the metabolite.}
	\end{eqnarray*} 
	The number of the metabolites followed by the model varies from compartment to compartment. To keep the computational complexity from becoming excessively large, we will consider a rather restricted metabolic model comprising 11 metabolites, some of them appearing only in the cellular compartments, astrocyte and neuron. The full list of the metabolites in the different compartments is given in Table~\ref{table:1}.
	
	\bigskip
	\begin{table*}[htbp]
		\caption{Metabolites and compartments: a cross indicates the presence of the metabolite indexing the column in the compartment indexing the row. The second row lists the indices of the metabolites. The concentrations of the metabolites are measured in mM.}
		\centering
		\begin{tabular}{l c c c c c c c c c c c}
			\hline
			& Glc & ${\rm O}_2$ & ${\rm CO}_2$ & Lac & Glu & Gln  & Pyr & ATP & ADP & NAD$^+$ & NADH \\ 
			\hline
			& 1 & 2 & 3 & 4 & 5 & 6  & 7 & 8 & 9 & 10 & 11 \\ 
			\hline
			Blood   &$\times$  &       $\times$            & $\times$   & $\times$      &          &                &                 &              &              &                 &
			\\
			ECS           & $\times$ &      $\times$             & $\times$   &  $\times$     &  $\times$         &$\times$   &   &              &              &                 &            \\
			Neuron      & $\times$ & $\times$     & $\times$   & $\times$      & $\times$          &$\times$   & $\times$  &$\times$&$\times$&$\times$    &$\times$\\
			Astrocyte   & $\times$ & $\times$     & $\times$   & $\times$      & $\times$          &$\times$   & $\times$  &$\times$&$\times$&$\times$    &$\times$\\
			
			\hline
		\end{tabular}
		\label{table:1}
	\end{table*}
	\bigskip
	
	The concentrations are collected into a single vector, first by stacking the metabolites in each compartment together, and subsequently by stacking the compartment vectors together. We define the compartment concentration vectors as
	\[
	\bU^1 = \left[\begin{array}{c} \bu^1_1 \\ \vdots \\ \bu^1_4\end{array}\right], \quad  \bU^2 = \left[\begin{array}{c} \bu^2_1 \\ \vdots \\ \bu^2_6\end{array}\right],
	\] 
	in blood ($j=1$) and ECS ($j=2$), and
	\[
	\bU^3 = \left[\begin{array}{c} \bu^3_1 \\ \vdots \\ \bu^3_{11}\end{array}\right], \quad  \bU^4 = \left[\begin{array}{c} \bu^4_1 \\ \vdots \\ \bu^4_{11}\end{array}\right],
	\]
	in neuron ($j=3$) and in astroyte ($j=4$), yielding the composite concentration vector
	\[
	\bU = \left[\begin{array}{c} \bU^1 \\ \vdots \\ \bU^4\end{array}\right] \in \R^{32\,N}.
	\] 
	In every compartment, each metabolite has its own sink/source term denoted by $\bq^\ell_k$, with the same indexing convention as for the metabolites. Following the same procedure used to define $\bU^\ell$, we collect the source terms of each metabolite in the $\ell$th compartment into the vector $\bQ^\ell$.
	
	In the finite element model, the mass matrix $\mM$ and the boundary mass matrix $\mB$ are the same for all equations, while, since the stiffness matrix $\mK$ depends on the diffusion coefficient characteristic to the metabolite and to the compartment, we index it as $\mK^\ell_k\in\R^{N\times N}$.  For simplicity, we assume that the parameter $\lambda$ in the Robin boundary condition is the same for all compartments and metabolites.
	Therefore, in the diffusive compartments $\ell=2$ (ECS) and $\ell=4$ (astrocyte) the vectors $\bU^\ell$ satisfy the equations
	\begin{equation}\label{eq2 4}
	\eta^\ell{\mathcal M}^\ell \frac{ d \bU^\ell}{dt} = \lambda \bV^\ell -\big({\mathcal K}^\ell + \lambda {\mathcal B}^\ell\big)\bU^\ell +{\,\mathcal M}^\ell \bQ^\ell,
	\end{equation}
	where the mass matrices are obtained through a diagonal replication by means of a Kronecker product,
	\[
	{\mathcal M}^2 = \mI_6\otimes \mM, \quad   {\mathcal M}^4 = \mI_{11}\otimes \mM,
	\]
	\[
	{\mathcal B}^2 = \mI_6\otimes \mB, \quad   {\mathcal B}^4 = \mI_{11}\otimes \mB,
	\] 
	and the stiffness matrices have block diagonal structure
	\[
	{\mathcal K}^2 = \left[\begin{array}{ccc} \mK^2_1 && \\ &\ddots & \\ && \mK^2_6\end{array}\right], \quad
	{\mathcal K}^4 = \left[\begin{array}{ccc} \mK^4_1 && \\ &\ddots & \\ && \mK^4_{11}\end{array}\right].
	\] 
	We combine the governing equations in the non-diffusive compartments $\ell=1$ (blood) and $\ell=3$ (neuron) in the same manner, to get 
	\begin{equation}\label{eq1 3}
	\eta^\ell{\mathcal M}^\ell \frac{ d \bU^\ell}{dt} =  {\,\mathcal M}^\ell \bQ^\ell,
	\end{equation}
	with obvious notations.
	
	To formulate a coupled diffusion-transport-reaction system for the composite vector $\bU$, we need to model the rates of the transports between compartments as well as for the reactions.  For later reference, we write
	\[
	\bQ = \left[\begin{array}{c} \bQ^1 \\ \vdots \\ \bQ^4\end{array}\right] =  \bQ_{\rm transport}+\bQ_{\rm reaction} + \bQ_{\rm flow},
	\]
	separating the contributions from the
	transports between compartments, reactions within the cellular compartments, and the blood flow.  
	
	\subsection{Source terms}
	
	The delivery of metabolites and the clearing of waste products through blood flow, transport of metabolites from one compartment to another, and the biochemical reactions inside the cellular domains are accounted for by the source terms.
	
	\subsubsection{Transport fluxes}
	All exchanges of metabolites occur through the ECS. In our model, only the first four metabolites (glucose, oxygen, lactate and carbon dioxide) are exchanged between the ECS, blood and the cellular compartments. In addition, glutamate and glutamine are exchanged between the cellular compartments and the ECS through the synaptic cleft, constituting a part of the ECS. The glutamate-glutamine exchange is referred to as the neurotransmitter cycle, or V-cycle. For bookkeeping's sake, we number the transport fluxes as indicated in Table~\ref{table:2}, where the notation $x\rightarrow y$  indicates the flux from compartment $x$ to compartment $y$.
	
	\bigskip
	\begin{table*}[htbp]
		\captionsetup{width=.97\textwidth}
		\caption{Transports between compartments: The ECS serves as passage for transport of metabolite from one compartment to another. The table lists the index of each of the transport flux.}
		\renewcommand{\arraystretch}{1.9}
		\centering
		\begin{tabular}{l c c c c c c}
			\hline
			Transport & $ \rm blood \rightarrow ecs $ & $ \rm ecs \rightarrow blood $ & $ \rm ecs \rightarrow neuron $ & $\rm neuron \rightarrow ecs $ & $ \rm ecs \rightarrow astrocyte $ & $ \rm astrocyte \rightarrow ecs $ \\ 
			\hline
			flux   & 1  & 2 & 3 & 4   & 5 & 6 \\  
			\hline
		\end{tabular}
		\label{table:2}
	\end{table*}
	\bigskip
	
	We are now ready to assemble the source terms considering the exchange of metabolites in each nodal point $p_k$. Since the formulas are invariant from point to point, we do not indicate the spatial dependency explicitly. We use the notation
	\[
	\bphi^{(j)} = \mbox{transport fluxes $x\to y$}, \quad 1\leq j\leq 6, 
	\]
	where the index $j$ refers to the numbering in Table~\ref{table:2}. The transport fluxes enter in the definition of the total transport flux which, in each compartment, takes into account the replenishment and depletion of each substance.  
	
	\textbf{Exchanges between blood and ECS}. All four metabolites tracked in the blood compartment, $\rm Glc$, $\rm O_2$, $\rm CO_2$ and $\rm Lac$,  are exchanged with ECS.  Hence, the transport fluxes from blood to ECS, $\bphi^{(1)}$,  and from ECS to blood, $\bphi^{(2)}$, comprise the  four $N$-vectors, 
	\[
	\bphi^{(1)} = \left[\begin{array}{c} \bphi^{(1)}_{\rm Glc} \\  \bphi^{(1)}_{\rm O_2}\\ \bphi^{(1)}_{\rm CO_2}\\\bphi^{(1)}_{\rm Lac}\end{array}\right]
	=\left[\begin{array}{c} \bphi^{(1)}_1 \\  \bphi^{(1)}_2\\ \bphi^{(1)}_3\\\bphi^{(1)}_4\end{array}\right], 
	\]
	where we use the metabolite numbering of Table~\ref{table:1}, and each vector $\bphi^{(1)}_j\in\R^N$ contains the transport fluxes at the $N$ nodal points in their components. Similarly, we write
	\[
	\bphi^{(2)} = \left[\begin{array}{c} \bphi^{(2)}_{\rm Glc} \\  \bphi^{(2)}_{\rm O_2}\\ \bphi^{(2)}_{\rm CO_2}\\\bphi^{(2)}_{\rm Lac}\end{array}\right] =
	\left[\begin{array}{c} \bphi^{(2)}_1 \\  \bphi^{(2)}_2\\ \bphi^{(2)}_3\\\bphi^{(2)}_4\end{array}\right].
	\] 
	The cross-membrane passage of glucose ($k=1$) and lactate ($k=4$) occurs through the action of specialized  glucose transporters (GLUT) and monocarboxylate transporters (MCT), respectively. We express the transport fluxes in Michaelis-Menten type  form,
	\begin{equation}\label{glucose_flux}
	\bphi^{(1)}_{k} = T^{(1)}_{k}\frac{{\bf u}^1_k}{{\bf u}^1_k +M^{(1)}_{k}}, \qquad \bphi^{(2)}_{k} = T^{(2)}_{k}\frac{{\bf u}^2_k}{{\bf u}^2_k +M^{(2)}_{ k}}
	\end{equation}
	where $T^{(\cdot)}_{k}$ and $M^{(\cdot)}_{k}$ are the maximum transport rates and affinity constants, respectively. Here, the division of a vector by a vector is to be understood as component-wise. For simplicity, we assume symmetry in the maximum transport rates and affinity constants, that is, $T^{(1)}_{k} = T^{(2)}_{k}$ and $M^{(1)}_{k} = M^{(2)}_{k}$, 
	where ${\bf u}^1_k$ and ${\bf u}^2_k$ are the concentration vectors of glucose ($k = 1$) or lactate ($k = 4$) in the blood and ECS, respectively.

	Special attention must be paid to the oxygen flux,  because in blood we must account for free oxygen dissolved in plasma, and for oxygen bound to hemoglobin. The total oxygen concentration $u^1_2$ at a given node can be written according to Hill's equation \cite{keener2009mathematical}  in terms of
	the free oxygen concentration $[{\rm O}_2^{\rm b}]_{\rm free}$ as
	\[
	u^1_2 = [{\rm O}_2^{\rm b}]_{\rm free} + \underbrace{4{\rm Hct[Hb]} \frac{[{\rm O}_2^{\rm b}]^n_{\rm free}}{K_H^n + [{\rm O}_2^{\rm b}]^n_{\rm free}}}_{[{\rm O}_2^{\rm b}]_{{\rm bound}}} = F\left([{\rm O}_2^{\rm b}]_{\rm free}\right),
	\]
	where Hct is the hematocrit, [Hb] is the hemoglobin concentration in plasma, $n = 5/2$ is Hill's constant, and $K_H$ is the hemoglobin affinity.
	
	We express the net oxygen flux from blood to ECS in terms of free oxygen, using a modified Fick's law of the form
	\begin{align}\label{oxygen_flux}
	\bphi^{(1)}_2 - \bphi^{(2)}_2 & = \lambda_{2}^{(1)} \left( [\rm O_2^b]_{\rm free} - {\bf u}^2_2\right)^\kappa \notag\\
	& = \lambda_{2}^{(1)}  \left(F^{-1}\left({\bf u}^1_2\right) - {\bf u}^2_2\right)^\kappa
	\end{align}
	with $\kappa = 0.1$. Here, $ F^{-1} $ is the inverse of the vectorized function mapping the free oxygen concentrations to the total concentrations, and ${\bf u}^1_2$ and ${\bf u}^2_2$ are the concentration vectors of the total oxygen in the blood and oxygen in the ECS. The above modification of the standard linear Fick's law was proposed in \cite{CALVETTI2018238} to account for the observed limited increase of oxygen uptake during activity, thus leading to the observed decrease rather than increase of the oxygen-glucose index (OGI) during neuronal activity. The physiological explanation for the somewhat unexpected OGI behavior could be related to the transport mechanism through gas channels: According to the classical theory of Overton \cite{overton}, the exchange of gases is modeled as passive diffusion through the lipid phase of the cell membrane, although there is evidence  \cite{boron} that the transport may take place, at least partly, through specified gas channels.
	However that further investigations are beyond the scope of the present paper.
	
	Finally, we model the $\rm CO_2$ ($k=3$) flux using Fick's law as
	
	\[
	\bphi^{(1)}_{3} -\bphi^{(2)}_{3}= \lambda_3^{(1)}  \left({{\bf u}^1_3} - {\bf u}^2_3\right) ,
	\]
	where ${\bf u}^1_3$ and ${\bf u}^2_3$ are the concentration of $\rm CO_2$ in blood and ECS. This model may need a modification similar to that of the oxygen transport, but for the purposes of the present model, Fick's standard formulation is adequate, as the carbon dioxide has no effect on the metabolic activity.

	\textbf{Exchanges between cellular compartments and ECS}. Six metabolites are exchanged between  ECS, and  neuron and astrocyte: $\rm Glc$, $\rm O_2$, $\rm CO_2$, $\rm Lac$, $\rm Glu$ and $\rm Gln$. The transport fluxes from ECS to neuron, $\bphi^{(3)}$, and neuron to ECS, $\bphi^{(4)}$, therefore consist of 6 $N$-vectors, 
	\[
	\bphi^{(3)} = \left[\begin{array}{c} \bphi^{(3)}_{\rm Glc} \\  \bphi^{(3)}_{\rm O_2}\\ \bphi^{(3)}_{\rm CO_2}\\ \bphi^{(3)}_{\rm Lac} \\
	\bphi^{(3)}_{\rm Glu} \\ \bphi^{(3)}_{\rm Gln}
	\end{array}\right]  = \left[\begin{array}{c} \bphi^{(3)}_1 \\  \bphi^{(3)}_2\\ \bphi^{(3)}_3\\ \bphi^{(3)}_4 \\
	\bphi^{(3)}_5 \\ \bphi^{(3)}_6
	\end{array}\right],
	\]
	the vector $\bphi^{(4)}$ having the same structure. Similarly, the transport fluxes from ECS to astrocyte, $\bphi^{(5)}$ and from astrocyte to ECS, $\bphi^{(6)}$, have the same form.
	The functional expressions for glucose and lactate are analogous to those of $\bphi^{(1)}$ and $\bphi^{(2)}$,
	\[
	\bphi^{(3)}_{k} = T^{(3)}_{k}\frac{{\bf u}^2_k}{{\bf u}^2_k +M^{(3)}_{\rm k}}, \qquad \bphi^{(4)}_{k} = T^{(4)}_{k}\frac{{\bf u}^3_k}{{\bf u}^3_k +M^{(4)}_{k}},
	\] 
	and
	\[
	\bphi^{(5)}_{k} = T^{(5)}_{k}\frac{{\bf u}^2_k}{{\bf u}^2_k +M^{(5)}_{\rm k}}, \qquad \bphi^{(6)}_{k} = T^{(6)}_{k}\frac{{\bf u}^4_k}{{\bf u}^4_k +M^{(6)}_{k}},
	\] 
	for $ k = 1,4$. For the gas exchanges of oxygen and carbon dioxide, we use the linear Fick's law
	\[
	\bphi^{(3)}_{k} =  \lambda_{k}^{(3)} {\bf u}^2_k, \qquad \bphi^{(4)}_{k} =  \lambda_{k}^{(4)} {\bf u}^3_k, \qquad k = 2,3,
	\] 
	and
	\[
	\bphi^{(5)}_{k} =  \lambda_{k}^{(5)} {\bf u}^2_k, \qquad \bphi^{(6)}_{k} =  \lambda_{k}^{(6)} {\bf u}^3_k, \qquad k = 2,3,
	\] 
	What remains to be described are the formulas for the glutamine-glutamate cycling between the neuron and the astrocyte.
	
	The glutamate-glutamine cycle, also known as the V-cycle, describes how glutamate is released by the glutamatergic neuron into the cleft, from where it needs to be removed rapidly by astrocytes via sodium-dependent excitatory amino acid transporters (EAAT), active transporters that take up the glutamate into the cell against its concentration gradient. 
	Once in astrocyte,  glutamate is synthesized to glutamine through the glutamine synthesis reaction. Glutamine is released into the cleft and taken up by  pre-synaptic neuron via amino acid transporter systems that depend on $\rm Na^+/K^+ - ATPase$. In pre-synaptic neuron, ammonium is released from glutamine via phosphate activated glutaminase producing glutamate. The released ammonium is transported to astrocyte, probably in the form of ammonia  \cite{10.3389/fendo.2013.00137}, and further recycled by astrocytes for the amidation of glutamate by glutamine synthetase to form glutamine, completing the cycle.
	
	Glutamate released into the cleft does not enter the neuron, and no glutamine is effluxed from neuron, therefore
	\[
	\bphi^{(3)}_{\rm Glu} =\bphi^{(3)}_{5} = \mO_{N\times 1}, \qquad \bphi^{(4)}_{\rm Gln}= \bphi^{(4)}_6 = \mO_{N\times 1},
	\]
	where $\mO_{N\times 1}$ denotes an $N$-vector of zeros.
	
	The rate at which glutamine enters the neuron is expressed in Michaelis-Menten type form,
	
	\[
	\bphi^{(3)}_{6}  = T^{(3)}_{6} \frac{{\bf u}^2_6}{{\bf u}^2_6 +M^{(3)}_{6}},
	\]  
	where ${\bf u}^2_6$ is the concentration of glutamine in ECS.
	Similarly, the mathematical formula for the rate at which glutamate flows out of the presynaptic neuron is
	\[
	\bphi^{(4)}_{5}  = T^{(4)}_{5} \frac{{\bf u}^3_5}{ {\bf u}^3_5 +M^{(4)}_{5} },
	\]
	where ${\bf u}^3_5$ is the concentration of glutamate in the neuron. A modification of this flux using an activity function will be introduced when we discuss how we simulate neuronal activity.
	
	Moreover, since we assume that there is no glutamate efflux from and glutamine influx into astrocyte, we write
	\[
	\bphi^{(5)}_{6} = \mO_{N\times 1},\qquad \bphi^{(6)}_{5} = \mO_{N\times 1},
	\]
	and for the glutamate flux into astrocyte, we use  the Michaelis-Menten type expression
	\[
	\bphi^{(5)}_{5}  = T^{(5)}_{5}\frac{{\bf u}^2_5}{{\bf u}^2_5 +M^{(5)}_{5}},
	\]     
	where ${\bf u}^2_5$ is the concentration of glutamate in ECS. 
	The expression for rate of efflux of glutamine from astrocyte is 
	\[
	\bphi^{(6)}_{6}  = T^{(6)}_{6}\frac{{\bf u}^4_6}{{\bf u}^4_6 +M^{(6)}_{6}},
	\]
	where ${\bf u}^4_6$ is the concentration of glutamine in astrocyte.
	
	We stack all the fluxes together into the combined transport flux vector,
	\[
	{\bphi} = \left[\begin{array}{c} {\bphi}^{(1)} \\ \vdots \\ {\bphi}^{(6)}\end{array}\right] \in\R^{(2\times 4+4\times 6) N} = \R^{32\,N}.
	\]
	and write the contributions of the transports to the source vector as
	\begin{equation}\label{transport term}
	\mQ_{\rm transport} = \mF \bphi \in \R^{\rm 32 N},
	\end{equation}
	where the matrix $\mF\in\R^{32 N\times 32 N}$ has the block structure shown in Table~\ref{matrix F}, with the matrices $\mE^{\rm b}$ and $\mE^{\rm ECS}$ defined as
	\[
	\mE^{\rm b} = \left[\begin{array}{c} \mI_{4N} \\ \mO_{2N\times 4N}\end{array}\right] , \qquad 
	\mE^{\rm ECS} = \left[\begin{array}{c} \mI_{6N} \\ \mO_{5N\times 6N}\end{array}\right].
	\]
	
	\begin{table*}[htbp!]
		\begin{align*}
		\mF &= \left[\begin{array}{llllll} \phantom{-} \mI_{4N} &  -\mI_{4N} & \phantom{-}\mO_{4N\times 6N} &\phantom{-}\mO_{4N\times 6N}  & \phantom{-}\mO_{4N\times 6N} & \phantom{-}\mO_{4N\times 6N}  \\
		-\mE^{\rm b} & \phantom{-}E^{\rm b} & \phantom{-}\mI_{6N} & -\mI_{6N} & \phantom{-}\mI_{6N} & -\mI_{6N} \\
		\phantom{-}\mO_{11N\times 4N} & \phantom{-}\mO_{11N\times 4N} & -\mE^{\rm ECS} & \phantom{-}\mE^{\rm ECS} & \phantom{-}\mO_{11N\times 6N}& \phantom{-}\mO_{11N\times 6N} \\
		\phantom{-}\mO_{11N\times 4N} & \phantom{-}\mO_{11N\times 4N}  &\phantom{-}\mO_{11N\times 6N} & \phantom{-}\mO_{11N\times 6N} &-\mE^{\rm ECS} & \phantom{-}\mE^{\rm ECS}\end{array}
		\right] \in\R^{32N\times 32 N},
		\end{align*}
		\caption{\label{matrix F} The matrix $\mF$ in formula (\ref{transport term}).}
	\end{table*}
	
	\begin{table*}[htbp]
		\captionsetup{width=1\textwidth}
		\caption{Transports: List of relevant parameters in the mathematical expression of transport rates of the six metabolites exchanged between compartments, their units and the values used in the computed experiments. $T^{(\ell)}_{\rm Met}$ and $M^{(\ell)}_{\rm Met}$ are the maximum transport rate and affinity constant, respectively. in the Michaelis-Menten equations for the transport $\ell$ of the metabolite ${\rm Met}$. $\lambda_{\rm Met}^{(\ell)}$ is the parameter in the Fick's law expression for the diffusion of oxygen and $\rm CO_2$ for transport flux $\ell$.}
		\centerline{
			\renewcommand{\arraystretch}{1.4}
			\begin{tabular}{ccccccccccccc}
				\hline
				\multirow{2}{*}{Transport}& &\multicolumn{3}{c}{$\rm Blood\leftrightarrow ecs$}& &\multicolumn{3}{c}{$\rm Neuron\leftrightarrow ecs$}& &\multicolumn{3}{c}{$\rm Astrocyte\leftrightarrow ecs$}\\
				\cline{3-5} \cline{7-9} \cline{11-13}
				& &Parameter & Value & Units & &Parameter & Value & Units & &Parameter & Value & Units\\
				\hline
				\rm Glc& & $T^{(1)}_{\rm Glc}$ & 1.02 & mM/min & &$T^{(3)}_{\rm Glc}$ & 5000 & mM/min & & $T^{(5)}_{\rm Glc}$ & 5000 & mM/min \\
				& & $M^{(1)}_{\rm Glc}$ & 4.7 & mM/min & &$M^{(3)}_{\rm Glc}$ & 0.4 & mM/min & & $M^{(5)}_{\rm Glc}$ & 12500 & mM/min \\\\
				$\rm O_2$& & $\lambda_{\rm O_2}^{(1)}$ & 2.43 & $\rm mM^{1-\kappa}/min$ & & $\lambda_{\rm O_2}^{(3)}$ & 56.63 & mM/min & & $\lambda_{\rm O_2}^{(5)}$ & 40.73 & mM/min \\\\
				$\rm CO_2$& & $\lambda_{\rm CO2}^{(1)}$ & 0.718 & mM/min & & $\lambda_{\rm CO2}^{(3)}$ & 0.224 & mM/min & & $\lambda_{\rm CO2}^{(5)}$ & 0.0275 & mM/min \\\\
				\rm Lac& & $T^{(1)}_{\rm Lac}$ & 10 & mM/min & &$T^{(3)}_{\rm Lac}$ & 4000 & mM/min & & $T^{(5)}_{\rm Lac}$ & 4000 & mM/min \\
				& & $M^{(1)}_{\rm Lac}$ & 5 & mM/min & &$M^{(3)}_{\rm Lac}$ & 0.4 & mM/min & & $M^{(5)}_{\rm Lac}$ & 0.4 & mM/min \\\\
				\rm Glu& &  &  &  & &$T^{(3)}_{\rm Glu}$ & 2.3614 & mM/min & & $T^{(5)}_{\rm Glu}$ & 1.348 & mM/min \\
				& &  &  &  & &$M^{(3)}_{\rm Glu}$ & 97.7431 & mM/min & & $M^{(5)}_{\rm Glu}$ & 3.57e-5 & mM/min \\\\
				\rm Gln& & &  &  & &$T^{(3)}_{\rm Gln}$ & 2.3560 & mM/min & & $T^{(5)}_{\rm Gln}$ & 2.3614 & mM/min \\
				& & & & & &$M^{(3)}_{\rm Gln}$ & 7e-5 & mM/min & & $M^{(5)}_{\rm Gln}$ & 0.0698 & mM/min \\
				\hline
			\end{tabular}
		}
		\label{table:6}
	\end{table*}
	
	\subsubsection{Reaction fluxes}
	Biochemical reactions only occur in neuron and astrocyte. Some of the reactions that we consider in our model are representative of sequences of reactions with intermediate species not included in the model. The complete list of reactions included in the model is given in Table~\ref{table:3}.
	
	\begin{table*}[htbp]
		\captionsetup{width=.95\textwidth}
		\caption{List of the reactions included in the model, and the corresponding abbreviations. Lactate dehydrogenase is the only reversible reaction in the model.}
		\centerline{
			\begin{tabular}{l l l l}
				\hline
				Name & Neuron & Astrocytes & Reaction \\
				\hline
				Glycolysis (Gcl) & $\bgamma^{3}_1$ & $\bgamma^{4}_1$ &
				${\rm Glc}+2\,{\rm NAD}^+ +2\, {\rm ADP} \rightarrow 2\,{\rm Pyr} +2\,{\rm NADH} +2\,{\rm ATP}$ \\
				Lactate dehydrogenase (LDH1) & $\bgamma^{3}_2$ & $\bgamma^{4}_2$ & ${\rm Pyr}+{\rm NADH}\rightarrow {\rm Lac} + {\rm NAD}^+$ \\
				Lactate dehydrogenase (LDH2) &  $\bgamma^{3}_3$ & $\bgamma^{4}_3$ & ${\rm Lac} + {\rm NAD}^+  \rightarrow {\rm Pyr}+{\rm NADH}$ \\
				Tricarboxylic acid cycle (TCA) & $\bgamma^{3}_4$ & $\bgamma^{4}_4$ & $ {\rm Pyr}+{\rm ADP}+5\,{\rm NAD}^+ \rightarrow 3\,{\rm CO}_2 +{\rm ATP} + 5\,{\rm NADH}$ \\
				Oxidative phosphorylation (OXPhos) &  $\bgamma^{3}_5$ & $\bgamma^{4}_5$ & $ {\rm O}_2+ 2\;{\rm NADH}+5\,{\rm ADP} \rightarrow 2\,{\rm NAD}^+ + 5\,{\rm ATP} + 2\;{\rm H_2O}$\\
				Phosphate activated glutaminase (PAG) & $\bgamma^{3}_6$ & - & $ {\rm Gln}\rightarrow {\rm Glu}$ \\
				Glutamine synthetase (GS) & - & $\bgamma^{4}_7$ & ${\rm Glu} + {\rm ATP}\rightarrow {\rm Gln}+{\rm ADP}$ \\
				ATP dehydrogenase (ATPase) & $\bgamma^{3}_8$ & $\bgamma^{4}_8$ & $ {\rm ATP}\rightarrow{\rm ADP}$\\
				\hline
			\end{tabular}
		}
		\label{table:3}
	\end{table*}
	
	The stoichiometric matrices for neuron and for astrocyte are almost identical $11\times 7$ matrices (one reaction is deleted in astrocyte (PAG), one in neuron (GS). We denote those matrices by $\mS^3 \in\R^{11\times 7}$ and $\mS^4 \in\R^{11\times 7}$, respectively, given by
	\[
	\mS^3 = \left[\begin{array}{rrrrrrrr}
	-1 & 0 & 0 & 0 & 0 & 0 & 0\\
	0 & 0 & 0 & 0 & -1 & 0 & 0\\
	0 & 0 & 0 & 3 & 0 & 0 & 0\\
	0 & 1 & -1 & 0 & 0 & 0 & 0\\
	0 & 0 & 0 & 0 & 0 & 1 & 0\\
	0 & 0 & 0 & 0 & 0 & -1 & 0\\
	2 & -1 & 1 & -1 & 0 & 0 & 0\\
	2 & 0 & 0 & 1 & 5 & 0 & -1\\
	-2 & 0 & 0 & -1 & -5 & 0 & 1\\
	-2 & 1 & -1 & -5 & 2 & 0 & 0\\
	2 & -1 & 1 & 5 & -2 & 0 & 0
	\end{array}\right]
	\]
	and
	\[
	\mS^4 = \left[\begin{array}{rrrrrrrr}
	-1 & 0 & 0 & 0 & 0 & 0 & 0\\
	0 & 0 & 0 & 0 & -1 & 0 & 0\\
	0 & 0 & 0 & 3 & 0 & 0 & 0\\
	0 & 1 & -1 & 0 & 0 & 0 & 0\\
	0 & 0 & 0 & 0 & 0 & -1 & 0\\
	0 & 0 & 0 & 0 & 0 & 1 & 0\\
	2 & -1 & 1 & -1 & 0 & 0 & 0\\
	2 & 0 & 0 & 1 & 5 & -1 & -1\\
	-2 & 0 & 0 & -1 & -5 & 1 & 1\\
	-2 & 1 & -1 & -5 & 2 & 0 & 0\\
	2 & -1 & 1 & 5 & -2 & 0 & 0\\
	\end{array}\right].
	\]
	We extend the stoichiometric matrices for the $N$ nodes through Kronecker products, defining
	\[
	\mSigma^3 = \mS^3\otimes\mI_N \in\R^{11N\times 7N}, \quad  \mSigma^4 = \mS^4\otimes\mI_N \in\R^{11N\times 7N}.
	\]
	
	Let $\bgamma^{3}_k\in\R^N$  denote the flux of the $k$th reaction in neuron, and $\bgamma^{4}_k\in\R^N$ the $k$th reaction in astrocyte, and define the combined reaction vectors as 
	\[
	\bgamma^{3} = \left[\begin{array}{c} \bgamma^{3}_1 \\ \vdots  \\ \bgamma^{4}_5 \\ \bgamma^{3}_6 \\ \bgamma^{3}_8\end{array}\right]\in\R^{7N}, \qquad \bgamma^{4} = \left[\begin{array}{c} \bgamma^{4}_1 \\ \vdots  \\ \bgamma^{4}_5 \\ \bgamma^{4}_7 \\ \bgamma^{4}_8\end{array}\right]\in\R^{7N},
	\]
	and write the reaction flux contribution to the ODE system as
	\[
	\bQ_{\rm reaction} = \left[\begin{array}{c} \mO_{4N\times 1} \\ \mO_{6N\times 1} \\  \mSigma^3\bgamma^{3} \\ \mSigma^4\bgamma^{4}\end{array}\right] =
	\mV \left[\begin{array}{c}\bgamma^3 \\ \bgamma^4\end{array}\right],
	\]
	where the parsing matrix $\mV$ is given by
	\[
	\mV = \left[\begin{array}{cc}  \mO_{4N\times 7N} &  \mO_{4N\times 7N} \\
	\mO_{6N\times 7N} &  \mO_{6N\times 7N}  \\
	\mSigma^3 & \mO_{11 N\times 7N} \\
	\mO_{11 N\times 7N}  & \mSigma^4
	\end{array}\right].
	\] 
	The functional forms of the reaction fluxes are listed in Table~\ref{table:4}. 
	
	\begin{table*}[htbp]
		\captionsetup{width=.97\textwidth}
		\caption{Functional form of reaction fluxes in neuron and astrocyte together with parameter values and units. In the table, $r^\ell$ denotes the phosphorylation rate, given as the fraction of $\rm ATP$ to $\rm ADP$, while $p^\ell$ is the redox rate, given as the ratio of $\rm NADH$ to $\rm NAD^+$,  $\ell=3$ is for neuron and $\ell=4$ is for astrocyte.}
		\centerline{
			\begin{tabular}{ p{8cm} p{2cm} p{2cm} p{2cm} p{2cm}}
				\hline
				Reaction & Parameter & Units & Neuron & Astrocyte \\
				\hline
				Gcl: $ V^\ell_{\rm Gcl} \frac{1/r^\ell}{1/r^\ell ~+~ \nu^\ell_{\rm Gcl}} 
				\frac{1/p^\ell}{1/p^\ell ~+~ \mu^\ell_{\rm Gcl}} 
				\frac{{\bf u}_{1}^\ell}{{\bf u}_{1}^\ell ~+~ K^\ell_{\rm Gcl}}$  & $V^\ell_{\rm Gcl}$ & mM/min & $15.30$ & $15.07$\\
				& $K^\ell_{\rm Gcl}$ & mM & $4.60$ & $3.10$\\
				& $\mu^\ell_{\rm Gcl}$ & & $0.09$ & $0.09$\\
				& $\nu^\ell_{\rm Gcl}$ & & $10.00$ & $10.00$\\
				\hline
				LDH1: $V^\ell_{\rm LDH1} \frac{r^\ell}{r^\ell ~+~ \nu^\ell_{\rm LDH1}} 
				\frac{{\bf u}_{7}^\ell}{{\bf u}_{7}^\ell ~+~ K^\ell_{\rm LDH1}}$ & $V^\ell_{\rm LDH1}$ & mM/min & $8.62e4$ & $2.50e5$\\
				& $K^\ell_{\rm LDH1}$ & mM & $2.15$ & $6.24$\\
				& $\nu^\ell_{\rm LDH1}$ & & $0.10$ & $0.10$\\
				\hline
				LDH2: $V^\ell_{\rm LDH2} \frac{1/r^\ell}{1/r^\ell ~+~ \nu^\ell_{\rm LDH2}} 
				\frac{{\bf u}_{4}^\ell}{{\bf u}_{4}^\ell ~+~ K^\ell_{\rm LDH2}}$ & $V^\ell_{\rm LDH2}$ & mM/min & $9.48e4$ & $1.95e5$\\
				& $K^\ell_{\rm LDH2}$ & mM & $23.70$ & $48.67$\\
				& $\nu^\ell_{\rm LDH2}$ & & $10.00$ & $10.00$\\
				\hline
				TCA: $V^\ell_{\rm TCA} \frac{1/r^\ell}{1/r^\ell ~+~ \nu^\ell_{\rm TCA}} 
				\frac{1/p^\ell}{1/p^\ell ~+~ \mu^\ell_{\rm TCA}} 
				\frac{{\bf u}_{7}^\ell}{{\bf u}_{7}^\ell ~+~ K^\ell_{\rm TCA}}$ & $V^\ell_{\rm TCA}$ & mM/min & $1.80$ & $0.56$\\
				& $K^\ell_{\rm TCA}$ & mM & $1.25e-2$ & $1.24e-2$\\
				& $\mu^\ell_{\rm TCA}$ & & $0.01$ & $0.01$\\
				& $\nu^\ell_{\rm TCA}$ & & $10.00$ & $10.00$\\
				\hline
				OxPhos: $V^\ell_{\rm OxPhos} \frac{r^\ell}{r^\ell ~+~ \nu^\ell_{\rm OxPhos}} \frac{1/p^\ell}{1/p^\ell ~+~ \mu^\ell_{\rm OxPhos}} 
				\frac{{\bf u}_{2}^\ell}{{\bf u}_{2}^\ell ~+~ K^\ell_{\rm OxPhos}}$ & $V^\ell_{\rm OxPhos}$ & mM/min & $491.04$ & $153.15$\\
				& $K^\ell_{\rm OxPhos}$ & mM & $1.00$ & $1.00$\\
				& $\mu^\ell_{\rm OxPhos}$ & & $0.01$ & $0.01$\\
				& $\nu^\ell_{\rm OxPhos}$ & & $0.1$ & $0.1$\\
				\hline
				PAG (neuron) : $V^\ell_{\rm PAG}\frac{{\bf u}_{6}^\ell}{{\bf u}_{6}^\ell ~+~ K^\ell_{\rm PAG}}$ & $V^\ell_{\rm PAG}$ & mM/min & $1.18$ & -\\
				& $K^\ell_{\rm PAG}$ & mM & $3e-3$ & -\\
				\hline
				GS (astrocyte): $V^\ell_{\rm GS} \frac{p^\ell}{p^\ell ~+~ \nu^\ell_{\rm GS}} \frac{{\bf u}_{5}^\ell}{{\bf u}_{5}^\ell ~+~ K^\ell_{\rm GS}}$ & $V^\ell_{\rm GS}$ & mM/min & - & $2.36$\\
				& $K^\ell_{\rm GS}$ & mM & - & $3e-2$\\
				& $\mu^\ell_{\rm GS}$ & & - & $100.00$\\
				\hline
				ATPase : $V^\ell_{\rm ATPase}\frac{{\bf u}_{8}^\ell}{{\bf u}_{8}^\ell ~+~ K^\ell_{\rm ATPase}}$ & $V^\ell_{\rm ATPase}$ & mM/min & $18.28$ & $2.92$\\
				& $K^\ell_{\rm ATPase}$ & mM & $2.00$ & $2.20$\\
				\hline
			\end{tabular}
		}
		\label{table:4}
	\end{table*}
	
	\subsubsection{Convection by blood flow}
	
	We assume that the blood flows through each node of the finite element mesh, replenishing the glucose and oxygen concentrations and removing the waste products, lactate and carbon dioxide. At a nodal point $p_j$,  we write the convection model of the $k$th metabolite in the blood compartment as
	\[
	c_k(p_j,t) = \frac{Q(p_j,t)}{F}\big(U_k^{\rm art}(p_j,t) - u_k^1(p_j,t) \big),
	\] 
	where $Q(p_j,t)$ is the blood flow, $F$ is the mixing ratio expressing the proportion of arterial to venous blood at $p_j$, and  $U^{\rm art}_k$ is the arterial concentration of the metabolite. As with the transports and reactions, we define an $N$-vector $\bc_k(t)$ for each of the four substances in blood, and stack them into the $4N$-vector $\bc(t)$. Since the convective component is present only in the blood compartment, we define
	\[
	\bQ_{\rm flow} = \left[\begin{array}{c} \bc \\  \mO_{6 N\times 1} \\ \mO_{11N\times 1} \\ \mO_{11N\times 1}\end{array}\right] \in \R^{32 N}.
	\]

	\begin{table*}[htbp]
		\caption{Volume fraction and blood flow parameters: We provide here the volume fraction of each of our four compartments. The cleft compartment is a sub-compartment of the ECS. We also provide the blood flow parameters and arterial concentrations.}
		\renewcommand{\arraystretch}{1.4}
		\centerline{
			\begin{tabular}{lccccccccc}
				\hline
				\multicolumn{2}{c}{Volume fraction}& &\multicolumn{3}{c}{Blood flow parameters}& &\multicolumn{3}{c}{Arterial concentration}\\
				\cline{1-2} \cline{4-6} \cline{8-10}
				Parameter & Value & & Parameter & Value & Units & &Parameter & Value & Units\\
				\hline
				Blood ($\eta_1$) & 0.04 & & Hct & 0.45 & & & $C_{\rm art, Glc}$ & 5 & mM\\
				ECS ($\eta_2$)& 0.25 & & Hb & 5.18 & & & $C_{\rm art, O_2}$ & 9.14 & mM\\
				Neuron ($\eta_3$) &0.45& & $K_H$ & $36.4e-3$ & mM & & $C_{\rm art, CO_2}$ & 23 & mM\\
				Astrocyte ($\eta_4$)&0.25&  & Q & 0.4 & 1/min & & $C_{\rm art, Lac}$ & 1.1 & mM \\
				Cleft ($\eta_{2c}$)& 0.01 & &  & & & & & & \\
				\hline
			\end{tabular}
		}
		\label{table:7}
	\end{table*}

	\subsection{The coupled system}
	
	We are now ready to combine equations (\ref{eq2 4}) and (\ref{eq1 3}) into a single coupled matrix equation. Denote by ${\bf M}$ the combined block diagonal mass matrix with blocks ${\mathcal M}^j$ on its diagonal, and by ${\bf \Psi}$  the diagonal matrix with appropriate volume fractions on its diagonal, and write the coupled ODE system as
	\begin{equation*}
	{\bf \Psi} {\bf M} \frac{d}{dt}\begin{bmatrix}
	{\bf U}_1\\ ----\\
	{\bf U}_2\\ ----\\
	{\bf U}_3\\ ----\\
	{\bf U}_4
	\end{bmatrix} = -
	\left({\bf K} + \lambda {\bf B}\right)\begin{bmatrix}
	\mO_{4N\times 1}\\ ----\\
	{\bf U}_2\\ ----\\
	\mO_{11N\times 1}\\ ----\\
	{\bf U}_4
	\end{bmatrix}  
	+
	\lambda \begin{bmatrix}
	\mO_{4N\times 1}\\ ----\\
	{\bf U_0^2}_{6N\times 1}\\ ----\\
	\mO_{11N\times 1}\\ ----\\
	{\bf U_0^4}_{11N\times 1}
	\end{bmatrix} + 	{\bf M} \mF\,\bphi +  {\bf M} \mV \bgamma +  {\bf M} \begin{bmatrix}
	\bc\\ ----\\
	\mO_{6N\times 1}\\ ----\\
	\mO_{11N\times 1}\\ ----\\
	\mO_{11N\times 1}
	\end{bmatrix},
	\end{equation*}
	
	\begin{table*}[!ht]
		\captionsetup{width=.90\textwidth}
		\caption{Steady state concentrations of the metabolites in units of mM. The model is first run with the following initial conditions without activation and using baseline blood flow value until steady state is reached. The steady state values obtained through this process and listed here are used as the initial values for different neuronal activation protocols.}
		\centerline{
			\begin{tabular}{l c c c c c c c c c c c}
				\hline
				& Glc & ${\rm O}_2$ & ${\rm CO}_2$ & Lac & Glu & Gln  & Pyr & ATP & ADP & NAD$^+$ & NADH \\ 
				\hline
				Blood & $4.60$ & $6.83$ & $25.15$ & $1.19$      &          &                &                 &              &              &                 &
				\\
				ECS & $1.13$ &      $0.034$ & $83.75$   &  $1.25$     &  $1e-5$         &$1e-5$   &   &              &              &                 &            \\
				Neuron      & $1.13$ & $9.8e-3$     & $85.46$   & $1.25$      & $14$          &$0.001$   & $0.12$  &$2.19$& $0.023$&$0.03$    &$0.003$\\
				Astrocyte   & $0.75$ & $9.8e-3$     & $85.46$   & $1.25$ & $0.01$    & $0.01$ & $0.12$   & $2.19$  &$0.023$&$0.03$&$0.003$ \\
				\hline
			\end{tabular}
		}
		\label{table:5}
	\end{table*}
	
	\begin{table*}[!ht]
		\captionsetup{width=.90\textwidth}
		\caption{Diffusion coefficients of the metabolites used in the model. The values represent the coefficient in a free medium such as water. These numbers are sourced from publications cited  in the third column. }
		\centerline{
			\begin{tabular}{ |p{5cm}|p{5cm}|p{5cm}|  }
				\hline
				Metabolite & Diffusive coefficient in free medium  & Source \\
				\hline
				Glucose & $6.73 \times 10^{-6} \rm cm^2/s$ & \cite{doi:10.1146/annurev.micro.50.1.317} \\
				Oxygen & $2.0 \times 10^{-5} \rm cm^2/s$   & \cite{doi:10.1021/jp952903y}, \cite{doi:10.1021/ja02220a002} \\
				CO2 & $1.88 \times 10^{-5} \rm cm^2/s$ & \cite{Maxarei} \\
				Lactate & $1.0 \times 10^{-5} \rm cm^2/s$ & \cite{C2AN36715G} \\
				Glutamate & $7.6 \times 10^{-6} \rm cm^2/s$ &  \cite{Rusakov} \\
				Glutamine & $7.6 \times 10^{-6} \rm cm^2/s$ & \cite{doi:10.1021/ja01118a065}   \\
				Pyruvate & $1.12 \times 10^{-5} \rm cm^2/s$ & \cite{C2AN36715G} \\
				ATP & $7.2 \times 10^{-6} \rm cm^2/s$ & \cite{BOWEN196430} \\
				ADP & $7.2 \times 10^{-6} \rm cm^2/s$ & \cite{BOWEN196430} \\
				NAD+ & $4.2 \times 10^{-6} \rm cm^2/s$ & \cite{A806004E} \\
				NADH & $3.9 \times 10^{-6} \rm cm^2/s$ & \cite{HASINOFF198753} \\
				\hline
			\end{tabular}
		}
		\label{tab:my_label}
	\end{table*}
	
	\section{Simulations and results}
	In the following numerical simulations, the domain $B$ is a disc of radius $0.25\;{\rm cm}$. The diffusion coefficients in the free medium for each metabolite are given in Table \ref{tab:my_label}. We assume that there is no diffusion in the blood and neuronal compartments, and since glutamate and glutamine are assumed to act locally only in the synaptic clefts, the diffusion of these metabolites is neglected. 
	Formula (\ref{tortuosity}) is used to calculate the diffusion coefficients in ECS based on the tortuosity, i.e., the effective diffusion coefficient in ECS compartment ($\ell = 2$) for metabolite $k$ is given by
	\[
	D_k^\ell = \frac{D}{\lambda^2},
	\]
	with $\lambda = 1.6$, corresponding to the experimental tortuosity of ECS in healthy brain tissue. We set $D_k^\ell = 0$ for $k = 5,6$, corresponding to glutamate and glutamine in ECS.
	
	The diffusion coefficients of the metabolites in astrocyte are poorly known. We make the assumption that the diffusion coefficients in astrocyte is smaller than in a free medium.  In lack of well established experimental or theoretical models, we set the diffusion coefficient of each metabolite in the astrocyte to be $70\%$ of that in free medium, that is, $D_{\rm a} = 0.7D$.  Furthermore, since the diffusion in the astrocyte syncytium depends on the gap junction strength $s$, we set
	\[
	D_k^\ell =sD_{\rm a}.\quad\mbox{for $\ell = 4$, $k = $ metabolite,}
	\]
	In our computer experiments, we will test different values of $s$ to better understand the effect of the gap junctions on the metabolism.
	We also assume that there is no diffusion of glutamate and glutamine in the astrocytic compartment. 
	
	\subsection{Simulation setup}
	We calibrate the model with a simulation of local normal neuronal activity to demonstrate the metabolic interaction between neuron and astrocyte in a spatially distributed framework. To this end, we introduce first an activity function that is a proxy for a detailed electrophysiologic activation activity.
	
	\subsubsection{Activity simulation}
	
	We simulate the increased release of glutamate from the small vesicles in the presynaptic neuron in response to intracellular calcium signaling writing a dynamic model for the glutamate transport from neuron to ECS of the form
	\begin{equation}\label{glu}
	\bphi^{(4)}_{\rm Glu}  = T^{(4)}_{\rm Glu}\left(1 + 2\xi(t)\right)  \frac{{\bf u}^3_5}{M^{(4)}_{\rm Glu}\left(1-0.5\xi(t)\right) + {\bf u}^3_5},
	\end{equation}
	where ${\bf u}^3_5$ is the concentration of glutamate in neuron, and $\xi(t)$ is a dimensionless activity function. To simulate  a local activation of 3 minutes starting after 2 minutes from the beginning of the run, we set
	\[
	\xi(t) =  \quad\left\{\begin{array}{lll} \rho, & & \mbox{$2 < t \leq 5$} \\ 0, & & \mbox{otherwise.}\end{array}\right.
	\]
	where the parameter $\rho$ is adjusted so that the peak glutamate concentration during the activity reaches the experimental value of $1.1\,{\rm mM}$. Numerical experiments suggest a value $\rho = 0.9099$. According to \cite{Dzubay5265}, the concentration of glutamate in the synaptic cleft following action potential mediated release exceeds $1$mM for $<10$ ms, and rapidly returns to $<20$ nM between release events due high affinity glutamate uptake by neurons and glia.

	\begin{figure}[htbp]
		\captionsetup{width=.97\textwidth}
		\centering
		\includegraphics[width=0.5\linewidth]{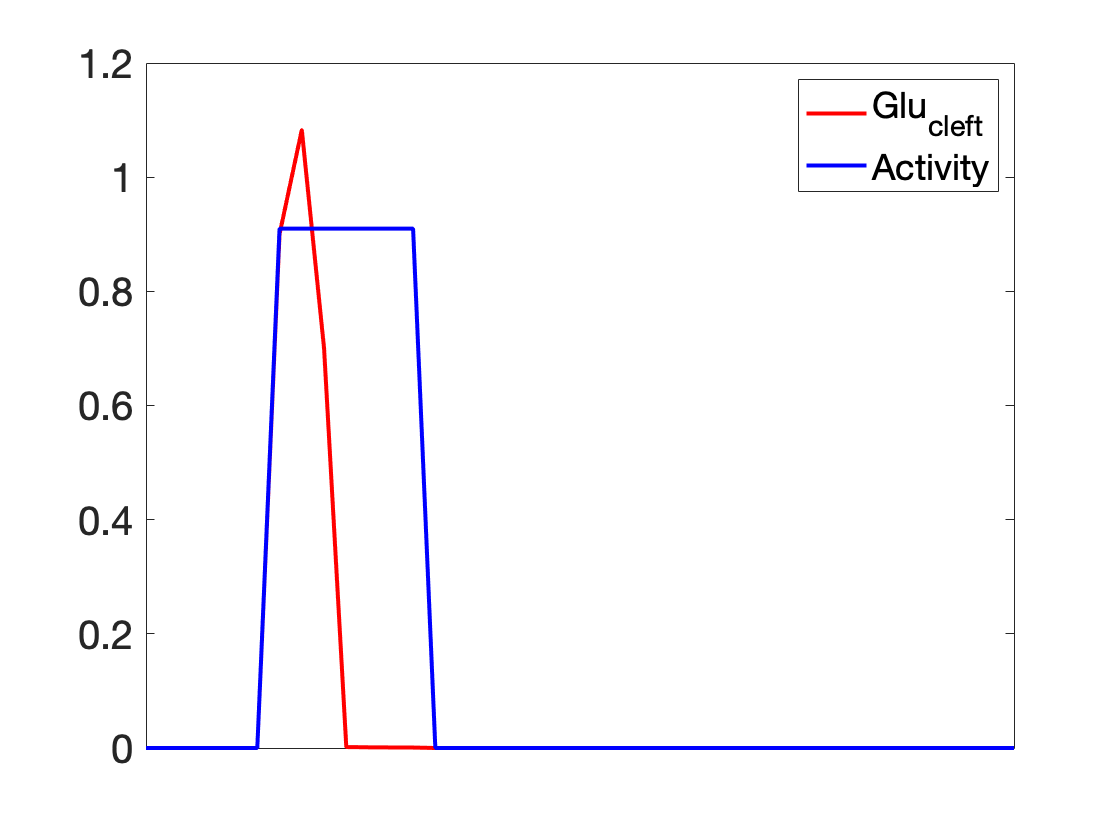}
		\label{fig:my_label2}
		\caption{Plot showing the activity function (blue) and the concentration of glutamate in the cleft (red) peaking at  about $1.1$ mM for one node of the FEM mesh using the modified form of glutamate flux from neuron to cleft.}
	\end{figure}
	
	Glutamate, an excitatory neurotransmitter which enables transmission of action potential from pre- to post-synaptic neuron needs to be removed rapidly from the cleft to prevent post-synaptic overexcitation which can lead to cell death. As long as glutamate concentration in the cleft is significant, glutamate sensitive ion channels in the postsynaptic neuron remain open, allowing influx of calcium and sodium ions,  thus triggering a transmission of the action potential in the postsynaptic neuron and potential loss of ionic balance. The energy required for this removal  by active pumps and exchangers is provided by an increase in ATP hydrolysis in neuron. 
	Following \cite{neurovascular}, we define ATPase in neuron as
	\[
	\bgamma^{3}_8 = \Phi_{\rm base} + V^3_{\rm ATPase}\cdot \mathcal{F}({\bf u}_8^3)\cdot\frac{{\bf u}_{8}^3}{{\bf u}_{8}^3 ~+~ K^3_{\rm ATPase}},
	\]
	where $\mathcal{F}$ is a sigmoid saturation function defined as
	\[
	\mathcal{F}({\bf u}_8^\ell) = \frac{({\bf u}_{5}^2)^2}{({\bf u}_{5}^2)^2 ~+~ k_{\rm Glu}}.
	\]
	This function triggers the increase in ATP hydrolysis in response to elevated glutamate level in the cleft. We set $k_{\rm Glu} = 1.67\times 10^{-10}\;{\rm mM}^2$, and $\Phi_{\rm base} = 3.9996\; {\rm mM}/{\rm min}$ is the baseline cost of maintaining steady state when no activity is taking place.
	
	\subsubsection{Blood flow simulation}
\begin{figure}[htbp]
		\centering
		\includegraphics[width=0.5\linewidth]{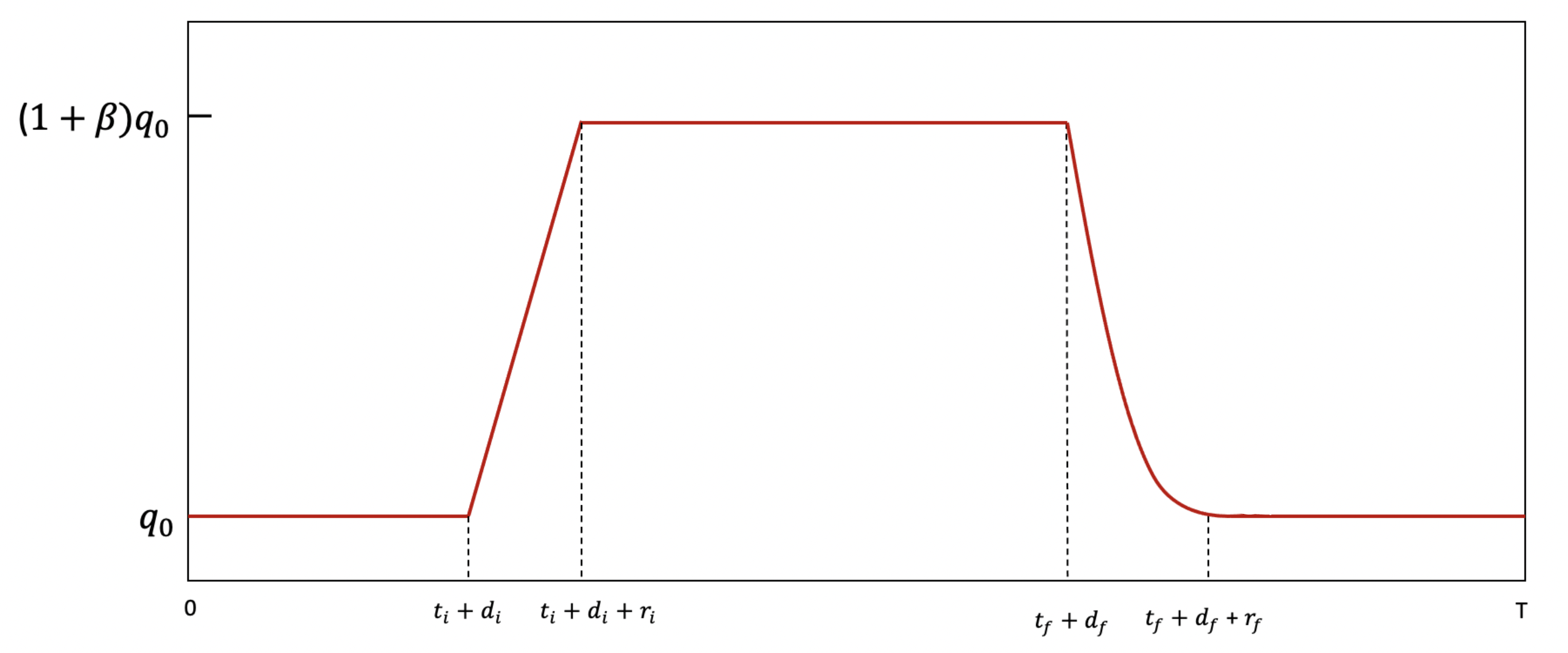}
		\captionsetup{width=.97\textwidth}
		\caption{Blood flow profile given as $q(t)$ as described in the equation above. After the beginning of the activation, the blood flow increases by $30\%$ from baseline, and remains at that level for $3$ minutes before returning gradually back to its baseline value.}
		\label{fig:my_label3}
	\end{figure}

	In an activation event, elevated values of $\xi(t)$ trigger an increase in cerebral blood flow.  The hemodynamic response to an elevated neuronal activity amounts to a local increase in arterial blood flow, thus increasing the amount of oxygen and glucose available to the tissue. The blood flow formula  used is given below:
	\[
	q(t) = \left\{\begin{array}{ll}
	q_0 & \mbox{$0\leq t < t_1 + d_1$,}  \\
	\left(1 + \beta\left(\frac{t - t_1 - d_1}{r_1}\right)\right)q_0  & t_1 + d_1 \leq t < t_1 + d_1 + r_1, \\
	(1 + \beta)q_0 & t_1 + d_1 + r_1 \leq t < t_f + d_f,\\
	\left(e^{-\alpha(x - t_f - d_f)}a + b\right)q_0 & t_f + d_f \leq t < t_f + d_f + r_f,\\
	q_0 & t_f + d_f + r_f \leq t < T,
	\end{array}\right.
	\]
	where $q_0$ is the baseline value.
	The blood flow responds to elevated neuronal activity with a delay of $d_1 = 2$ seconds, and it increases $30\%$ ($\beta = 0.3$) above its baseline value $q_0 = 0.4\; {\rm mL}/{\rm min}$ per one gram tissue, remaining elevated until  $d_f = 5$ seconds after the end of the activation. In the model the ramping time response of the blood flow is of $r_1$ and $r_f$ seconds at the start and end of the activity event.  In our computations, we use $t_1 = 2$ and $t_f = 5$, expressed in the units of minutes, as the initial and final times of the activation event. The blood flow  profile is displayed in Figure \ref{fig:my_label3}.
	
	\subsection{Results}
	
	The first computed experiment simulates an elevated neuronal activity event, accompanied by an increase of the blood flow. The activated area is a circular patch of radius $r = 0.03$ cm, corresponding to roughly $10\%$ of the radius of the domain, centered at the node point with coordinates $[-0.075, -0.05]$. The
	activity and the increased blood flow are limited to the nodal points inside this patch.
	
\begin{figure*}[htbp]
		\centering
		\subfigure[Glc in Neuron at $t=2.4$ min]{\includegraphics[width=0.32\textwidth]{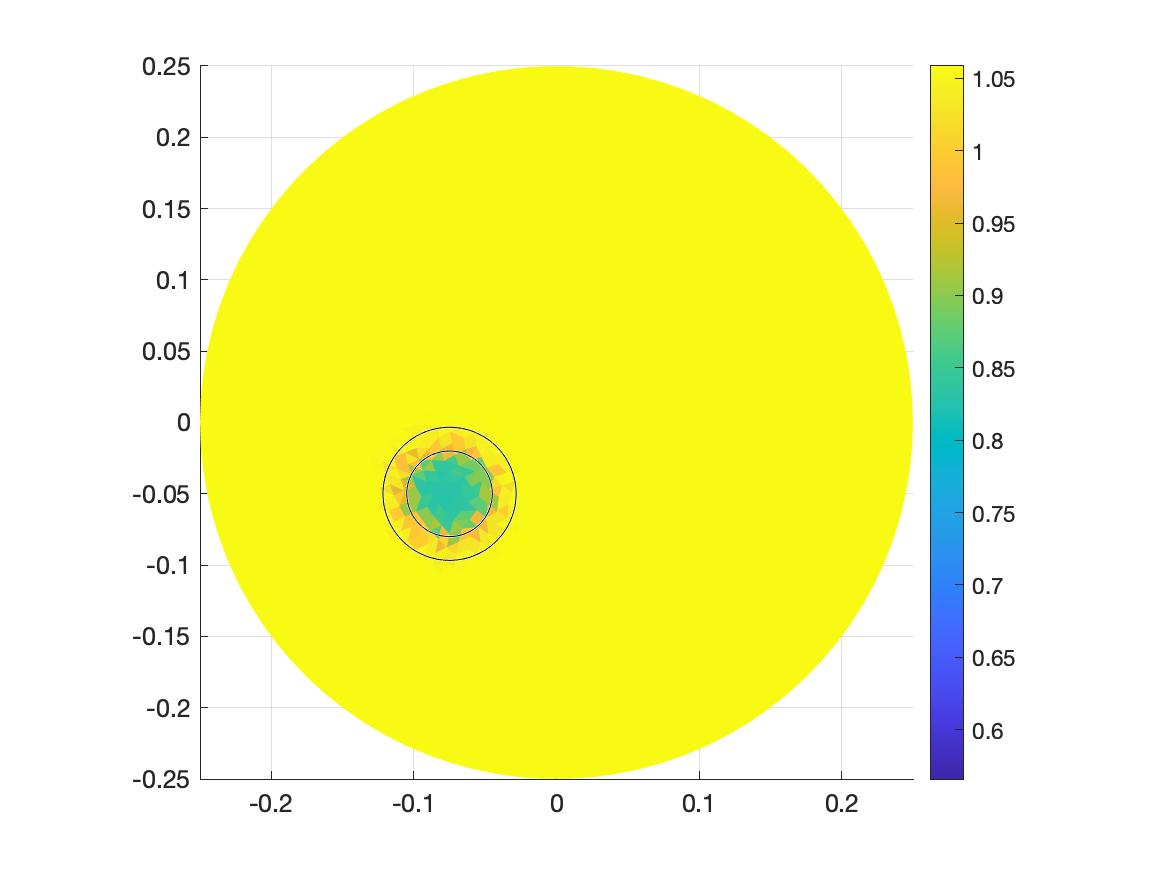}}
		\subfigure[Glc in Neuron at $t=4.8$ min]{\includegraphics[width=0.32\textwidth]{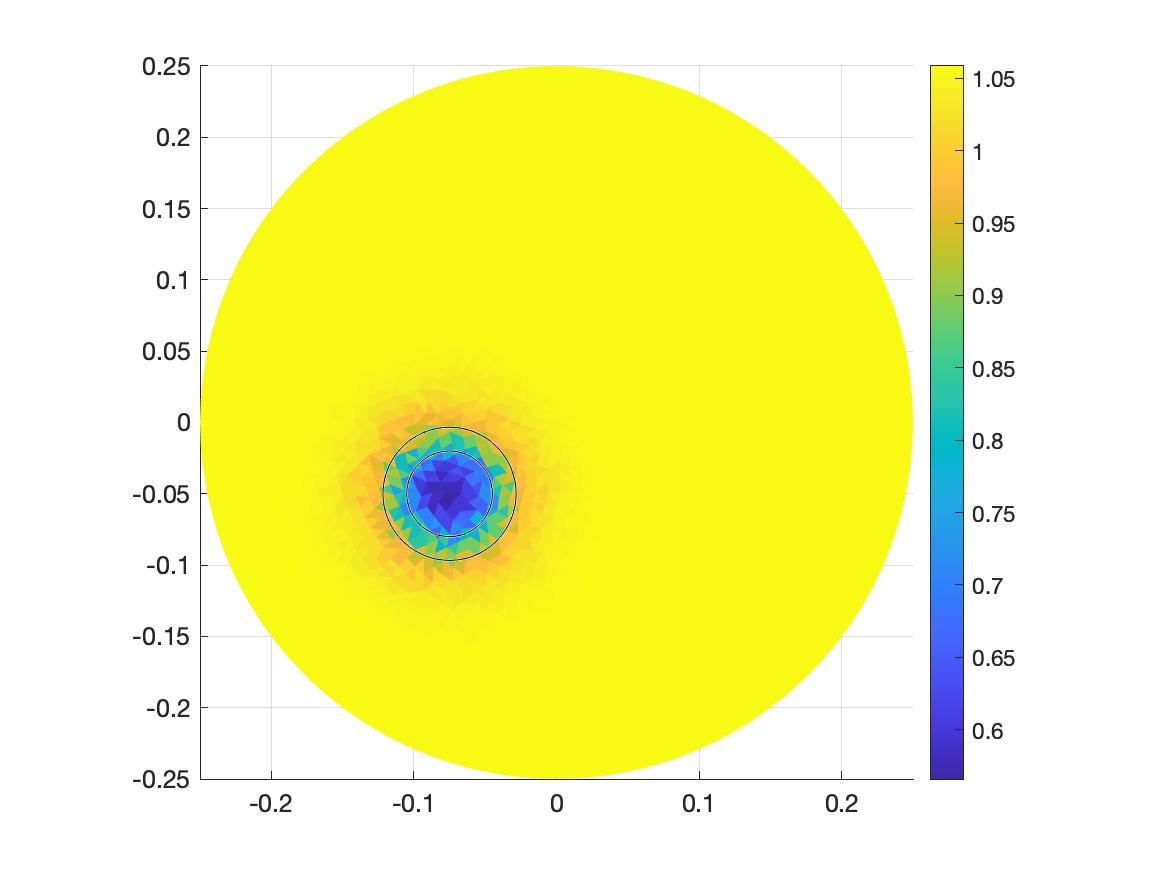}}
		\subfigure[Glc in Neuron at $t=6.8$ min]{\includegraphics[width=0.32\textwidth]{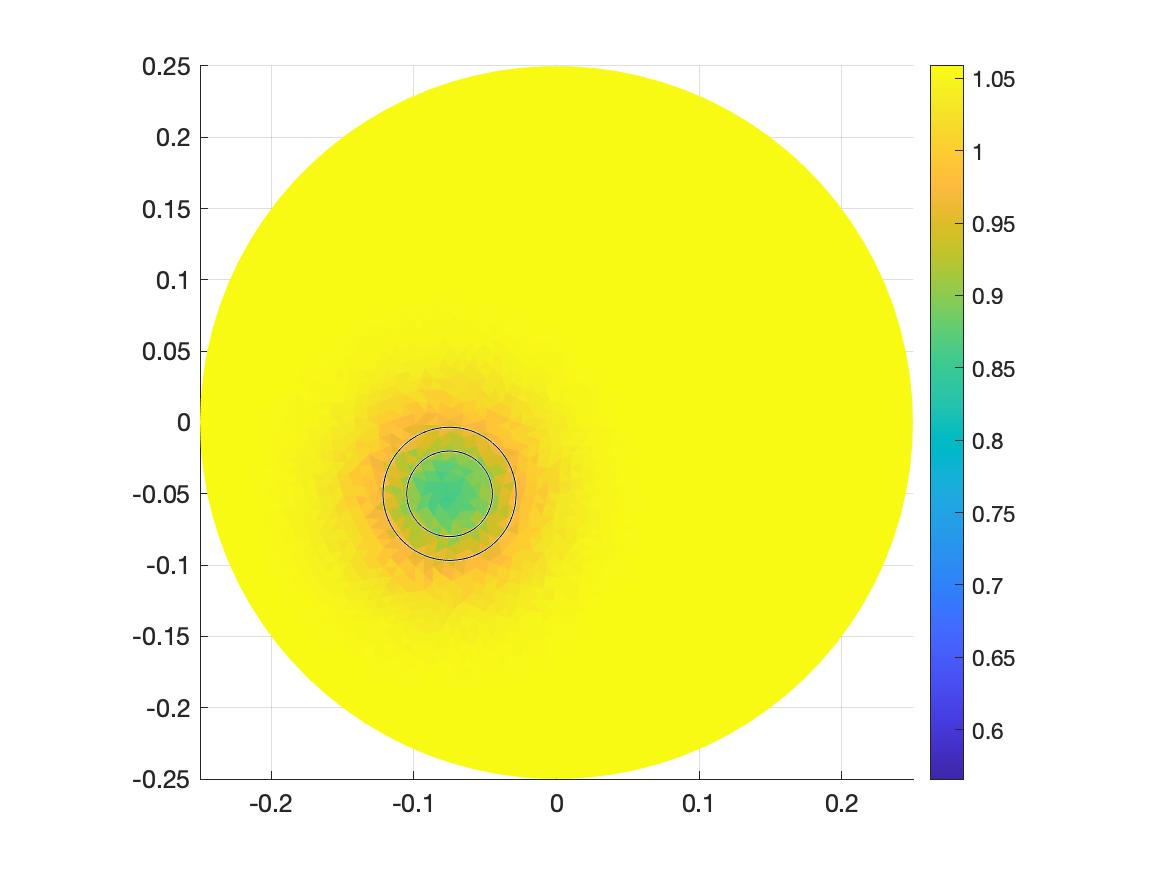}}
		\subfigure[$\rm O_2$ in Neuron at $t=2.4$ min]{\includegraphics[width=0.32\textwidth]{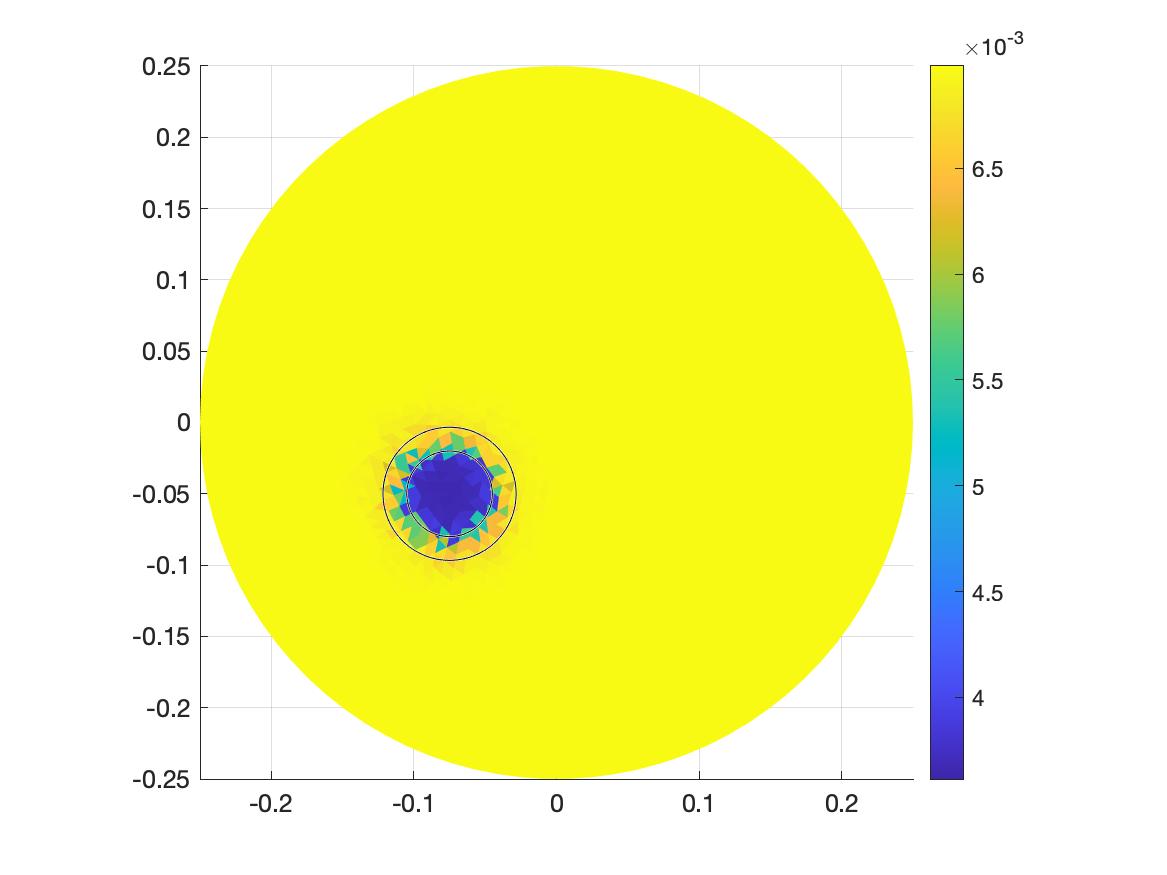}}
		\subfigure[$\rm O_2$ in Neuron at $t=4.8$ min]{\includegraphics[width=0.32\textwidth]{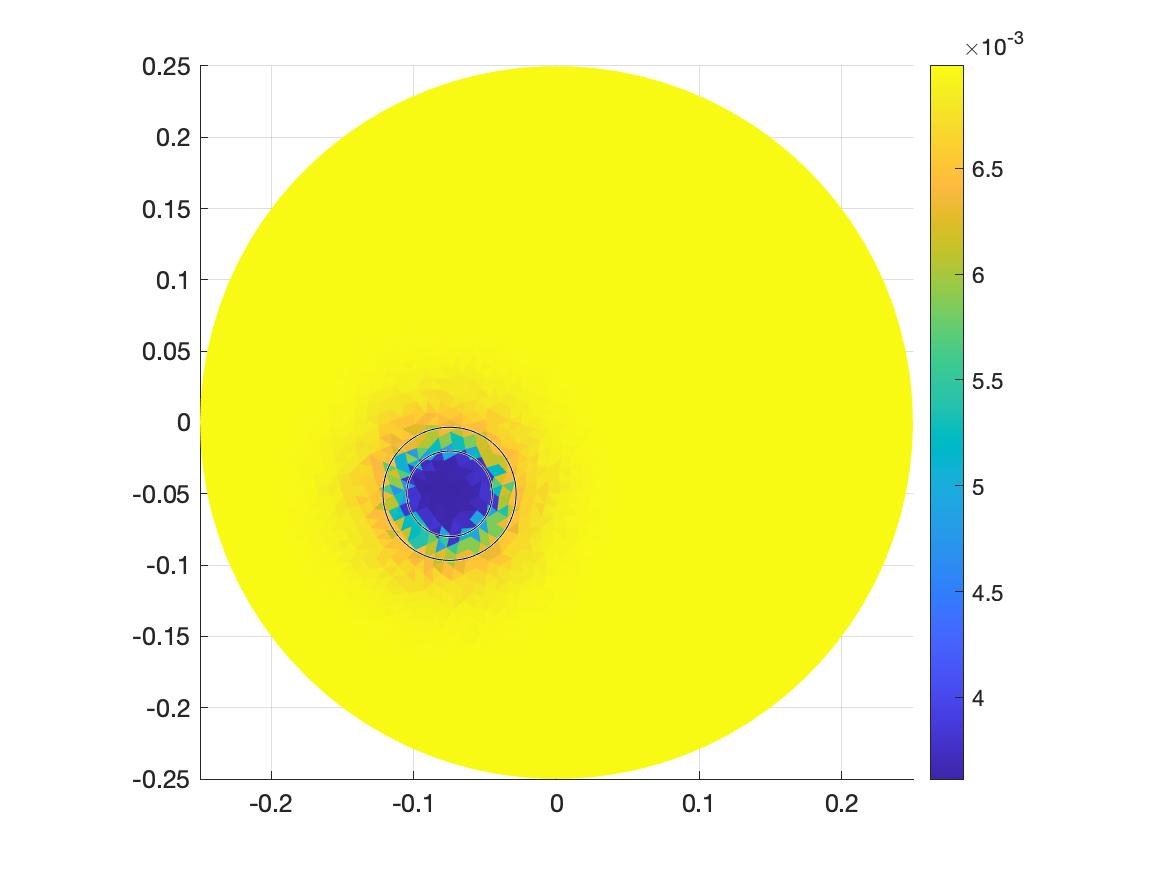}}
		\subfigure[$\rm O_2$ in Neuron at $t=6.8$ min]{\includegraphics[width=0.32\textwidth]{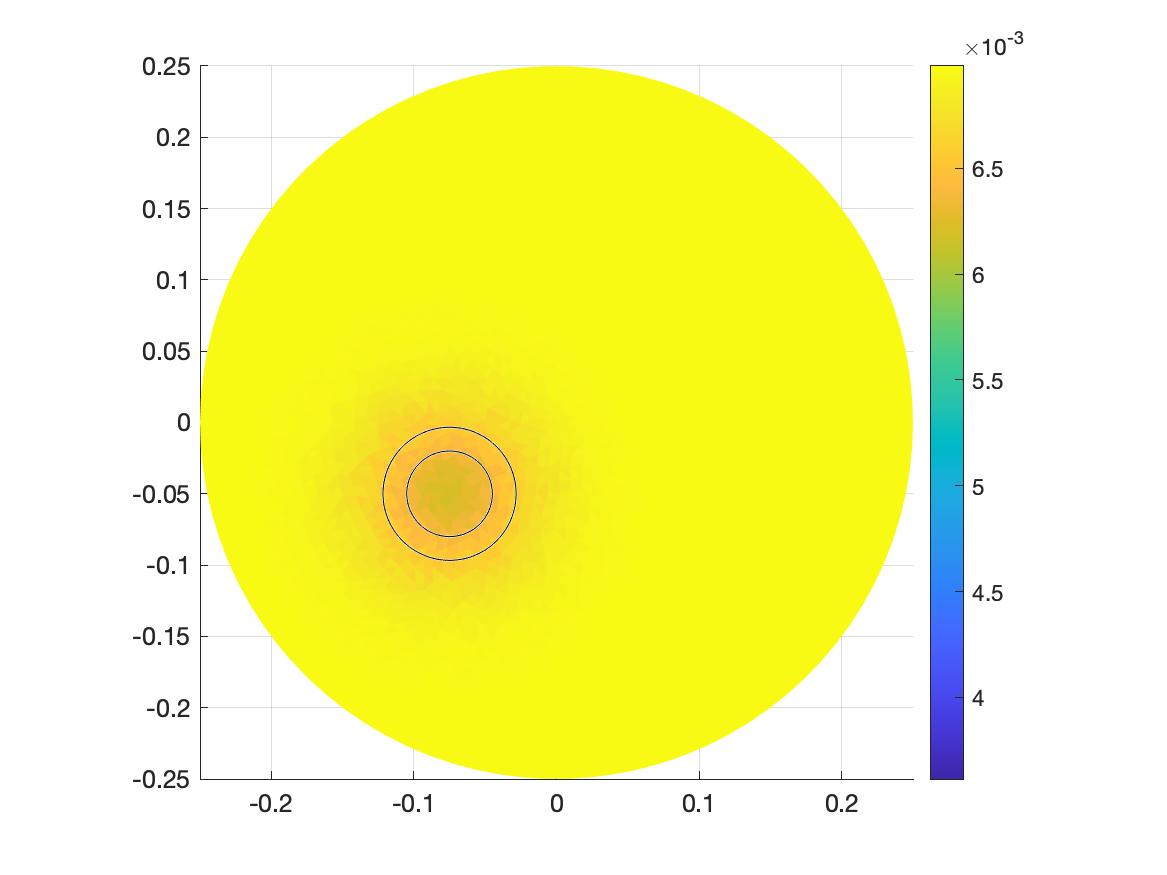}}
		\subfigure[Lac in Neuron at $t=2.4$ min]{\includegraphics[width=0.32\textwidth]{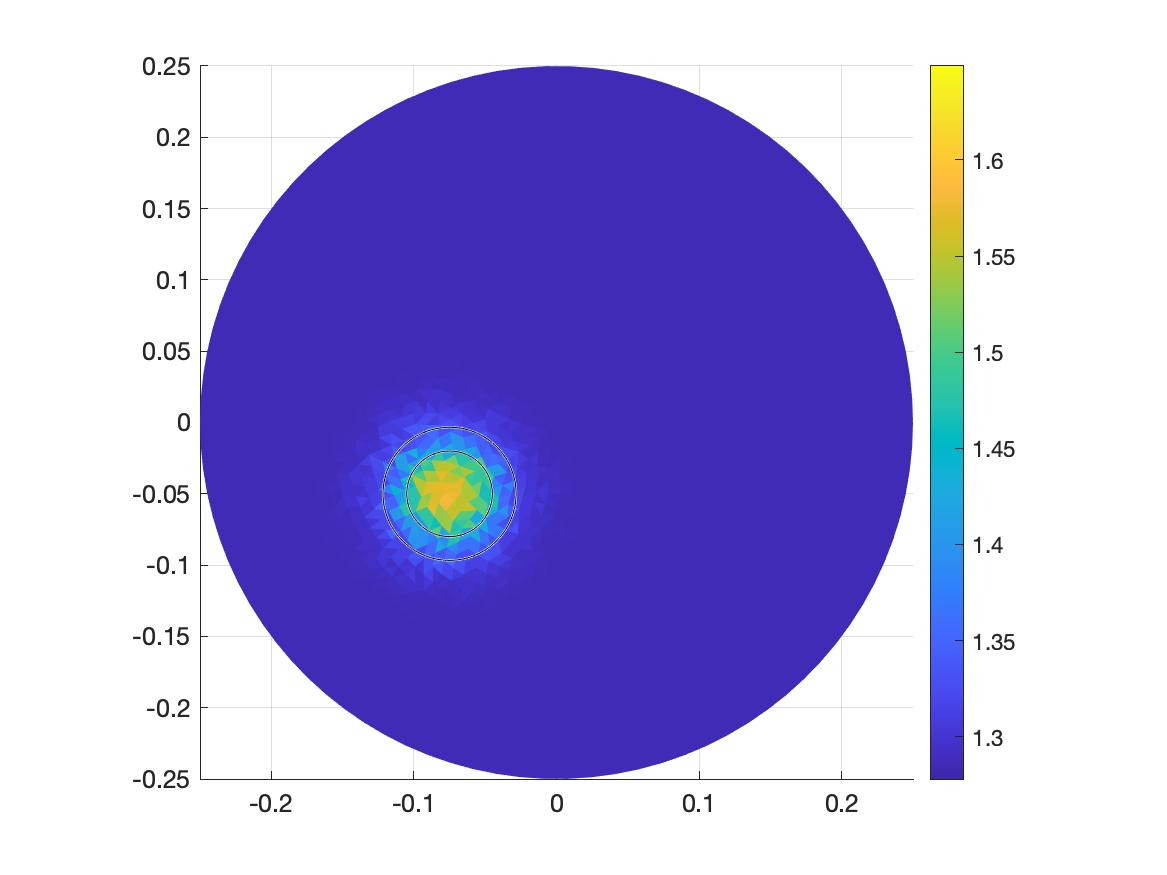}}
		\subfigure[Lac in Neuron at $t=4.8$ min]{\includegraphics[width=0.32\textwidth]{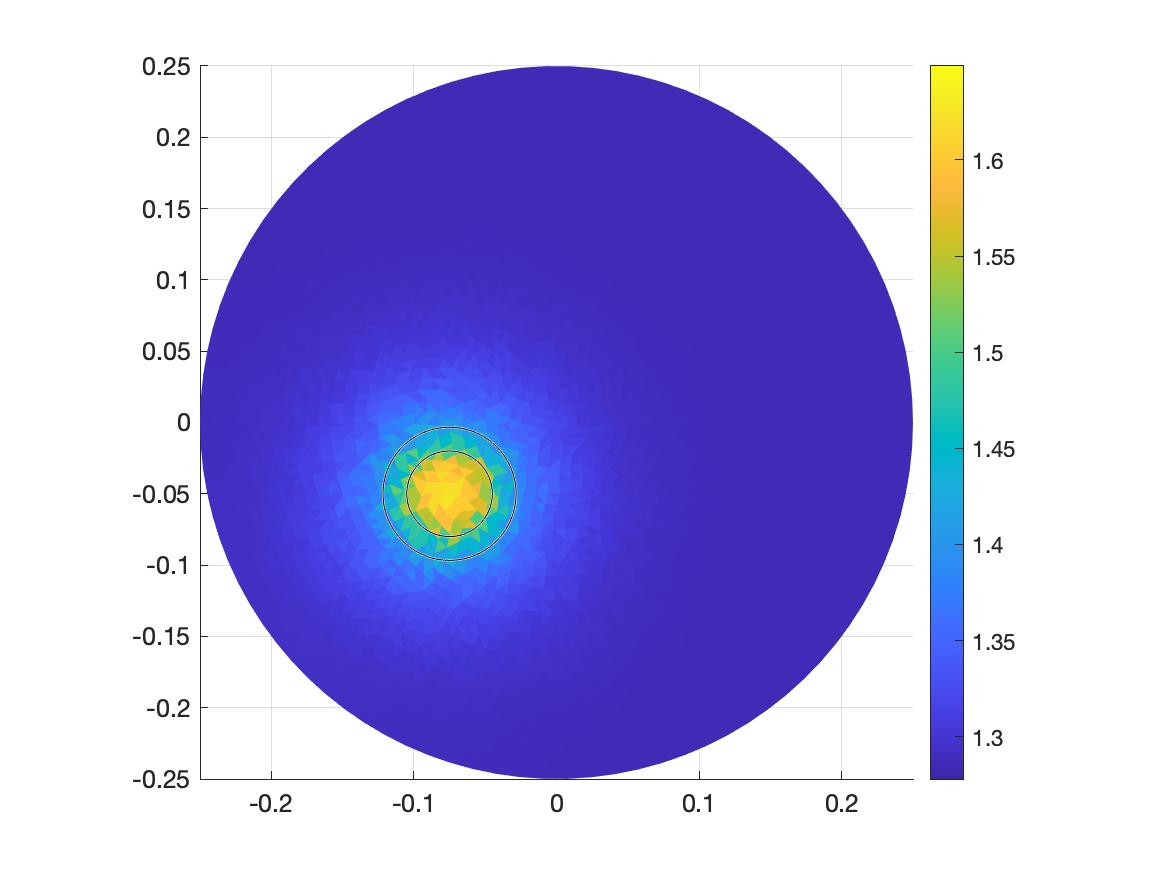}}
		\subfigure[Lac in Neuron at $t=6.8$ min]{\includegraphics[width=0.32\textwidth]{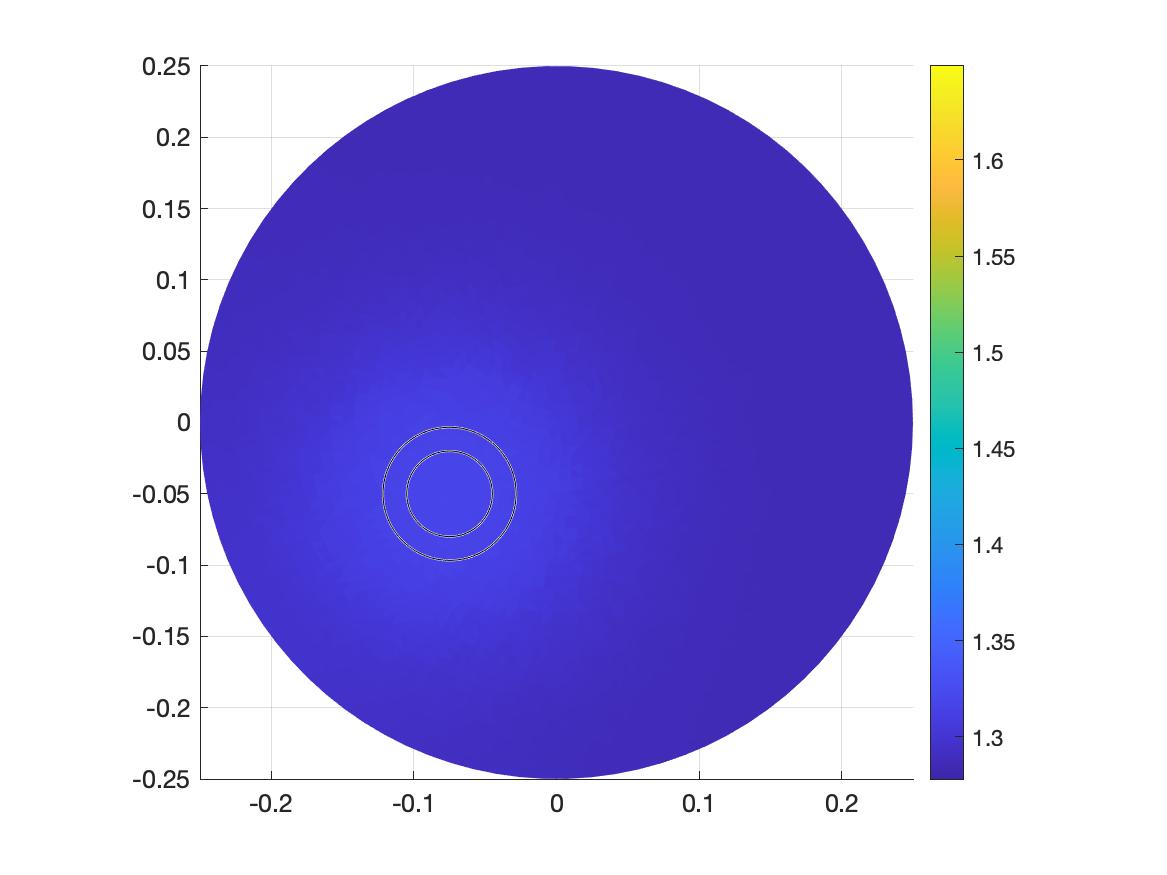}}
		\subfigure[Pyr in Neuron at $t=2.4$ min]{\includegraphics[width=0.32\textwidth]{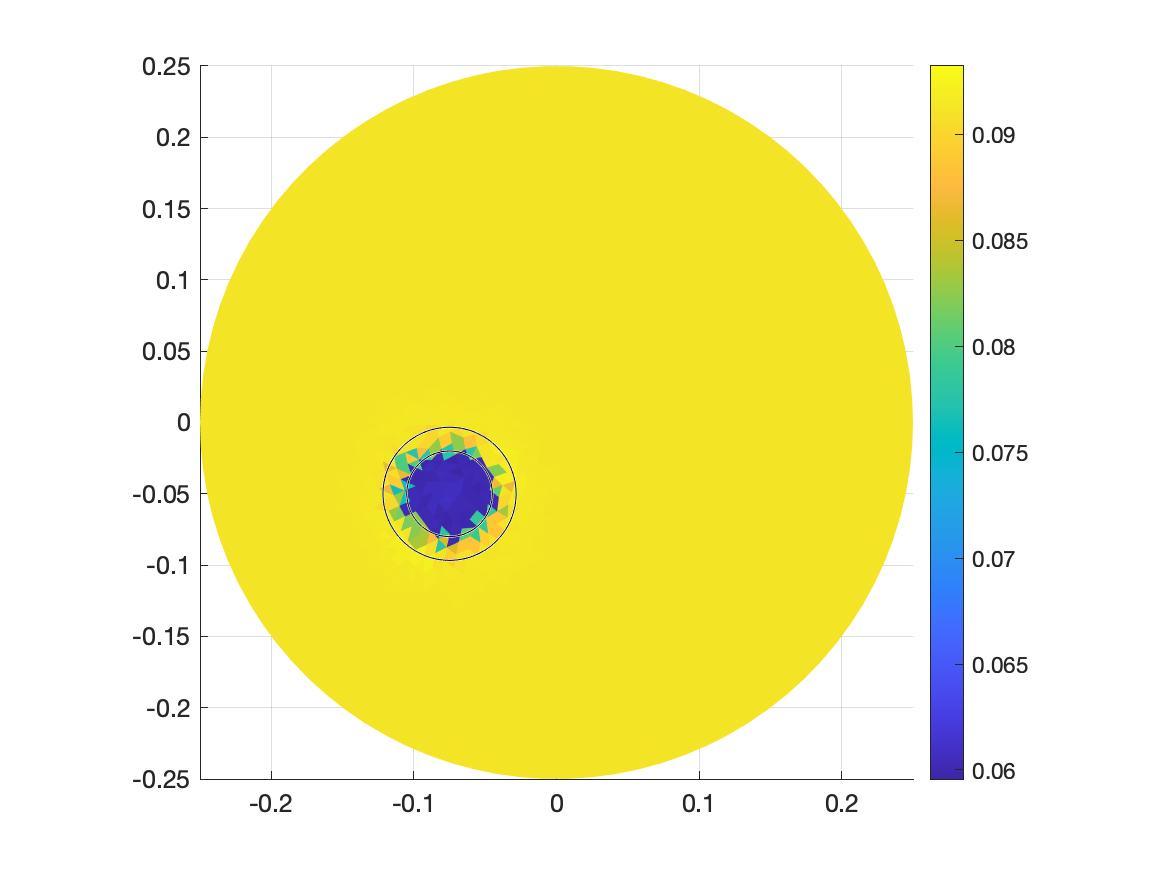}}
		\subfigure[Pyr in Neuron at $t=4.8$ min]{\includegraphics[width=0.32\textwidth]{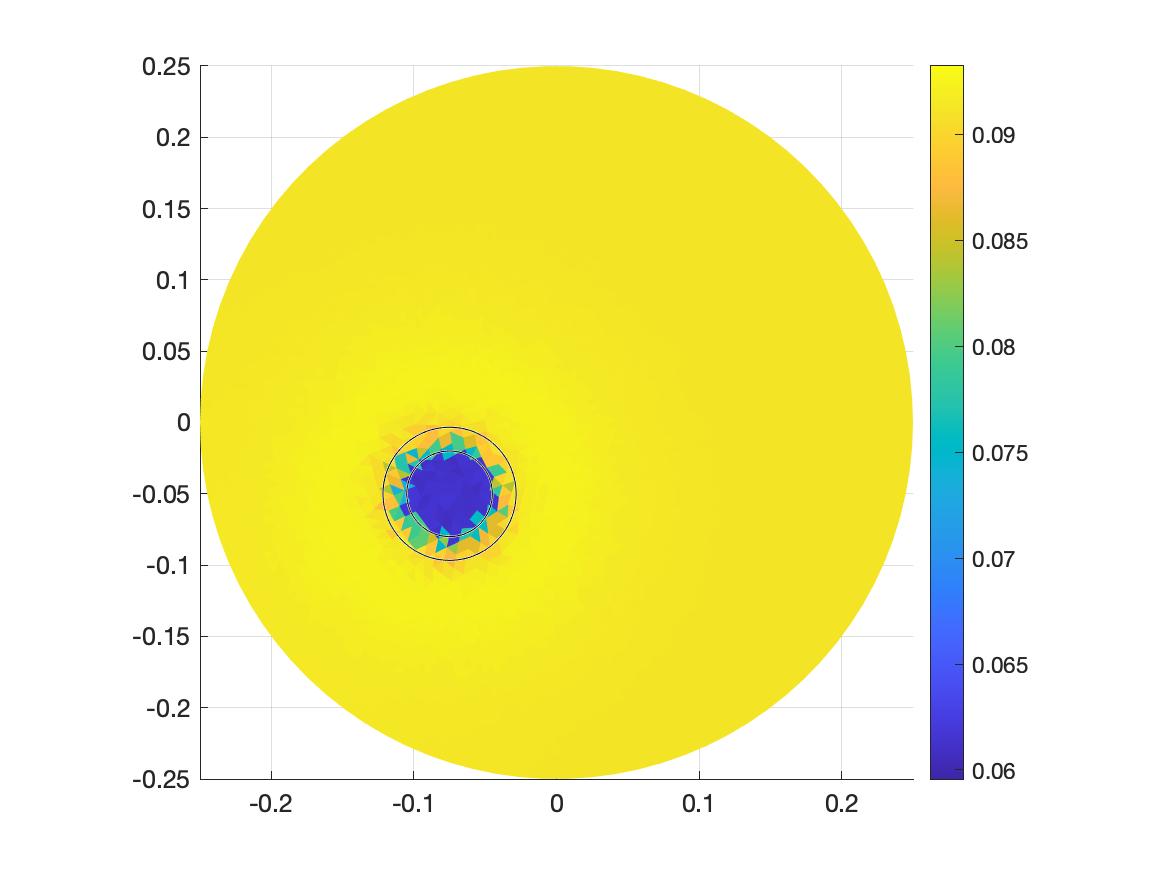}}
		\subfigure[Pyr in Neuron at $t=6.8$ min]{\includegraphics[width=0.32\textwidth]{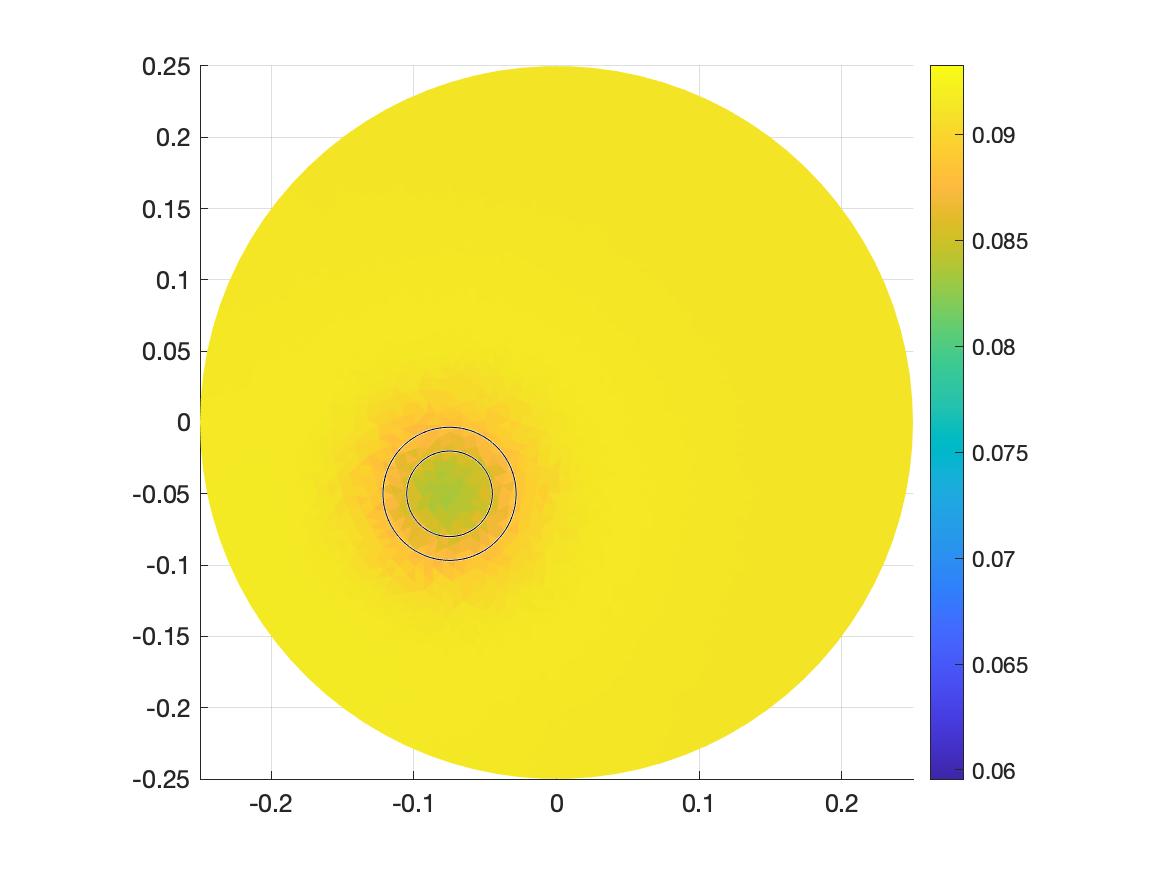}}
		\captionsetup{width=.95\textwidth}
		\caption{Snapshots of the concentrations of glucose (row 1), oxygen (row 2), lactate (row 3), and pyruvate (row 4)  in the neuron compartment.
			In this simulation, the gap junction strength was set as $s=1$ and tortuosity of the ECS at the normal value $\lambda =1.6$. The snapshots correspond to times  24 seconds after the activation onset ($t = 2.4$ minutes), 12 seconds before the end of the activation ($t = 4.8$ minutes), and at 108 seconds after the end of the activation ($t = 6.8$ minutes). The smaller circles in black enclose the region of elevated glutamate secretion and blood flow, referred to as the core of the activity. The region in between the smaller and larger circle is referred to as the margin of the activity.}
		\label{fig:result_4}
	\end{figure*}

\begin{figure*}[htbp]
		\centering
		\subfigure[Glc in Astrocyte at $t=2.4$ min]{\includegraphics[width=0.32\textwidth]{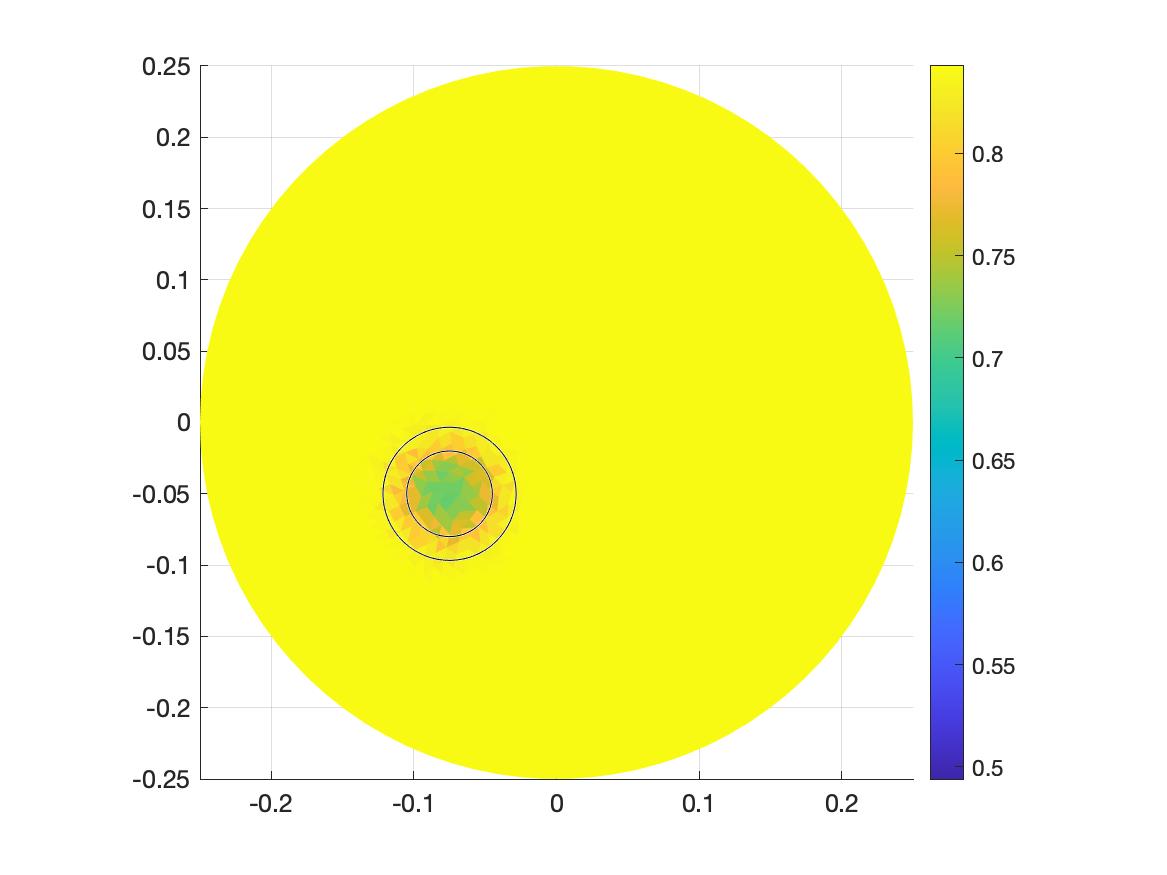}}
		\subfigure[Glc in Astrocyte at $t=4.8$ min]{\includegraphics[width=0.32\textwidth]{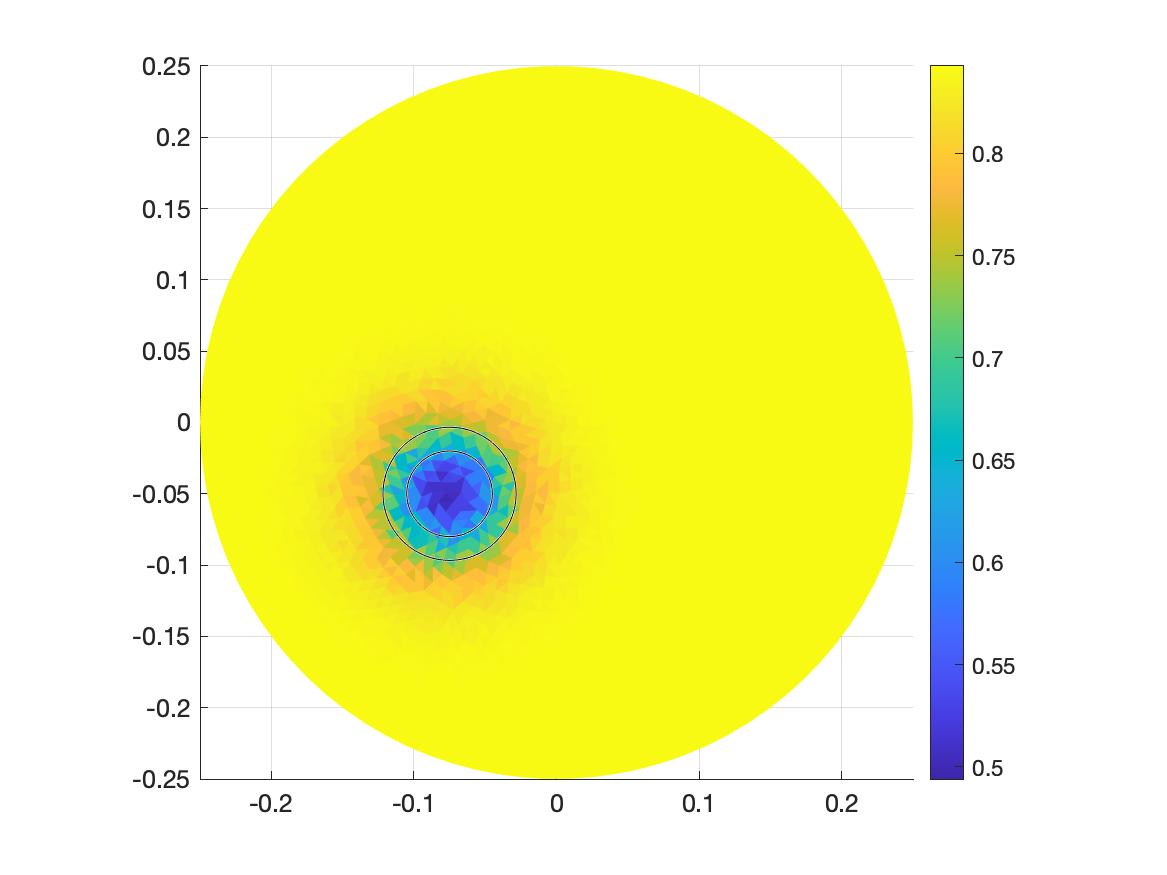}}
		\subfigure[Glc in Astrocyte at $t=6.8$ min]{\includegraphics[width=0.32\textwidth]{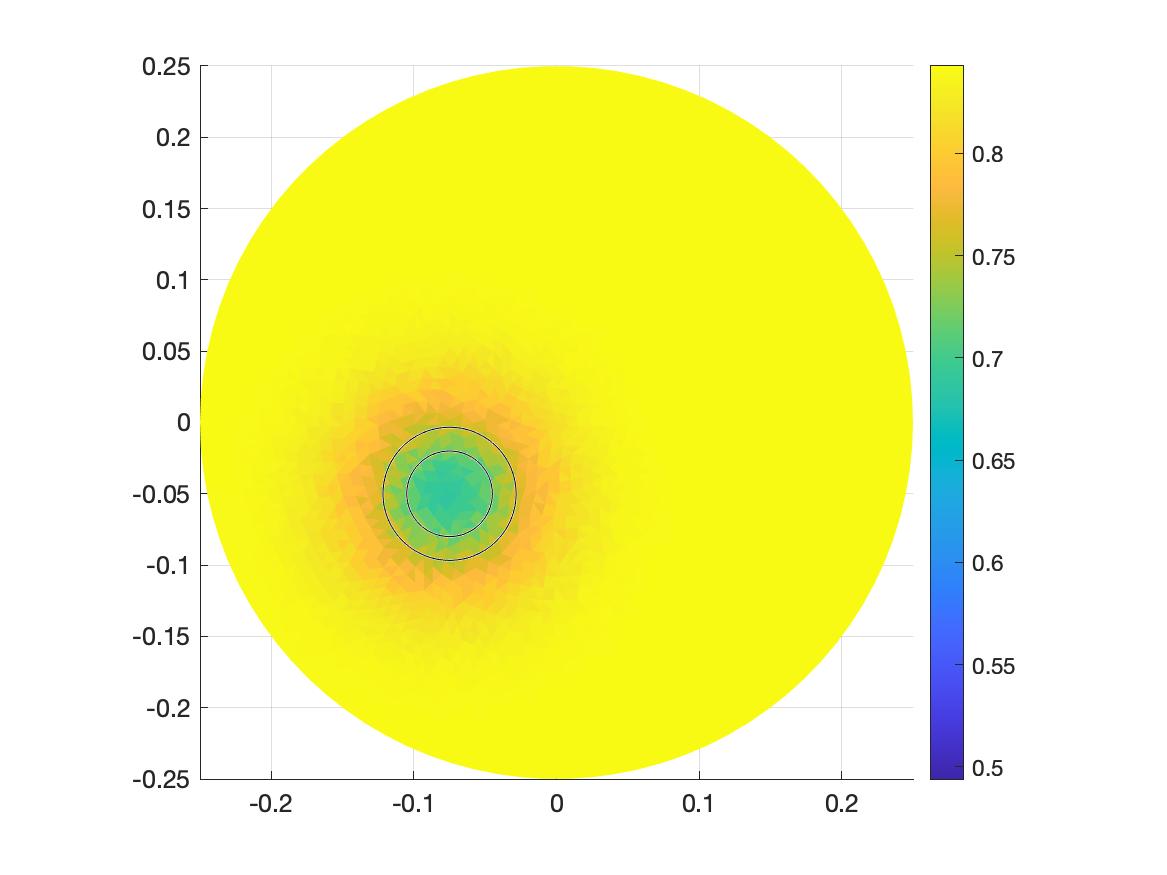}}
		\subfigure[$\rm O_2$ in Astrocyte at $t=2.4$ min]{\includegraphics[width=0.32\textwidth]{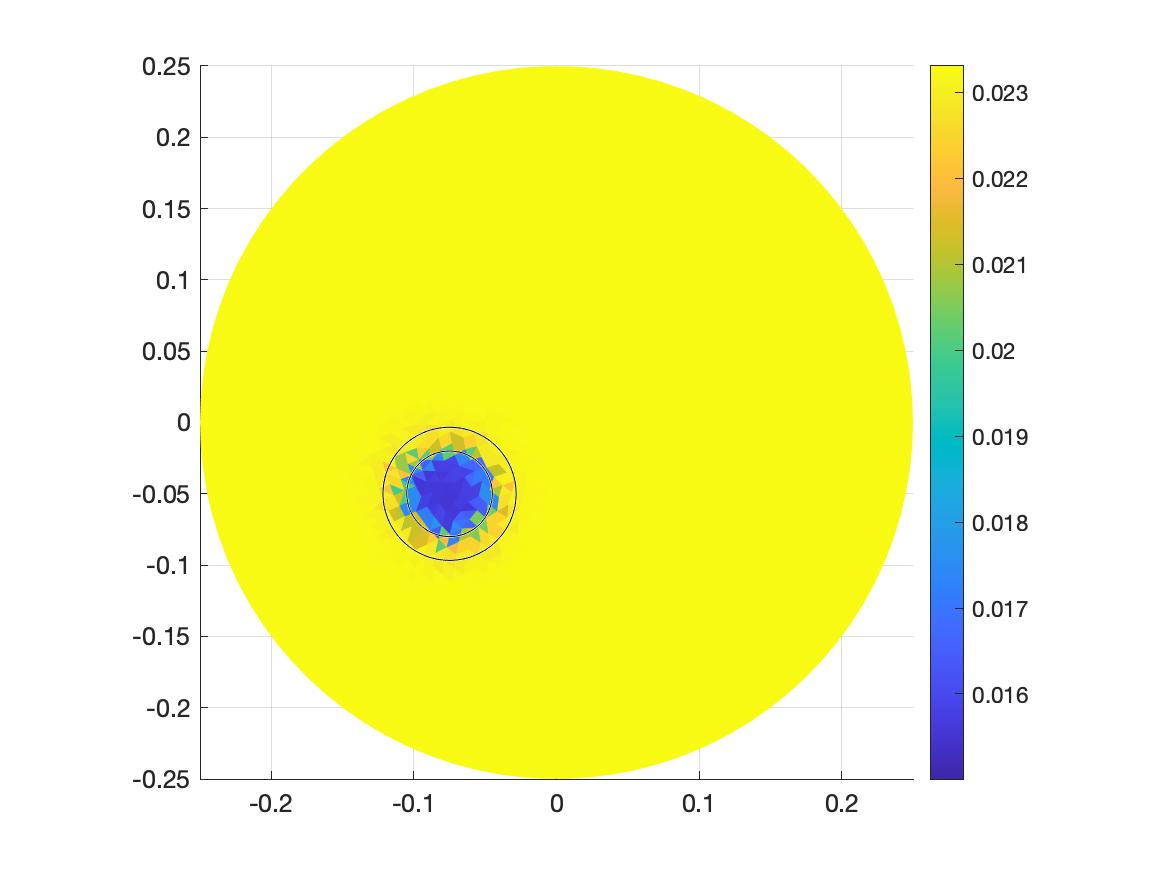}}
		\subfigure[$\rm O_2$ in Astrocyte at $t=4.8$ min]{\includegraphics[width=0.32\textwidth]{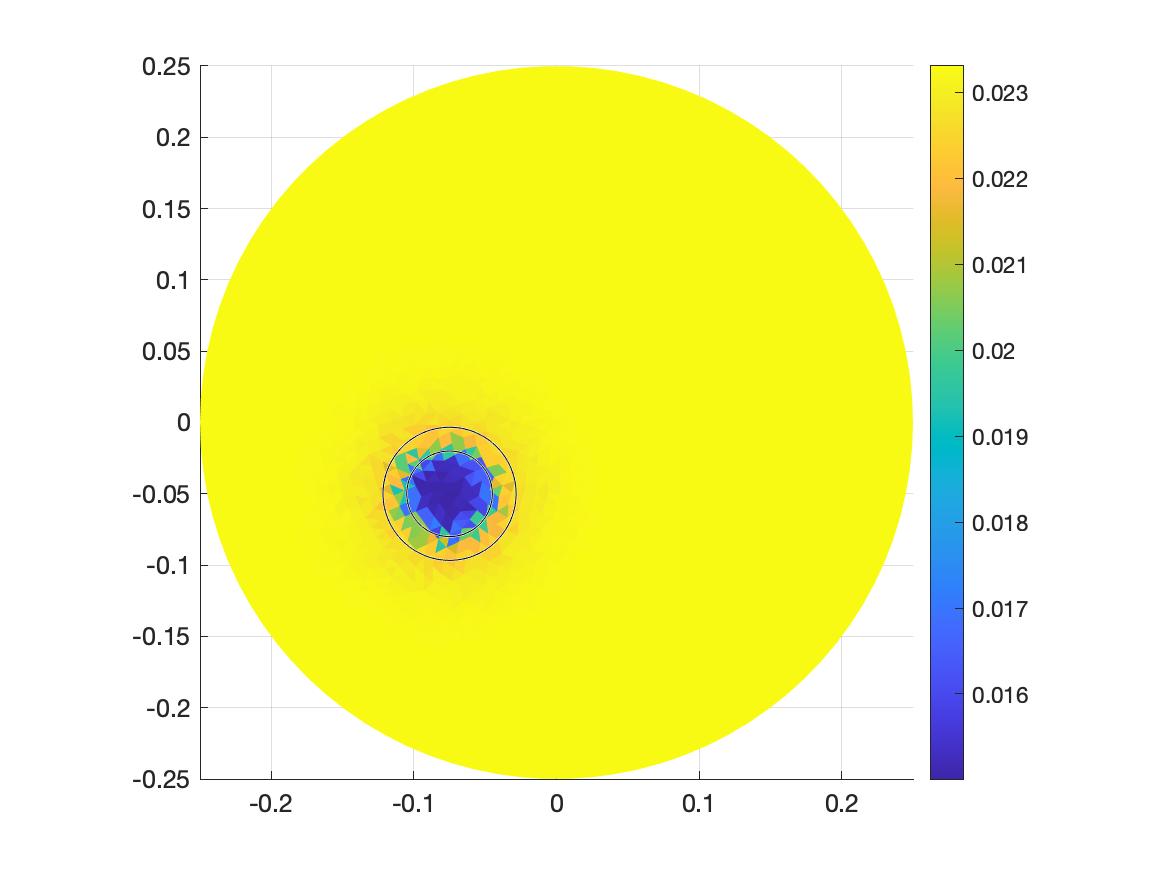}}
		\subfigure[$\rm O_2$ in Astrocyte at $t=6.8$ min]{\includegraphics[width=0.32\textwidth]{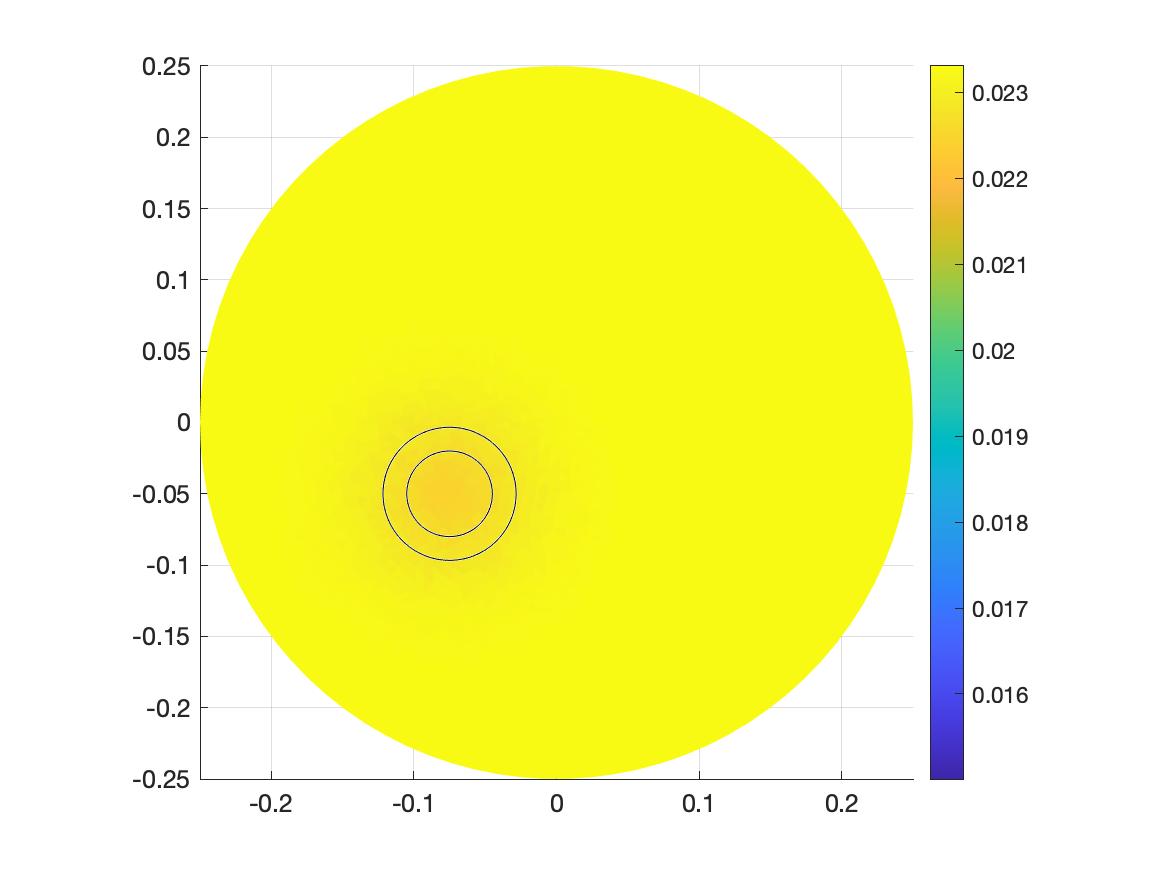}}
		\subfigure[Lac in Astrocyte at $t=2.4$ min]{\includegraphics[width=0.32\textwidth]{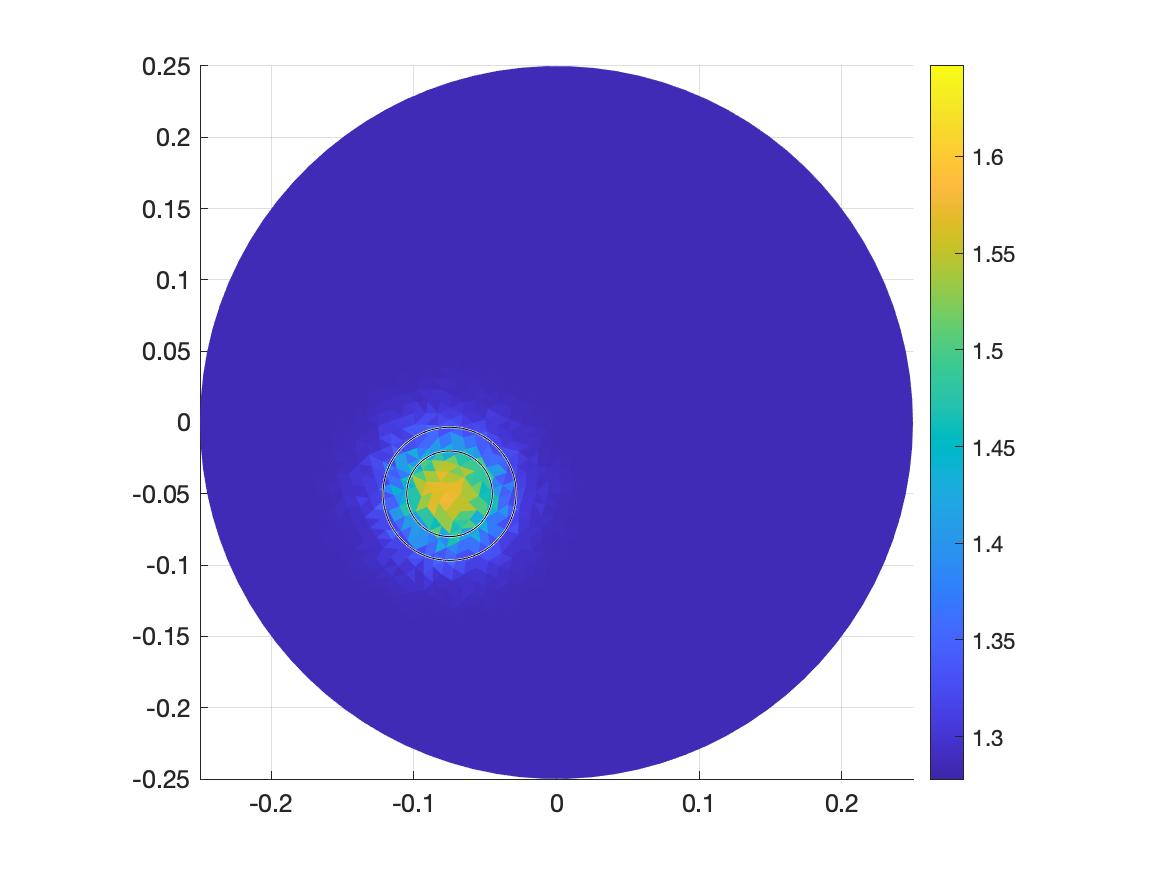}}
		\subfigure[Lac in Astrocyte at $t=4.8$ min]{\includegraphics[width=0.32\textwidth]{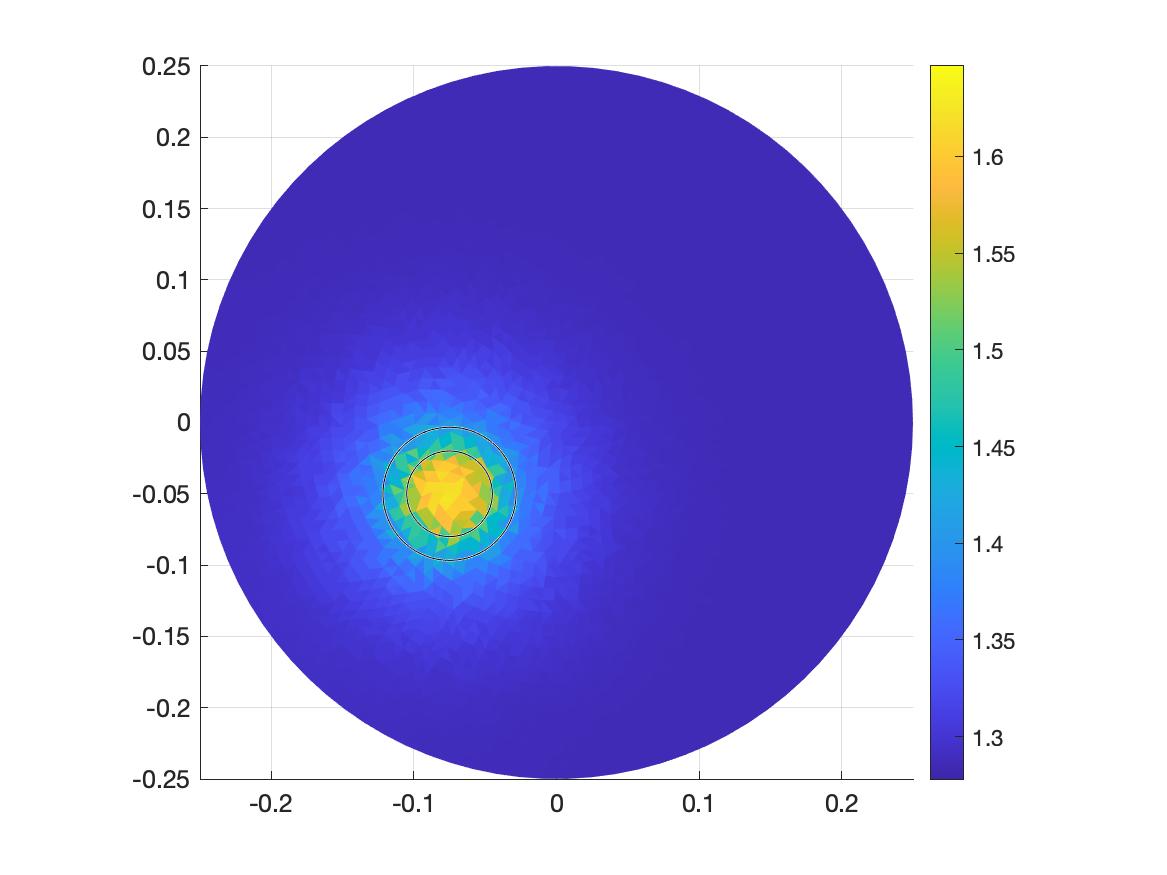}}
		\subfigure[Lac in Astrocyte at $t=6.8$ min]{\includegraphics[width=0.32\textwidth]{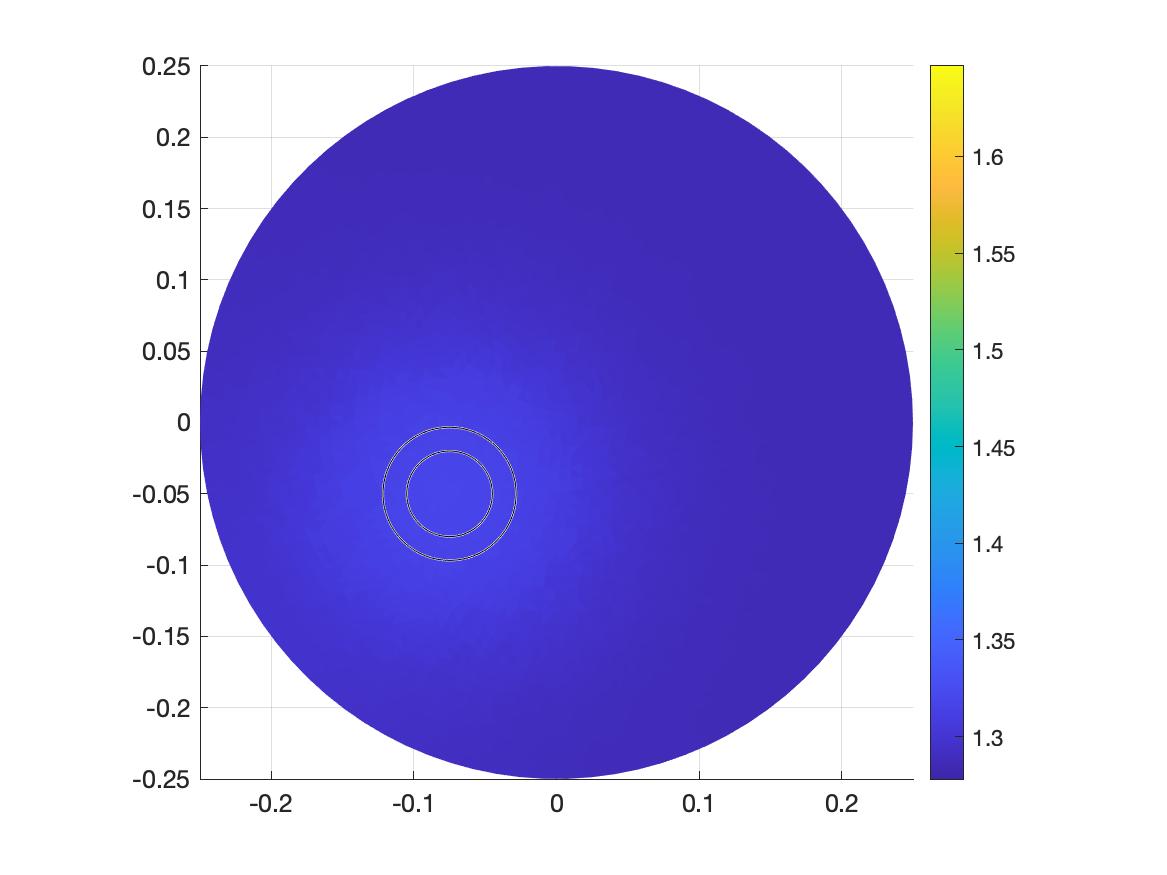}}
		\subfigure[Pyr in Astrocyte at $t=2.4$ min]{\includegraphics[width=0.32\textwidth]{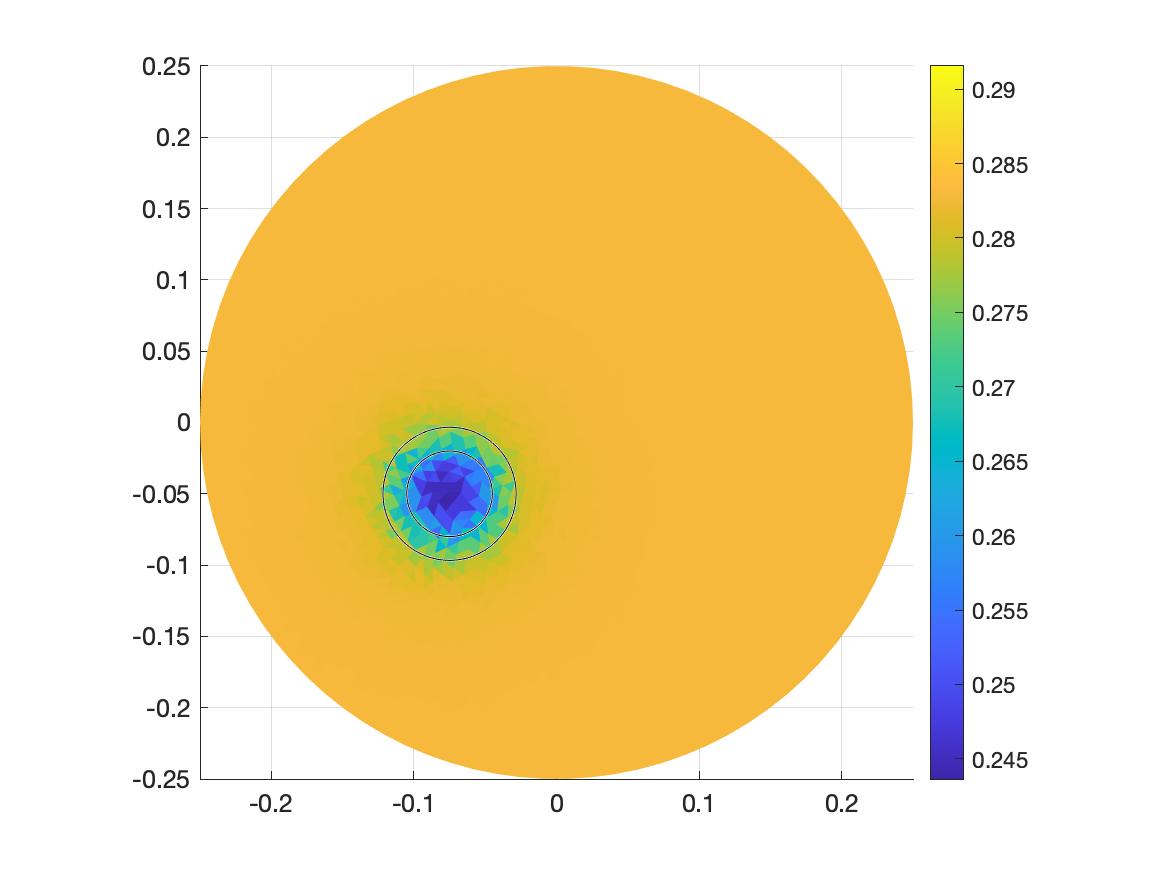}}
		\subfigure[Pyr in Astrocyte at $t=4.8$ min]{\includegraphics[width=0.32\textwidth]{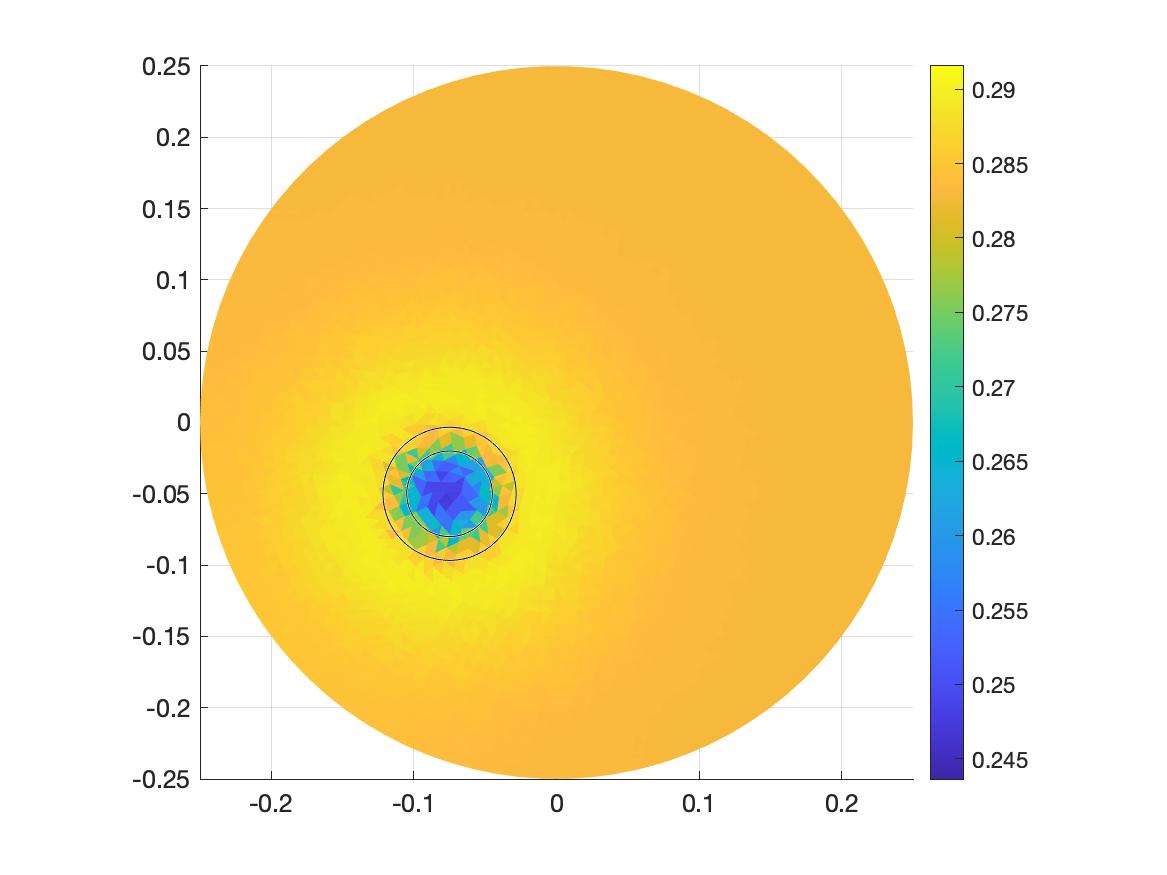}}
		\subfigure[Pyr in Astrocyte at $t=6.8$ min]{\includegraphics[width=0.32\textwidth]{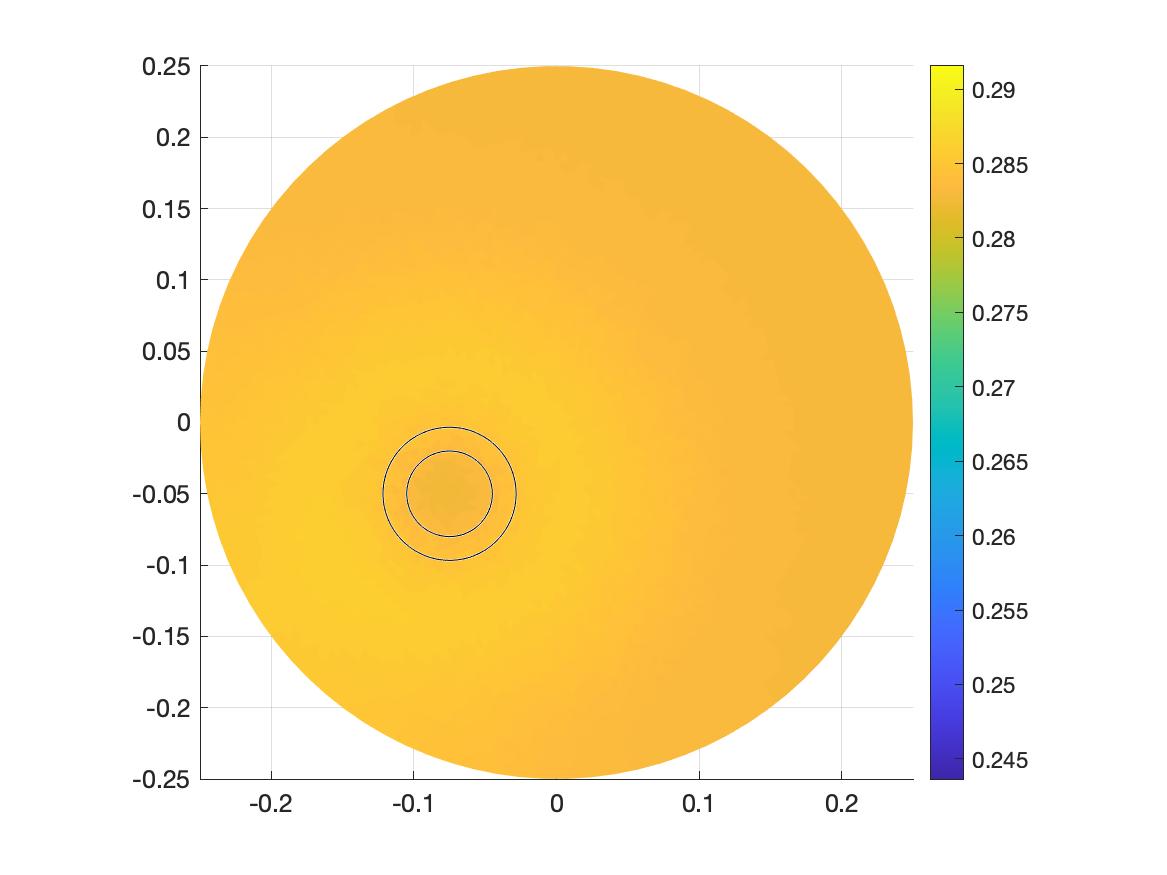}}
		\captionsetup{width=.95\textwidth}
		\caption{Snapshots of the concentrations of glucose (row 1), oxygen (row 2), lactate (row 3), and pyruvate (row 4)  in the astrocyte compartment. As in Figure~\ref{fig:result_4}, the gap junction strength was set as $s=1$ and tortuosity of the ECS at the normal value $\lambda =1.6$, and the snapshots correspond to times  24 seconds after the activation onset ($t = 2.4$ minutes), 12 seconds before the end of the activation ($t = 4.8$ minutes), and at 108 seconds after the end of the activation ($t = 6.8$ minutes). The core and the margin of the activity are again indicated by the black circles.}
		\label{fig:result_5}
	\end{figure*}

	During the simulation, we follow the  time courses of metabolites in all compartments of the model through a time window of $t = 30$ minutes. In this experiment, the  gap junction strength is set to $s = 1$, and tortuosity to $\lambda = 1.6$. Figures~\ref{fig:result_4} and \ref{fig:result_5} show snapshots of the concentration distribution of glucose, oxygen, lactate and pyruvate in neuron and the astrocyte, respectively, at times $t = 2.4$ minutes, or $24$ seconds after the onset of the elevated activity level, $t=4.8$ minutes, or 12 seconds before the end of the activity, and $t = 6.8$ minutes, or 108 seconds after the end of the activity. 
	The plots show that diffusion causes a change in the concentrations of metabolites in an area beyond the activity patch, or core of the activity, indicated by a circle in the plots. In both intracellular compartments, we observe a decrease in glucose, oxygen and pyruvate concentrations and an increase of lactate concentration in the core of the activity, as well as in the margin of the activity.
	
	In neuron, lactate returns to equilibrium much faster than the other metabolites, while glucose shows a rather slow return to equilibrium. In astrocyte, the concentration of metabolites others than glucose return to an equilibrium state faster than in neuron. Due to the diffusion in astrocyte, the area with a higher concentration of glucose and pyruvate is wider than in neuron. We remark that, even though the model does not assume diffusion in the neuron compartment, diffusion in ECS and astrocyte paired with the interaction between the cell type, indirectly affects the neuron compartment, too.
	
	After the baseline simulation with normal neuronal activity, we test the effect of different parameters by runnin three different simulation protocols. The first test is designed to investigate the effect of gap junction strength in astrocyte, the second one addresses the role of tortuosity, and the third one investigates anisotropic diffusion in astrocyte.

	\subsubsection{Protocol 1: Gap junction strength}
	Experimental results reported in the literature have rather mixed outcomes about the role of gap junctions, with some suggesting that gap junction communication is beneficial to cells, and others warning that they may extend damaged areas in some pathological conditions. In our first series of computed experiments, we test different strengths of the gap junctions by varying 
	$s$ in  equation (\ref{gap_junction}), setting $s=0$ corresponding to a block in astrocytic diffusion, $s= 0.25$ for a partial block, $s=1$ assuming free diffusion between astrocytes, and $s=4$, which is a hypothetical scenario of amplification of diffusion by gap junctions. 
	
		\begin{figure*}[!ht]
		\centering
		\subfigure{\includegraphics[width=0.42\textwidth]{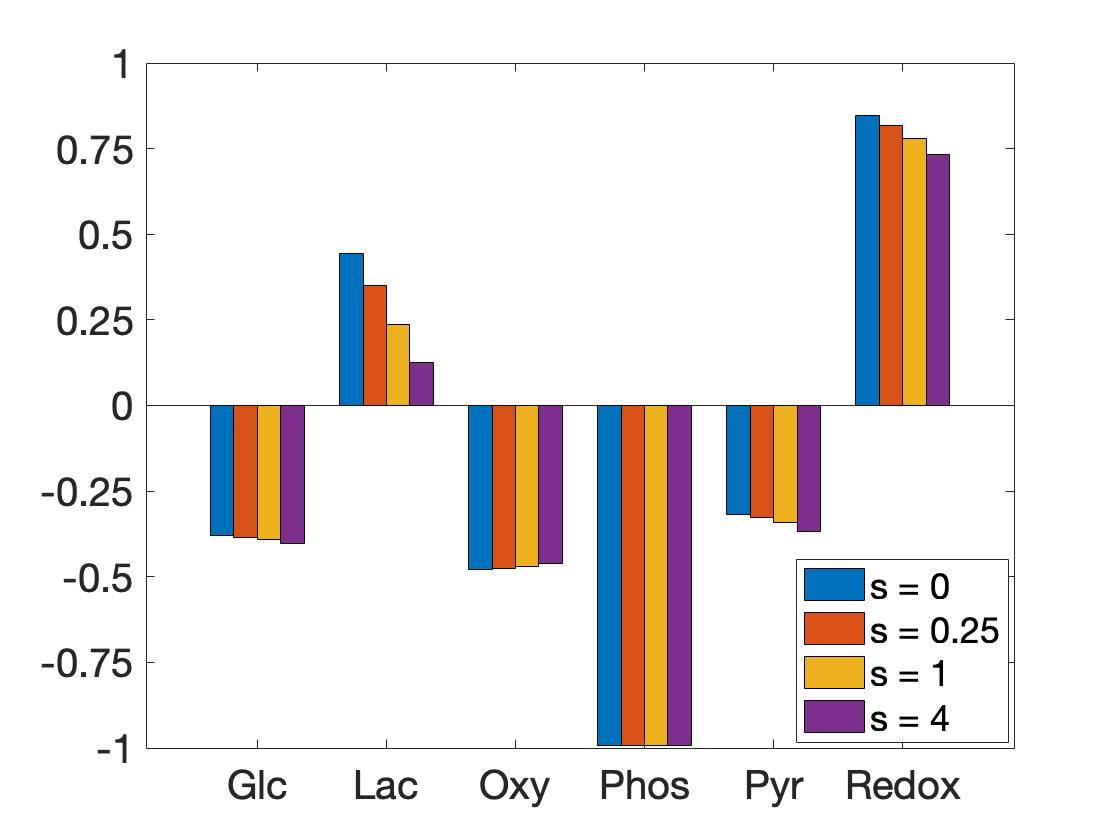}}
		\subfigure{\includegraphics[width=0.42\textwidth]{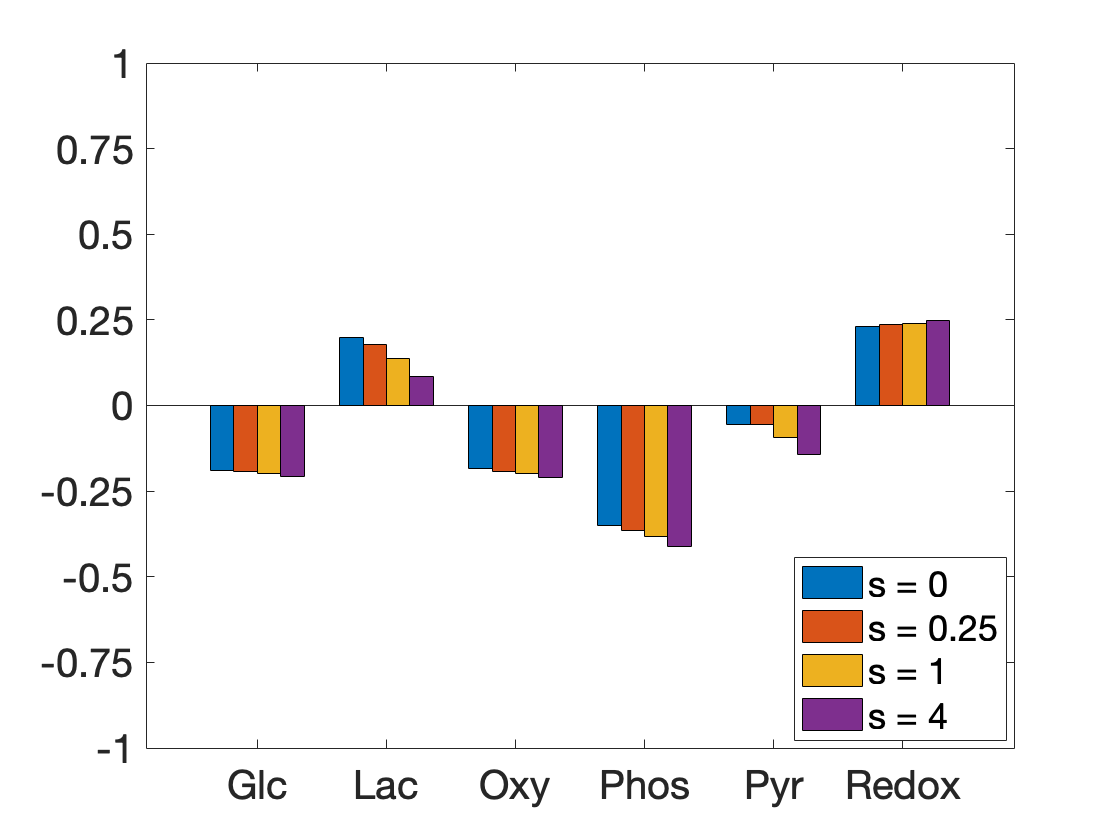}}
		\subfigure{\includegraphics[width=0.42\textwidth]{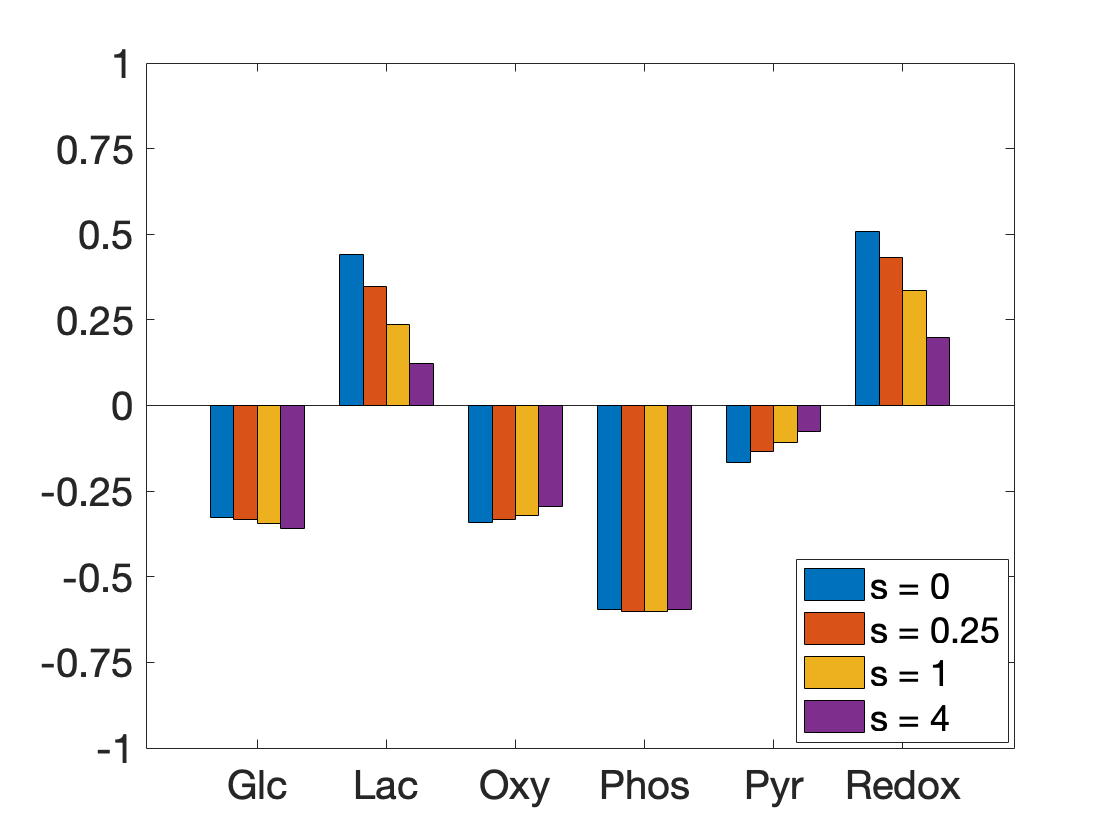}}
		\subfigure{\includegraphics[width=0.42\textwidth]{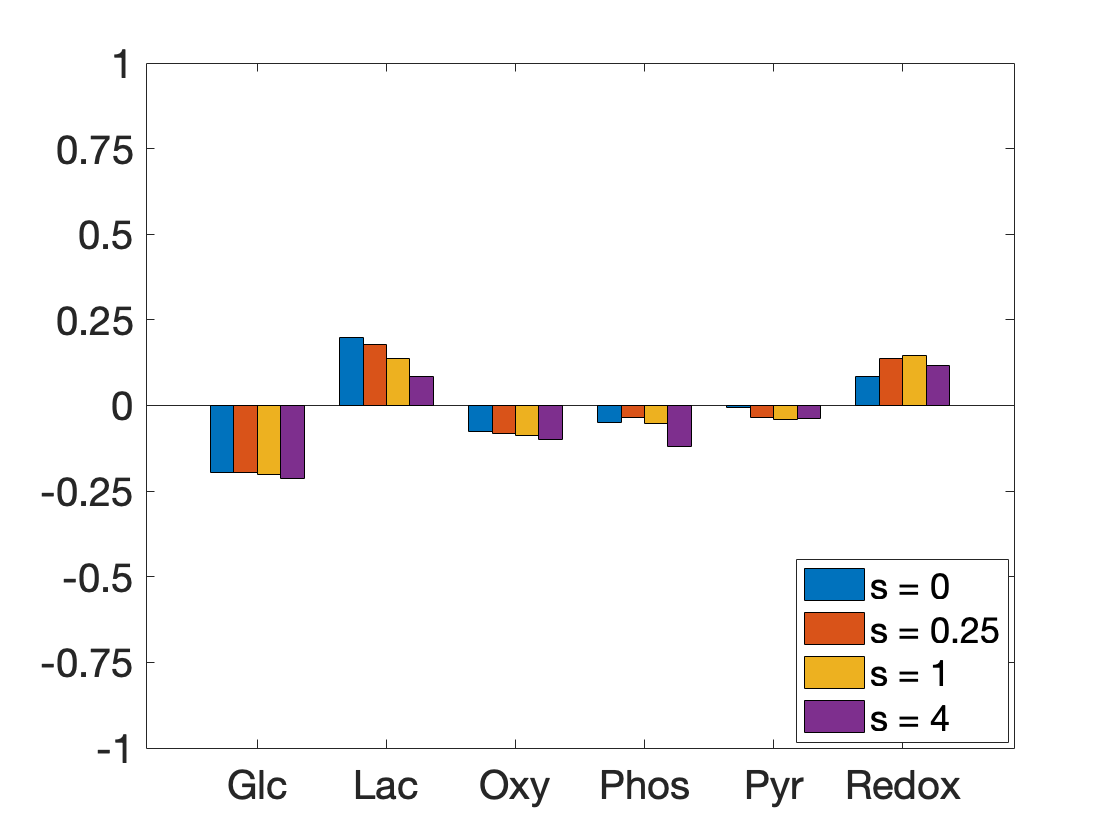}}
		\captionsetup{width=.95\textwidth}
		\caption{Relative maximum changes of metabolite concentrations for different values of the gap junction strength. The panels in the first column correspond to the core activity region, while those in the second column are for the marginal activity region. The top row corresponds to neuron compartment, and the bottom row to the astrocyte compartment.	
		}
		\label{fig:protocol 1}
	\end{figure*}
	
	The results of the simulations indicate that in the cellular domains in the core activity region, the diffusion process in general reduces the drop of glucose and oxygen concentration and the increase of the lactate concentration by distributing the metabolites more evenly into the margin area. This supports the hypothesis that stronger gap junctions will distribute the metabolic changes in response to neuronal activation more evenly over a wider region.  To test this hypothesis and to quantify the results, we consider the average concentrations over the core and the margin regions. Thus, if $u(x,t)$ denotes any particular metabolite concentration, the average over a domain $S$ is given by 
	\[
	\overline u_S(t) = \frac{1}{|S|}\int_S u(x,t) dt.
	\]
	For each metabolite, we compute the signed relative maximum change over the simulation period compared to the baseline steady state concentration $u^0$,
	\[ 
	u_S^* = \pm \frac{\max\{ |\overline u_S(t) - u^0| : 0\leq t\leq T\}}{u^0},
	\]

	where the sign is chosen according to whether the concentration has increased (+) or decreased (-) as a result of neuronal activity. Hence, for instance, lactate concentration assumes a positive sign, and oxygen a negative sign. The results, summarized in Figure~\ref{fig:protocol 1}, indicate that in both neuron (first row) and astrocyte (second row), the maximum average lactate concentrations are most strongly affected by the gap junction strength. At the core of activity in neuron, lactate increases up to about $44\%$ from its baseline values when $s = 0$, while  when  $s =  0.25$, the increase is $35\%$. When $s=1$, lactate increase is only up to $24\%$, and at $s=4$, the maximal increase in concentration at the core is $13\%$. Similar results hold in the astrocyte compartment. Observe that the redox state that is known to follow the lactate/pyruvate ratio shows similar behavior, while the effect on glucose and oxygen, as well as on the phosphorylation states is significantly less prominent.
	Relative change in redox state in the neurons of the core area goes from about $85\%$ increase at $s=0$ to a $73\%$ increase at $s=4$. The plot also shows that the relative change in oxygen concentration is higher rather than lower with increasing $s$, suggesting that the margin region is supplying oxygen to the core area by diffusion more effectively with increasing $s$.
	
	\begin{figure*}[!ht]
		\centering
		\subfigure{\includegraphics[width=0.24\textwidth]{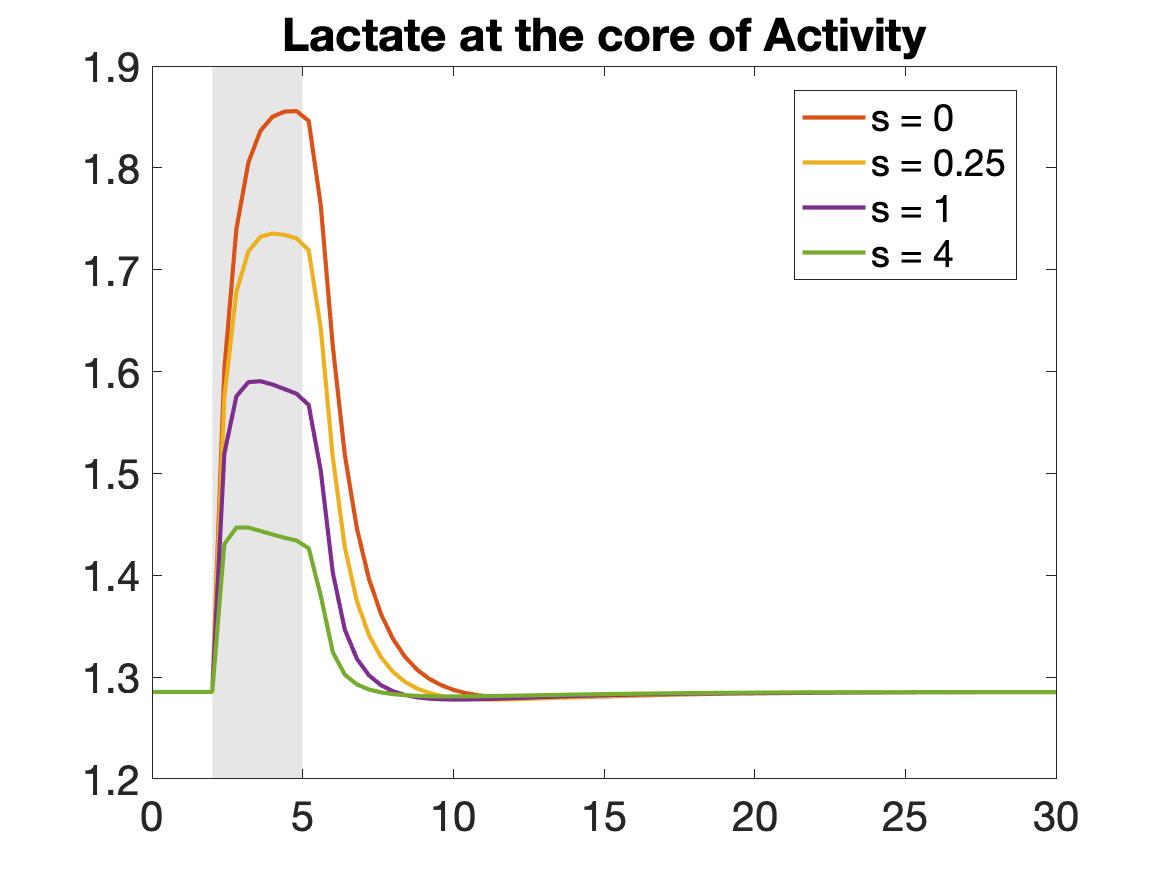}}
		\subfigure{\includegraphics[width=0.24\textwidth]{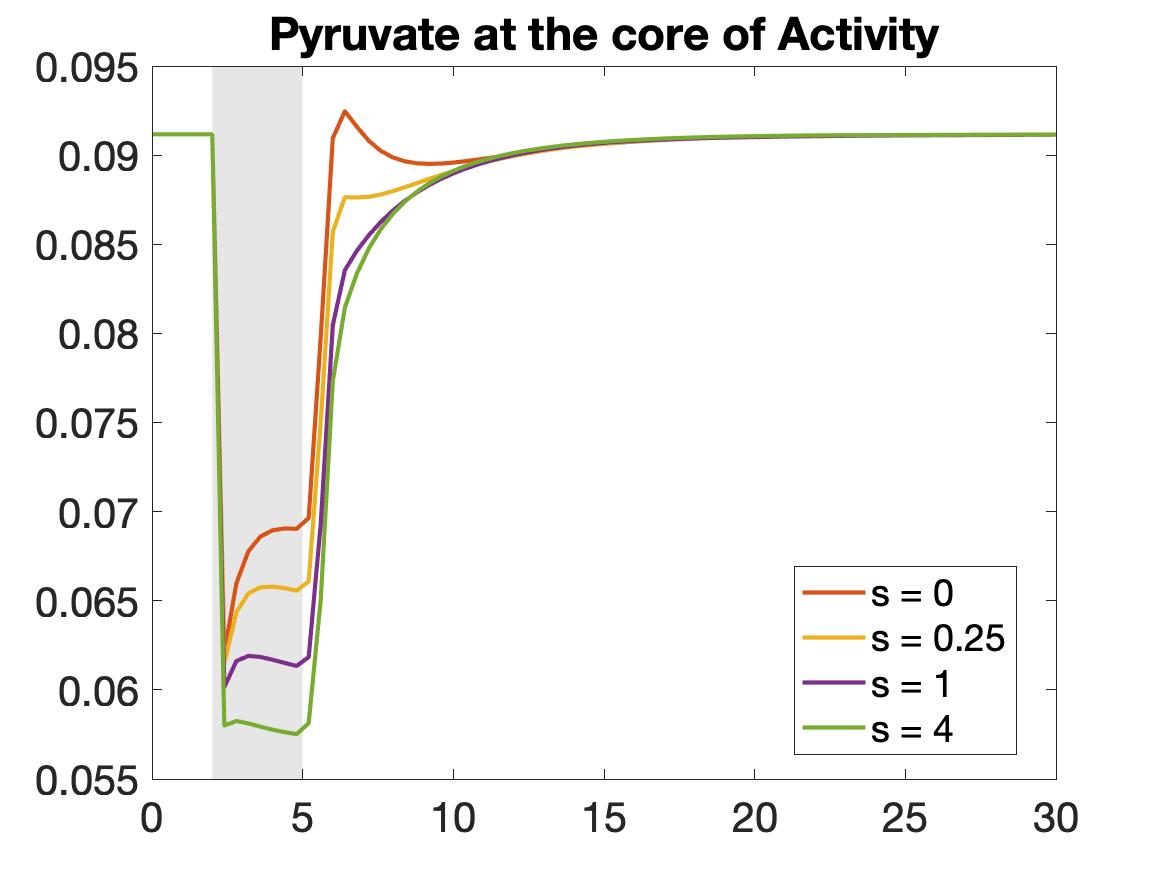}}
		\subfigure{\includegraphics[width=0.24\textwidth]{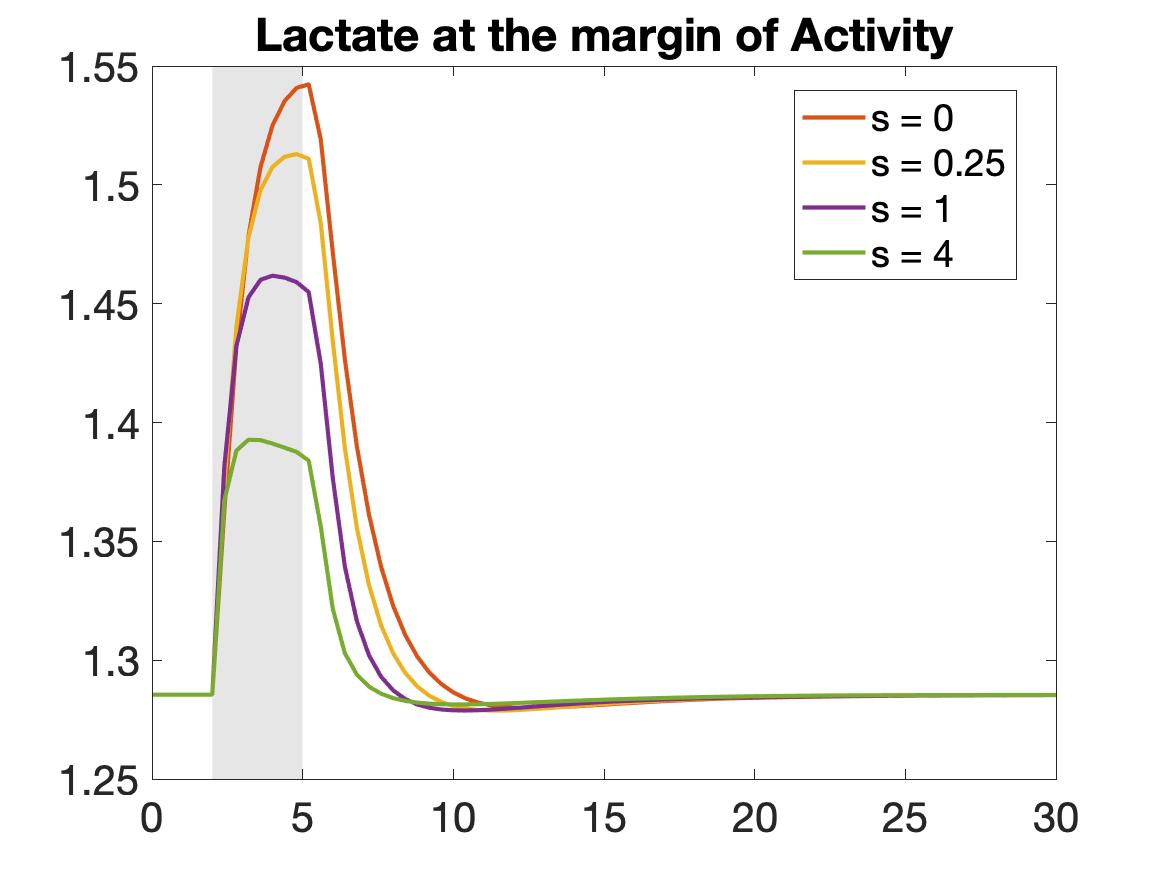}}
		\subfigure{\includegraphics[width=0.24\textwidth]{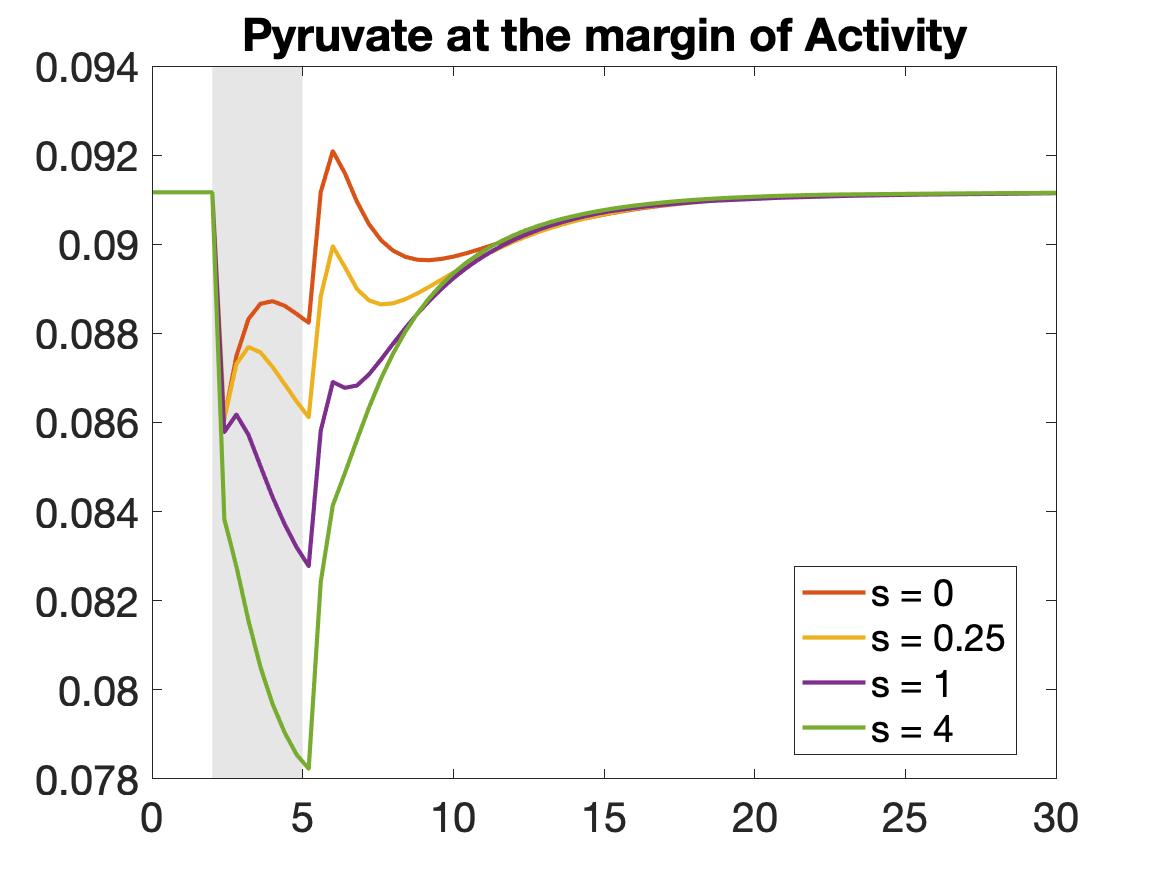}}
		\subfigure{\includegraphics[width=0.24\textwidth]{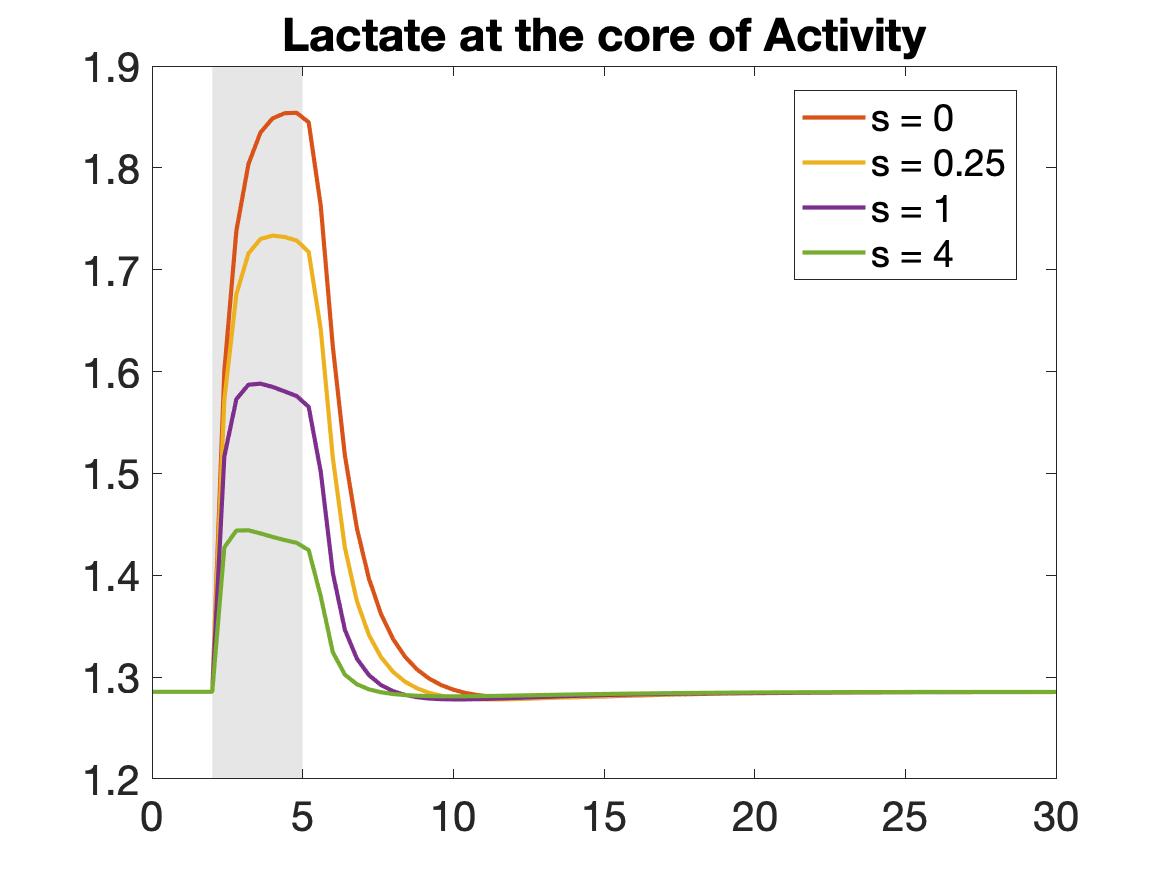}}
		\subfigure{\includegraphics[width=0.24\textwidth]{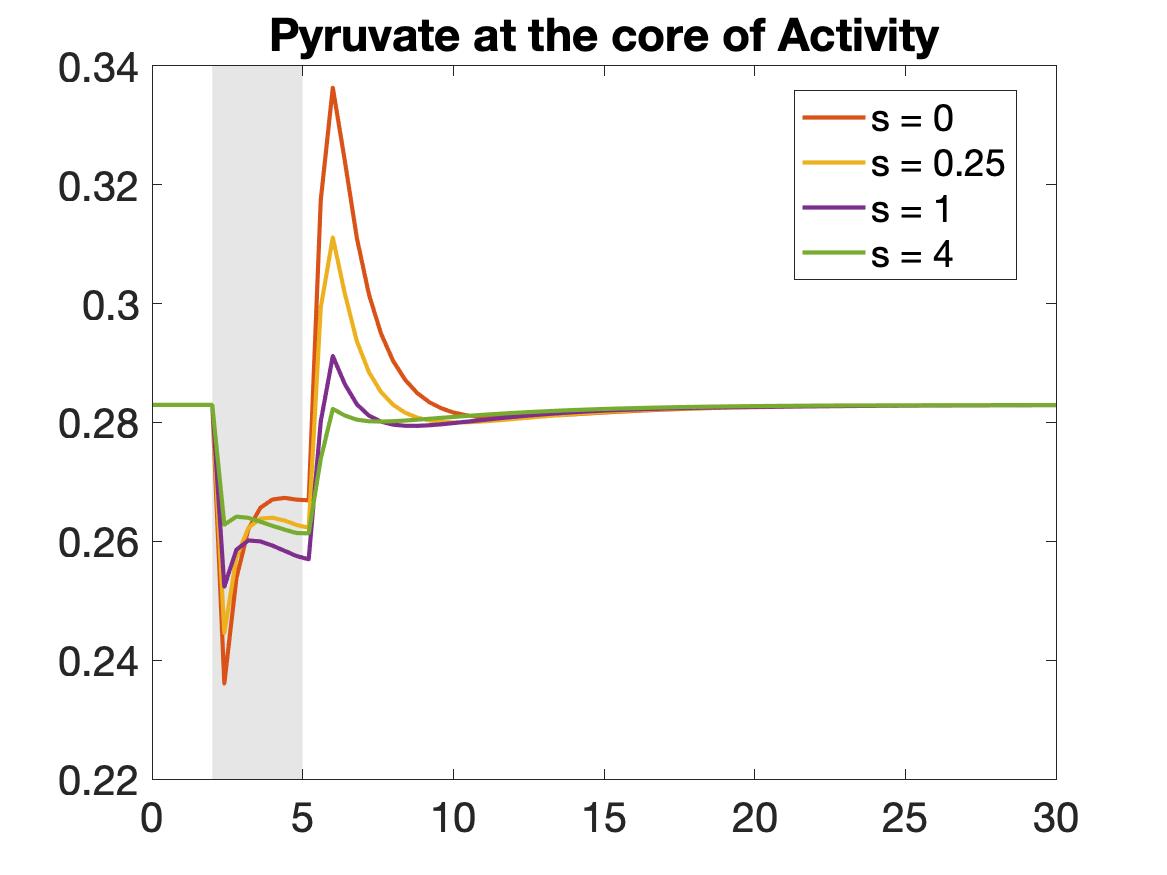}}
		\subfigure{\includegraphics[width=0.24\textwidth]{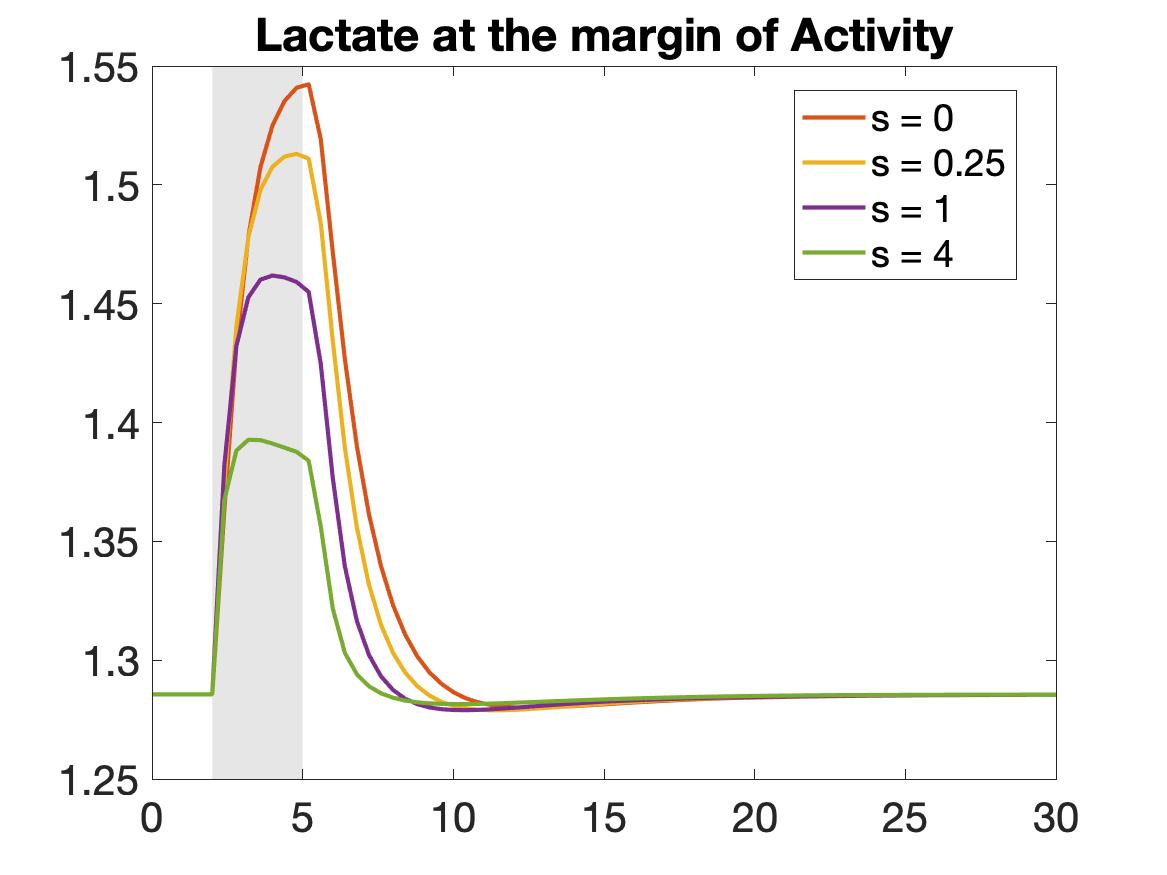}}
		\subfigure{\includegraphics[width=0.24\textwidth]{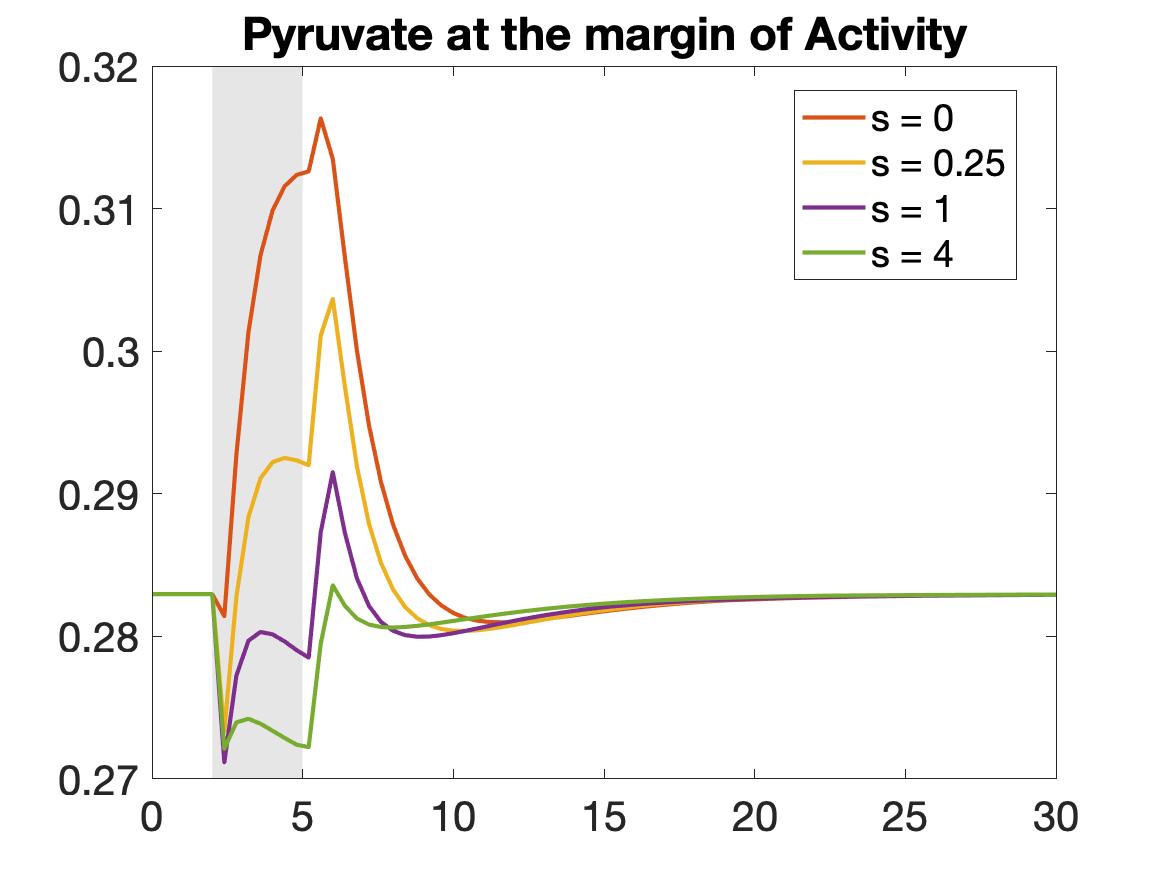}}
		\captionsetup{width=.95\textwidth}
		\caption{\label{fig:protocol 1 bis}Time courses $u_S(t)$ of the averaged concentrations of lactate and pyruvate for different values of the gap junction strength $s$. The panels in the top row refer to neuron, and those in the bottom row to astrocyte. In the first two columns, the region of interest $S$ is the core activity area, and in the last two the activity margin. For each figure, the activity period is indicated by a gray shade.}
	\end{figure*}
	
	The time courses of the averaged concentrations $u_S(t)$ are also of interest, as they indicate the rate at which the metabolite concentrations return to baseline values. Figure~\ref{fig:protocol 1 bis} shows selected time courses of selected metabolites for different values of $s$.
	We observe that increasing $s$ speeds up significantly the return to the baseline level of lactate concentration in both astrocyte and in neuron, not only in the core but also in the margin area,  where the blood flow does not change. In the core area, the recovery time in both neuron and astrocyte decreases from approximately 6 minutes when $s=0$ to only one minute when $s=4$. Moreover, while the peak values of pyruvate are minimally affected by the gap junction strength, temporal profiles of pyruvate concentration change significantly with $s$, indicating that the gap junction strength affects the balance between aerobic and anaerobic metabolism. It is worth noting how the pyruvate spike in the astrocyte, appearing in the core area at the end of the activation period, and after the onset of the activation in the marginal area, is dampened by the increased gap junction activity, indicating that the decreased lactate concentration slows down the lactate oxidation by LDH.

	\subsubsection{Protocol 2: Tortuosity}
	In this protocol, we investigate the effect of increased tortuosity of the ECS, which could be related, e.g., to pathological cell swelling.  Experimental evidence suggests that the value of the tortuosity parameter $\lambda$ of a healthy brain tissue is approximately $1.6$, whereas in pathologies such as stroke or cortical spreading depolarization, the value of the tortuosity parameter is believed to be much higher. In the following, we run the baseline model with three different tortuosity values, $\lambda = 1.6$, $2.0$, and  $3.0$. 
	
\begin{figure*}[!ht]
		\centering
		\subfigure{\includegraphics[width=0.24\textwidth]{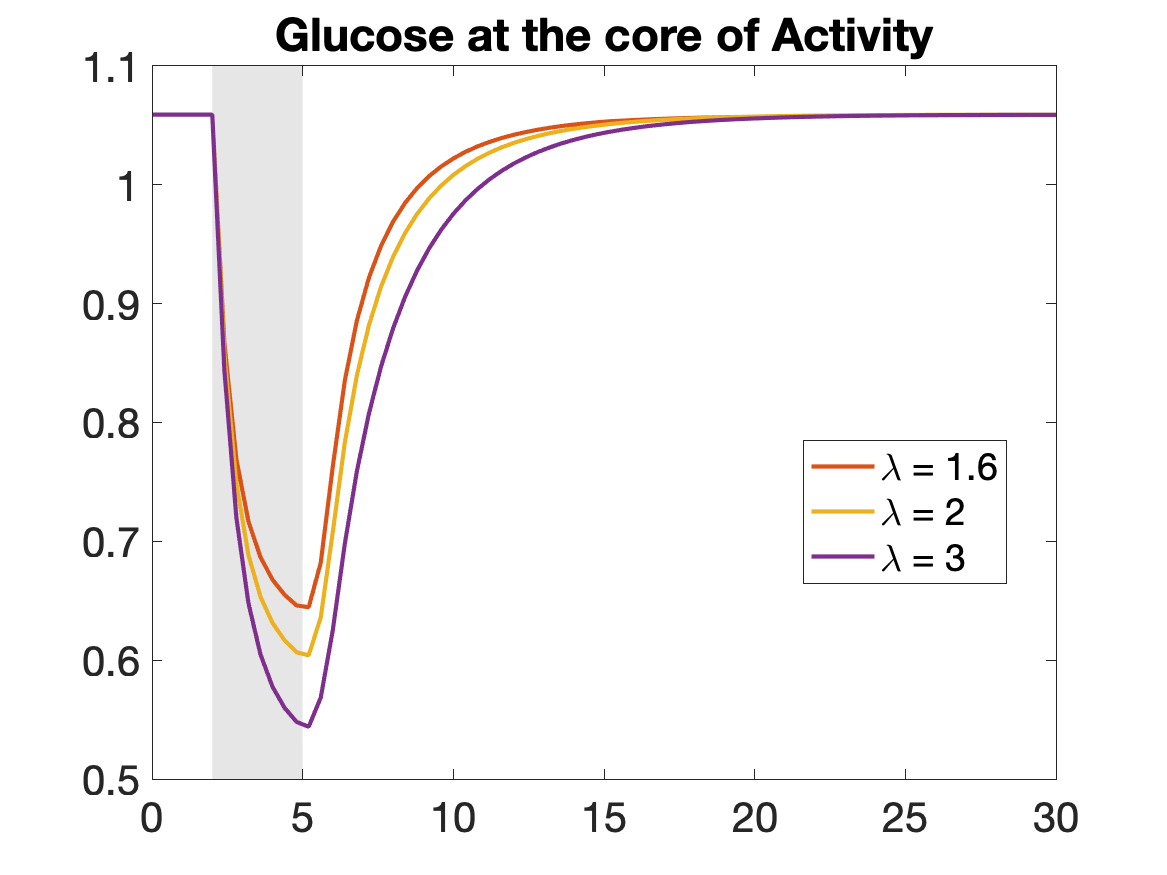}}
		\subfigure{\includegraphics[width=0.24\textwidth]{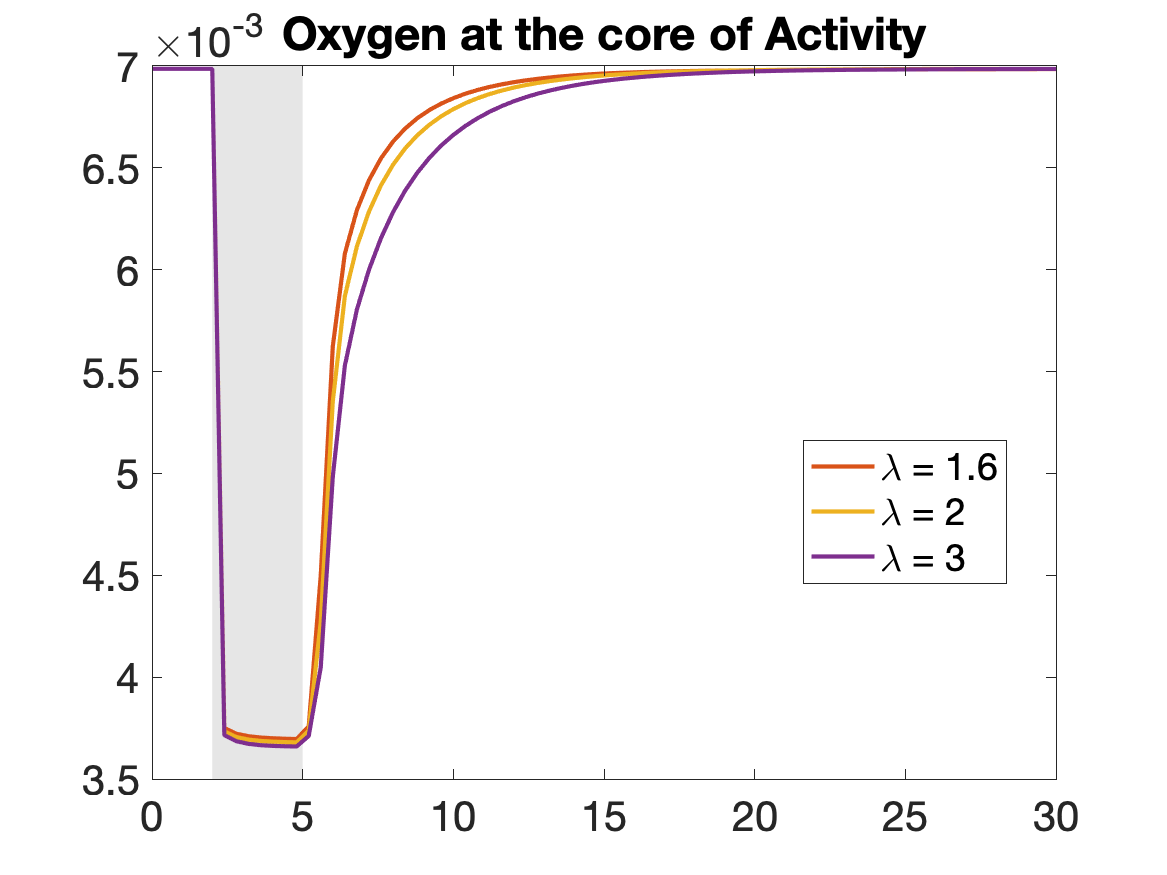}}
		\subfigure{\includegraphics[width=0.24\textwidth]{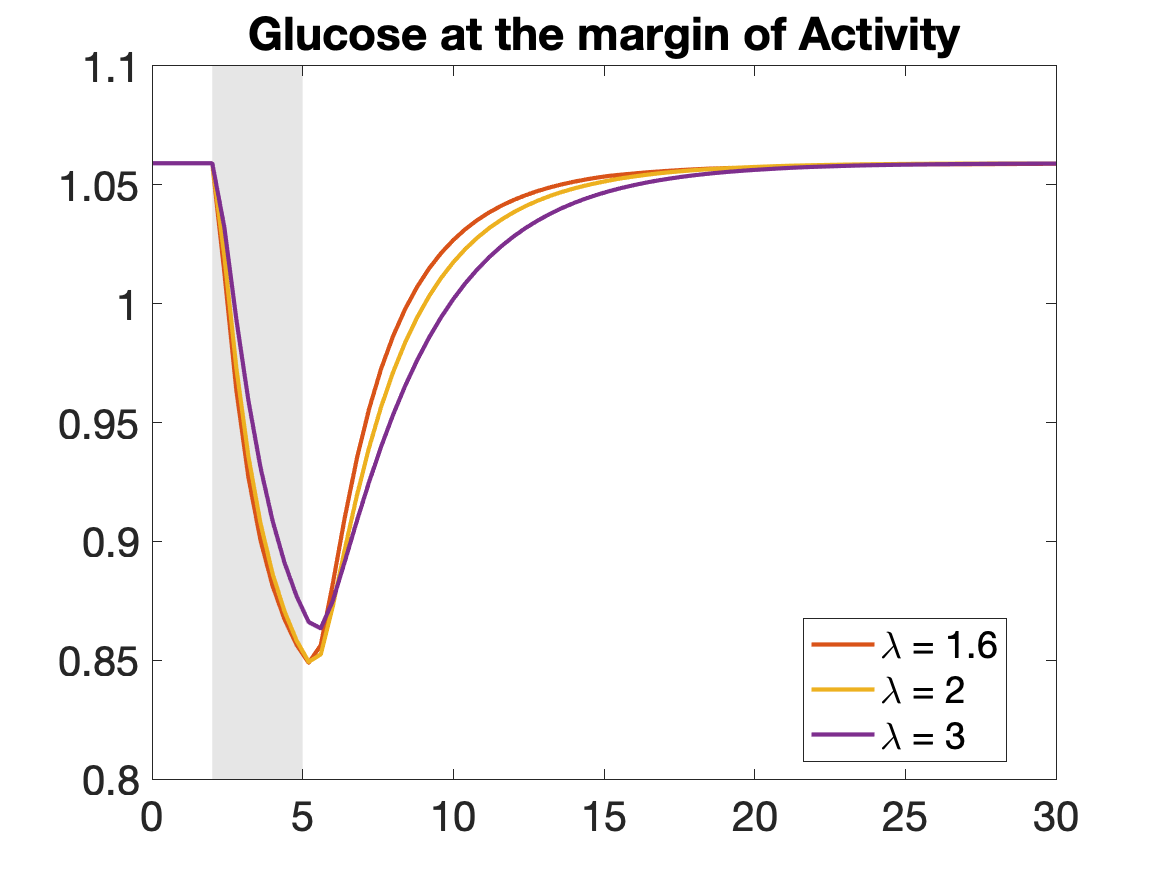}}
		\subfigure{\includegraphics[width=0.24\textwidth]{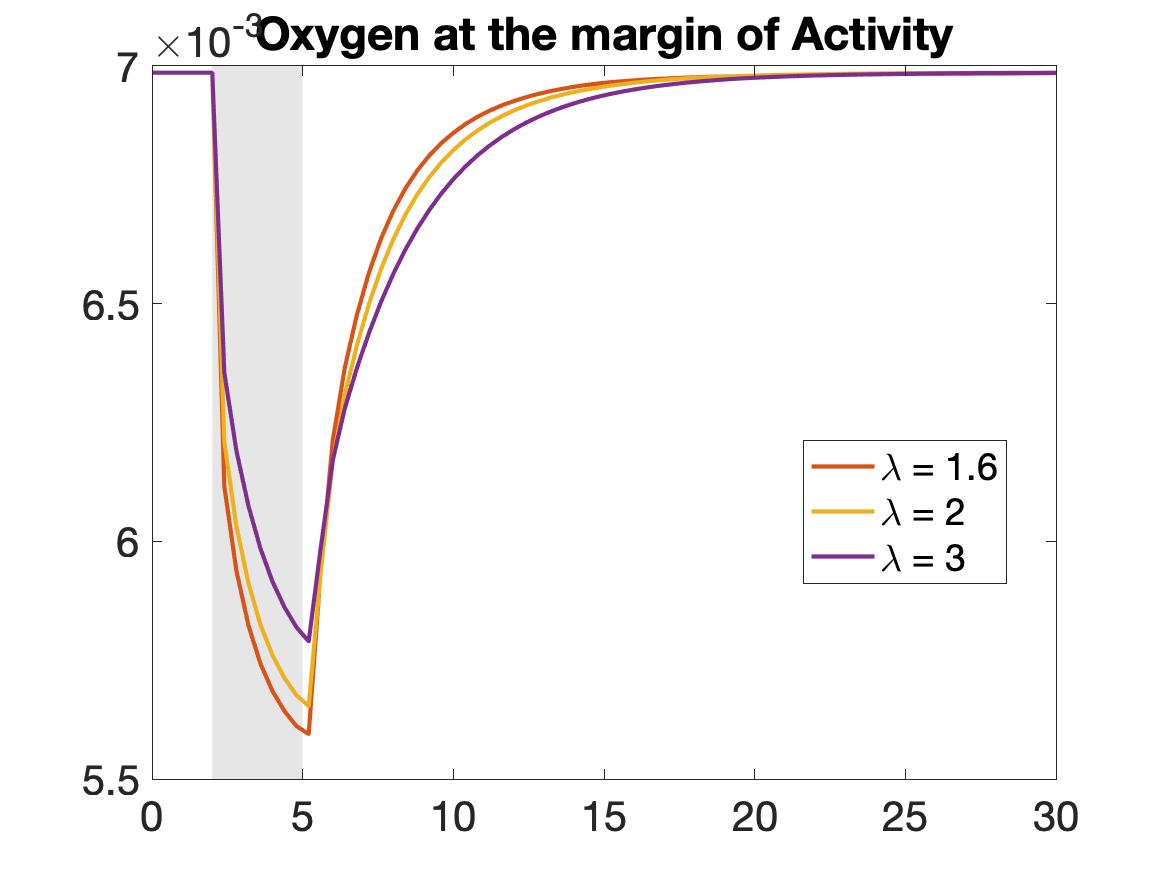}}
		\subfigure{\includegraphics[width=0.24\textwidth]{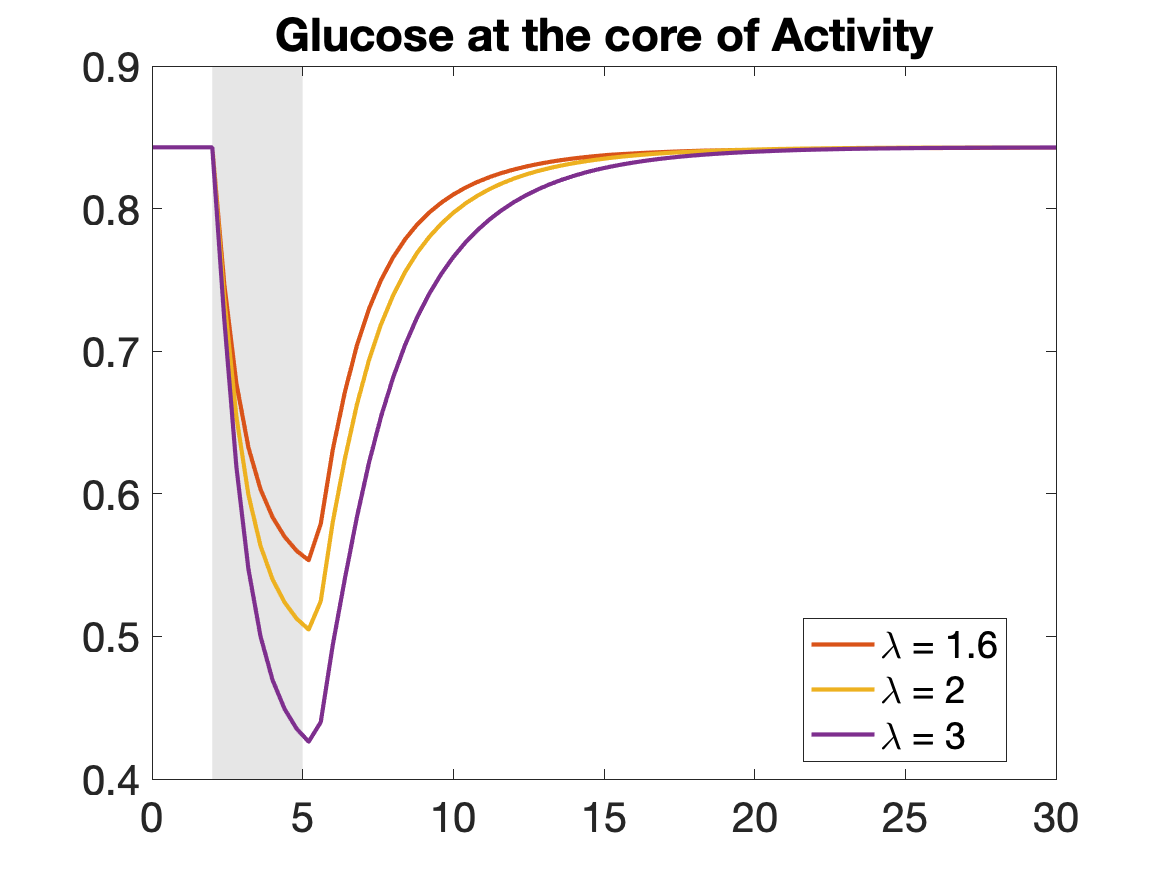}}
		\subfigure{\includegraphics[width=0.24\textwidth]{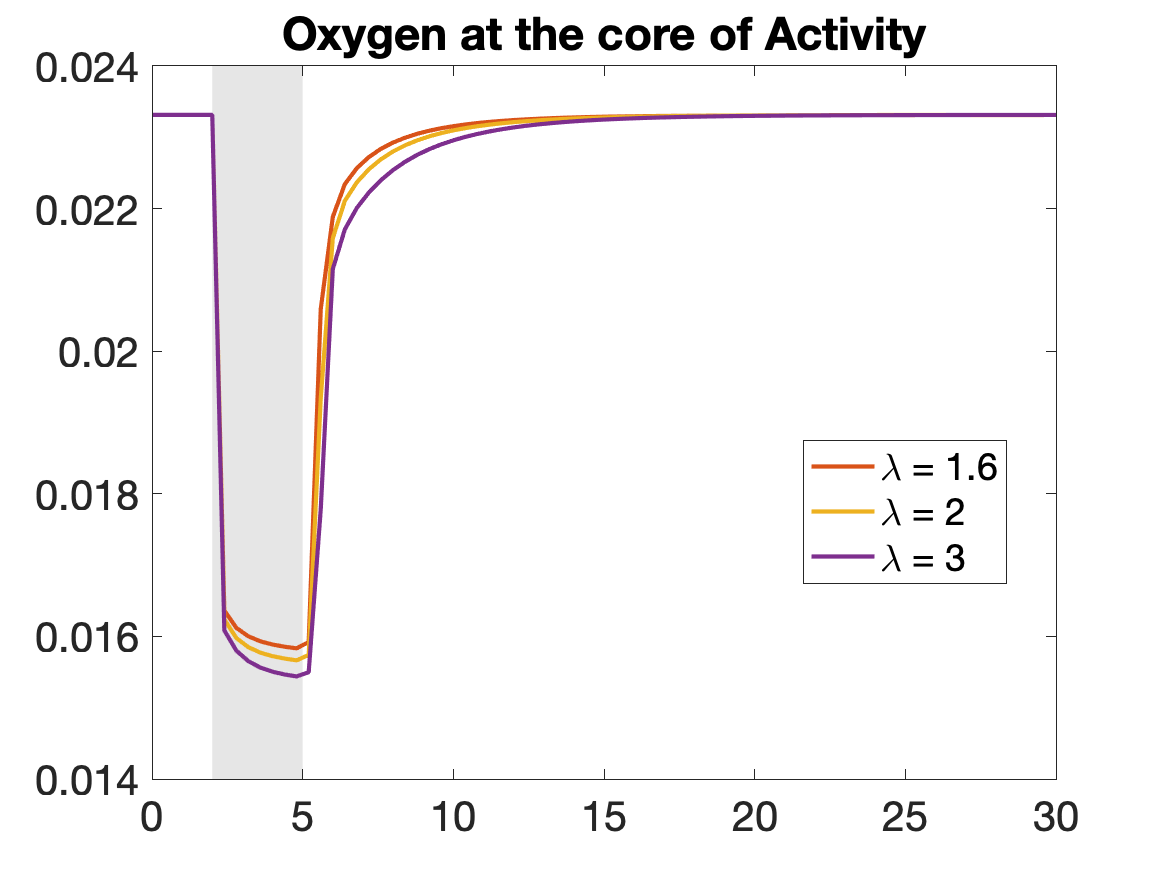}}
		\subfigure{\includegraphics[width=0.24\textwidth]{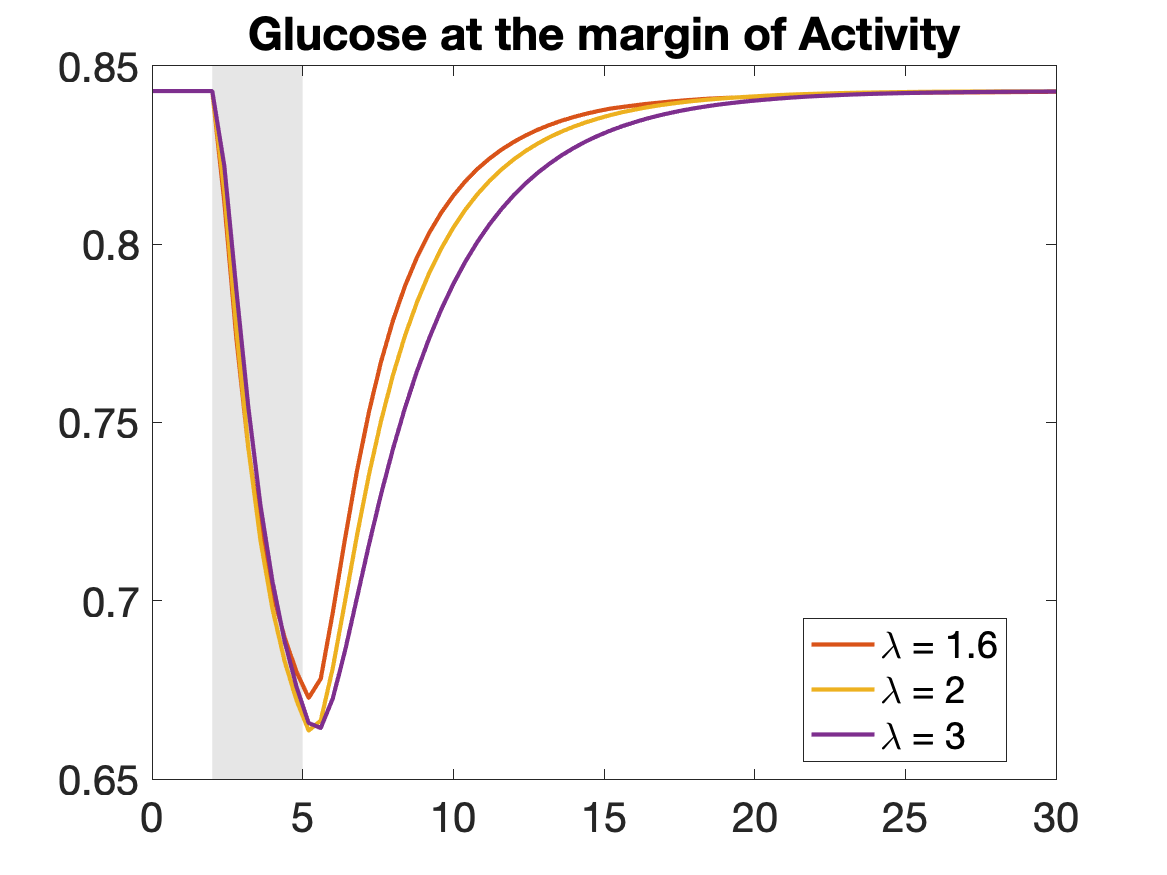}}
		\subfigure{\includegraphics[width=0.24\textwidth]{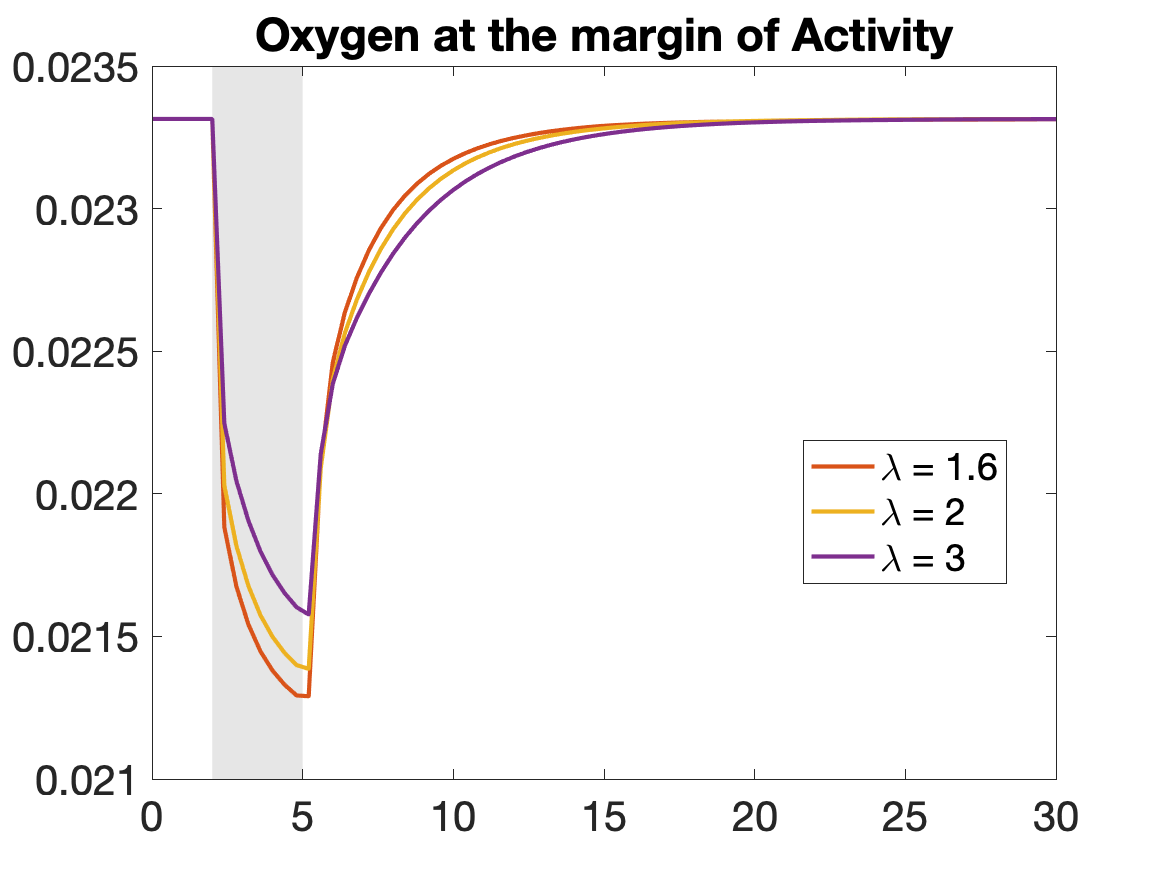}}
		\captionsetup{width=.95\textwidth}
		\caption{Snapshots of the  the averaged concentrations of glucose and oxygen in neuron (top row) and in astrocyte (bottom row). The first two columns are averages over the core activity region, and the other two refer to the margin activity region. The three curves in each plot correspond to different values of tortuosity $\lambda$ in the ECS compartment. The activation period is indicated by the gray shade.
		}
		\label{fig:protocol 2}
	\end{figure*}

	The numerical simulations indicate that, unlike the gap junction strength that most prominently affected the pyruvate and lactate concentrations, the tortuosity has a notable effect on the glucose and oxygen levels. The averaged time courses $u_S(t)$ of glucose and oxygen in both cell types in the core and at the margin areas are shown in Figure~\ref{fig:protocol 2}.

	We observe that increasing tortuosity does not  change significantly the oxygen profile in the core activity area, while the glucose concentration drops.  Interestingly, the opposite is true for the marginal activity area, where the drop of the oxygen concentration is less pronounced as tortuosity increases, and the glucose concentration is only marginally affected. A plausible explanation for this specular behavior is that during the period of the activation, the marginal area provides oxygen to the core area through diffusion. When diffusion is hindered by increased tortuosity, the oxygen remains segregated in the margin area, which therefore shows increased oxygen availability. In turn, the core area has less oxygen available, hence must rely more on wasteful anaerobic metabolism, with a consequent significant dip in glucose concentration.
	
		\begin{figure*}[htbp]
		\centering
		\subfigure{\includegraphics[width=0.3\textwidth]{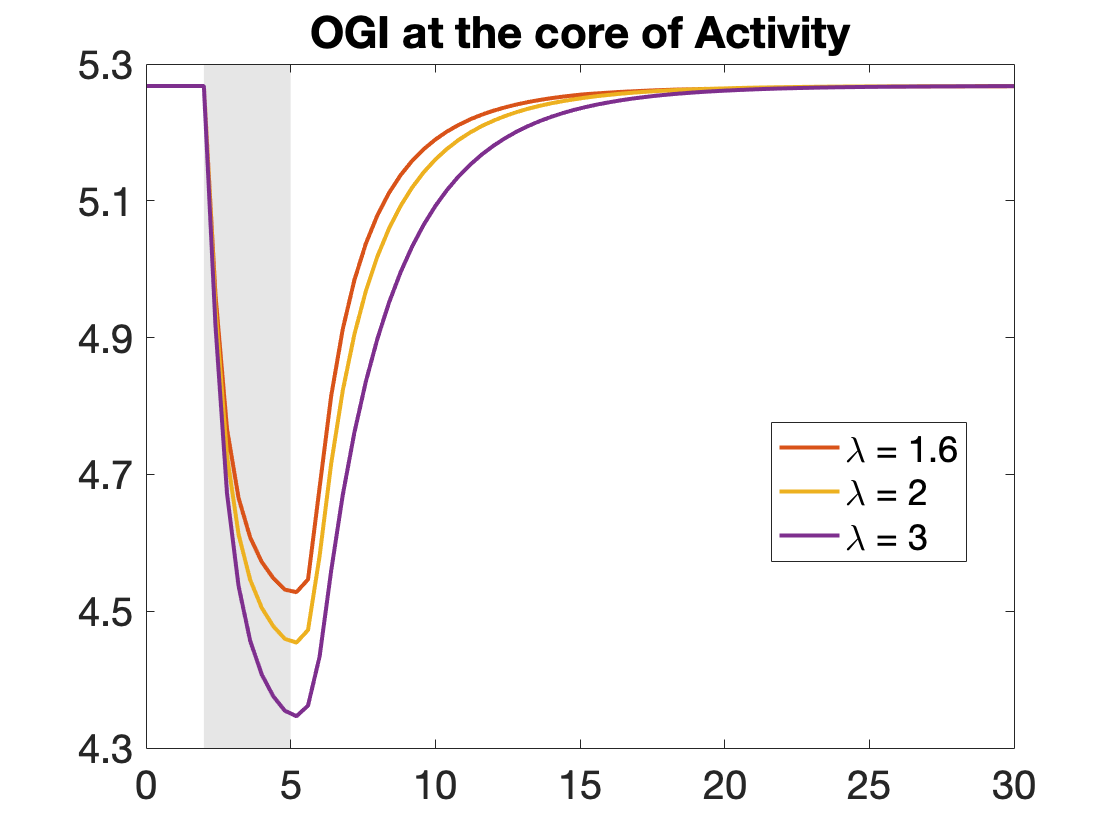}}
		\subfigure{\includegraphics[width=0.3\textwidth]{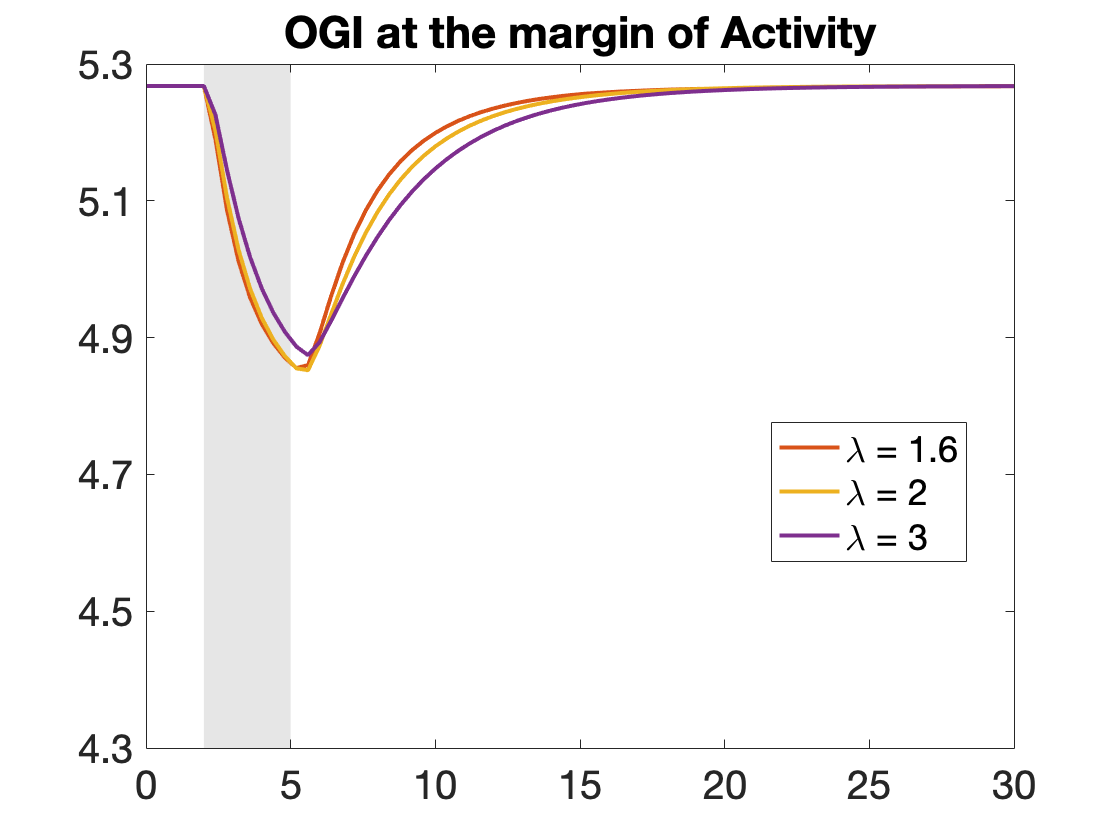}}
		\captionsetup{width=.95\textwidth}
		\caption{Time courses of average OGI for protocols 2 at core and margin of activity areas, respectively, for varying tortuosity values in the ECS. The activation period is indicated by the gray shade.}
		\label{fig:OGI}
	\end{figure*}
	
	To test the robustness of the proposed explanation, we consider the Oxygen-Glucose Index (OGI) which is a reliable measure of the level of aerobic/anaerobic metabolism state in tissue. The OGI is defined as the ratio between the flux of the oxygen and glucose from the blood compartment to the extracellular space \cite{Massucci}, \cite{Shulman2001LactateEA},
	\[
	\rm OGI = \frac{\bphi^{(\rm blood)}_{\rm O_2}}{\bphi^{(\rm blood)}_{\rm Glc}},
	\]
	where $\bphi^{(\rm blood)}_{\rm Glc} = \bphi^{(1)}_{1}  -  \bphi^{(2)}_{1}$ and $\bphi^{(\rm blood)}_{\rm O_2} = \bphi^{(1)}_2 - \bphi^{(2)}_2$ are calculated as in (\ref{glucose_flux}) and (\ref{oxygen_flux}), respectively. It has been reported in literature (\cite{Massucci}, \cite{Shulman2001LactateEA}, \cite{doi:10.1097/00004647-199807000-00005}) that OGI values under normal resting conditions are between $5$ and $5.5$, and between $4$ and $4.5$ for sustained neuronal activation.  Figure~\ref{fig:OGI} shows the OGI values in the core activity area (left) and in the marginal area (right), computed by using the averaged fluxes over the region of interest. As expected, the increased tortuosity lowers the OGI during the activity in the core area as a result of limited diffusive oxygen supply, while in the marginal area, the OGI remains essentially unaltered, indicating that the cells do not profit from the excess oxygen remaining in situ, but rather maintain the balance between aerobic and anaerobic metabolism remarkably stable. Let us point out that the gap junction strength parameter $s$ has essentially no effect on the computed OGI (data not shown here).
	
	\subsubsection{Protocol 3: Anisotropy}

	In the third simulation protocol, we test the effect of anisotropic orientation of the astrocyte network on diffusion and metabolism.
	The astrocytic communication pathways are enabled by junctional proteins called connexins (Cxs). Several reports have demonstrated that gap junctional communication in astrocytes does not involve all astrocytes and that subpopulations of glial cells with a specific phenotype are not mutually coupled \cite{Wallraff}, \cite{houades_rouach_ezan_kirchhoff_koulakoff_giaume_2006}, \cite{Schools}. In \cite{Houades5207}, the authors determined the coupling properties and spatial organization of gap junction-mediated astrocytic networks in layer IV of the primary somatosensory cortex, supporting the hypothesis that  the orientation of the coupling plays a significant role. Within a discrete astrocyte cluster, also known as a barrel, dye coupling is oriented toward the  barrel center, indicating that intercellular exchanges are favored within a cluster rather than between two adjacent barrels.  In \cite{Anders}, it was found that coupling length constant and anisotropy were sensitive spatial measures in astrocytic gap junction coupling. Moreover, the authors report that astrocyte coupling is temperature-dependent and anisotropic in the stratum radiatum of the hippocampal CA1 region. 
	
		\begin{figure*}[!ht]
		\centering
		\subfigure[Lac in Astrocytes at $t=2.4$ min]{\includegraphics[width=0.32\textwidth]{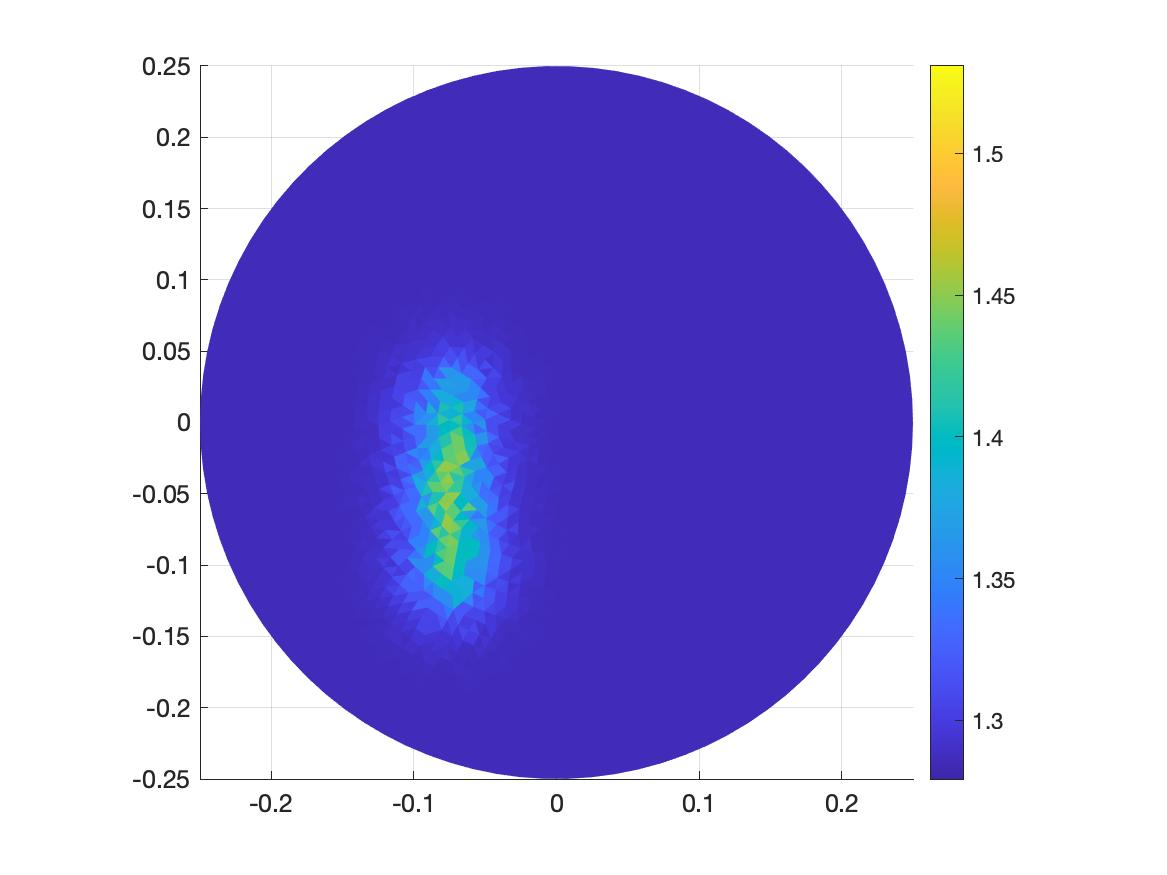}}
		\subfigure[Lac in Astrocytes at $t=4.8$ min]{\includegraphics[width=0.32\textwidth]{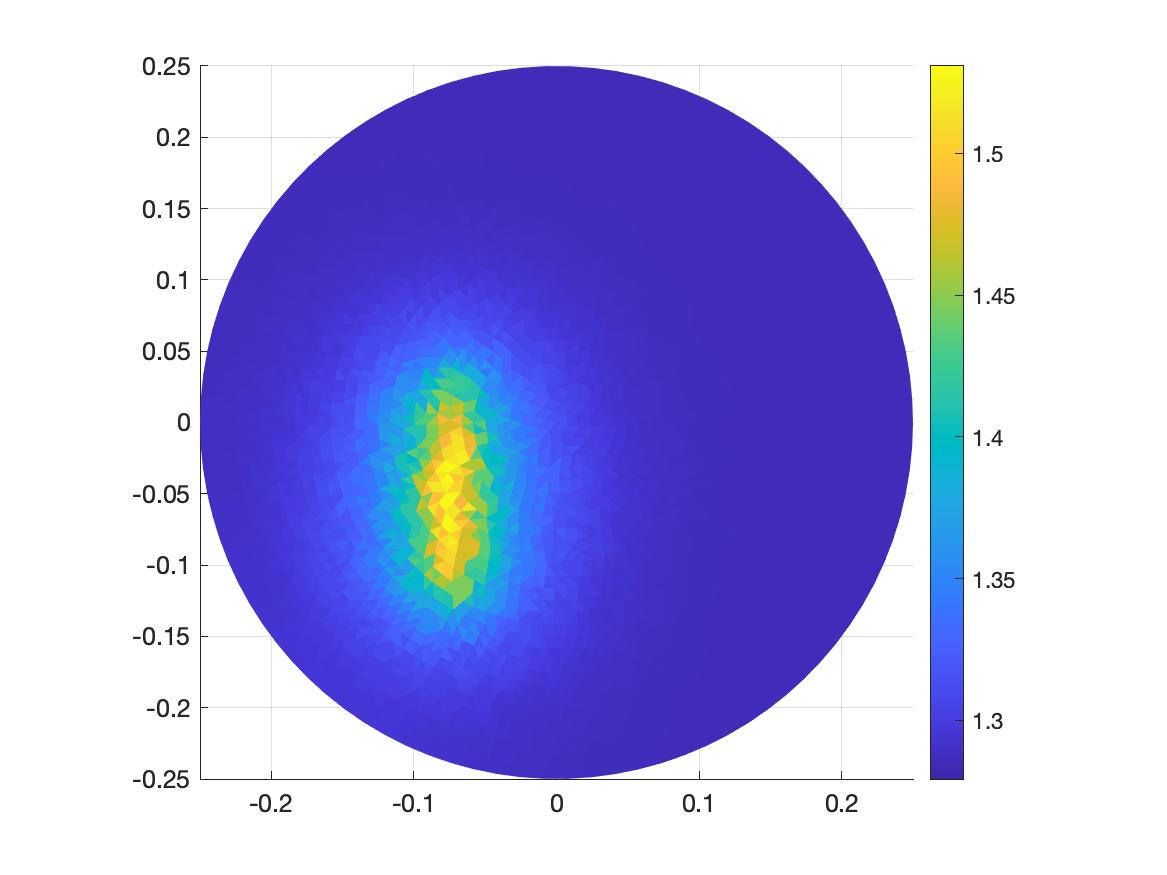}}
		\subfigure[Lac in Astrocytes at $t=6.8$ min]{\includegraphics[width=0.32\textwidth]{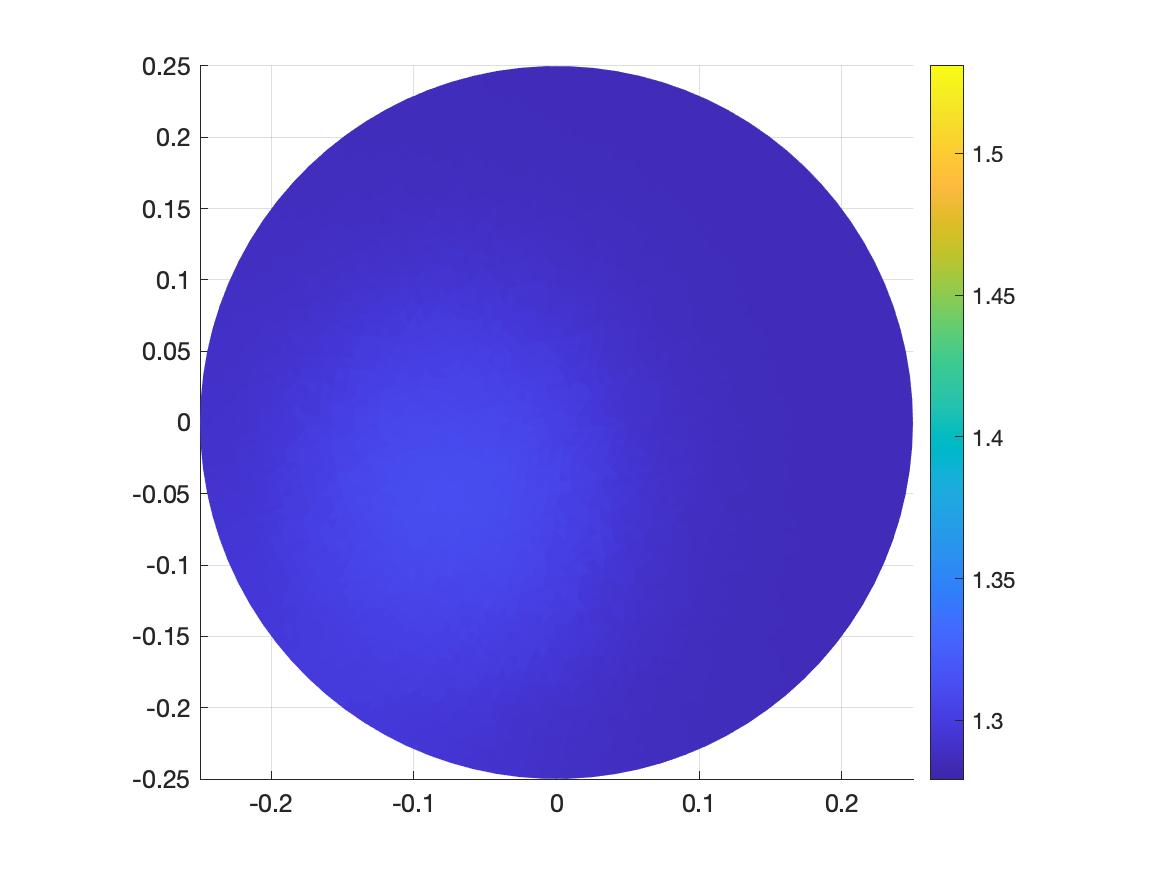}}
		\subfigure[Lac in Astrocytes at $t=2.4$ min]{\includegraphics[width=0.32\textwidth]{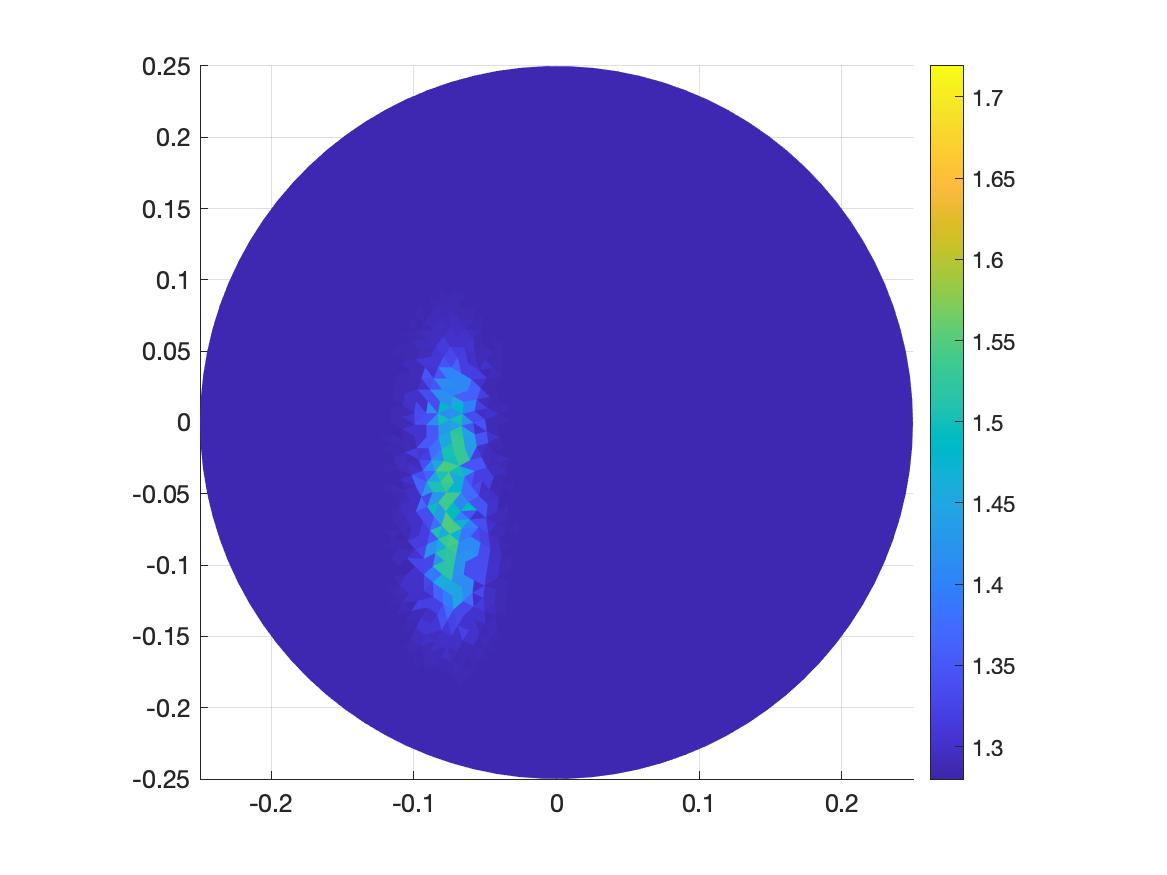}}
		\subfigure[Lac in Astrocytes at $t=4.8$ min]{\includegraphics[width=0.32\textwidth]{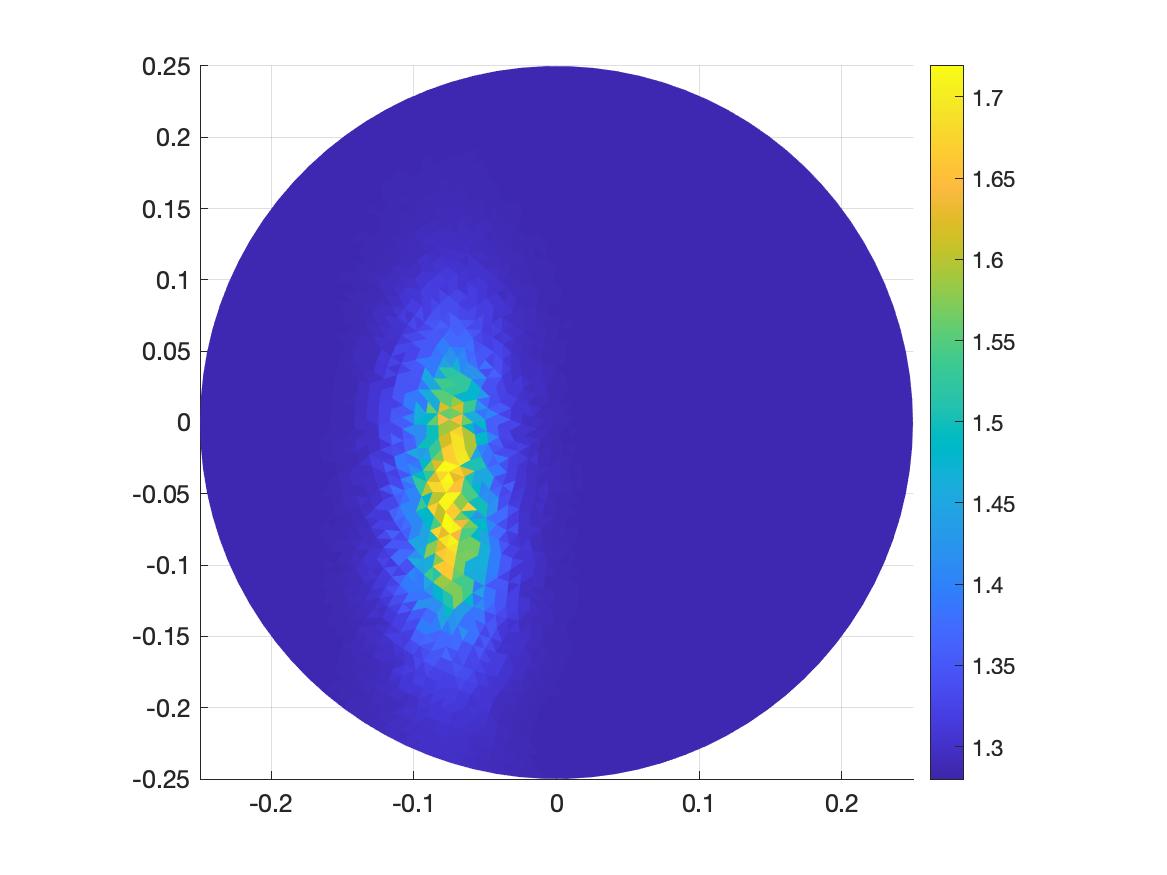}}
		\subfigure[Lac in Astrocytes at $t=6.8$ min]{\includegraphics[width=0.32\textwidth]{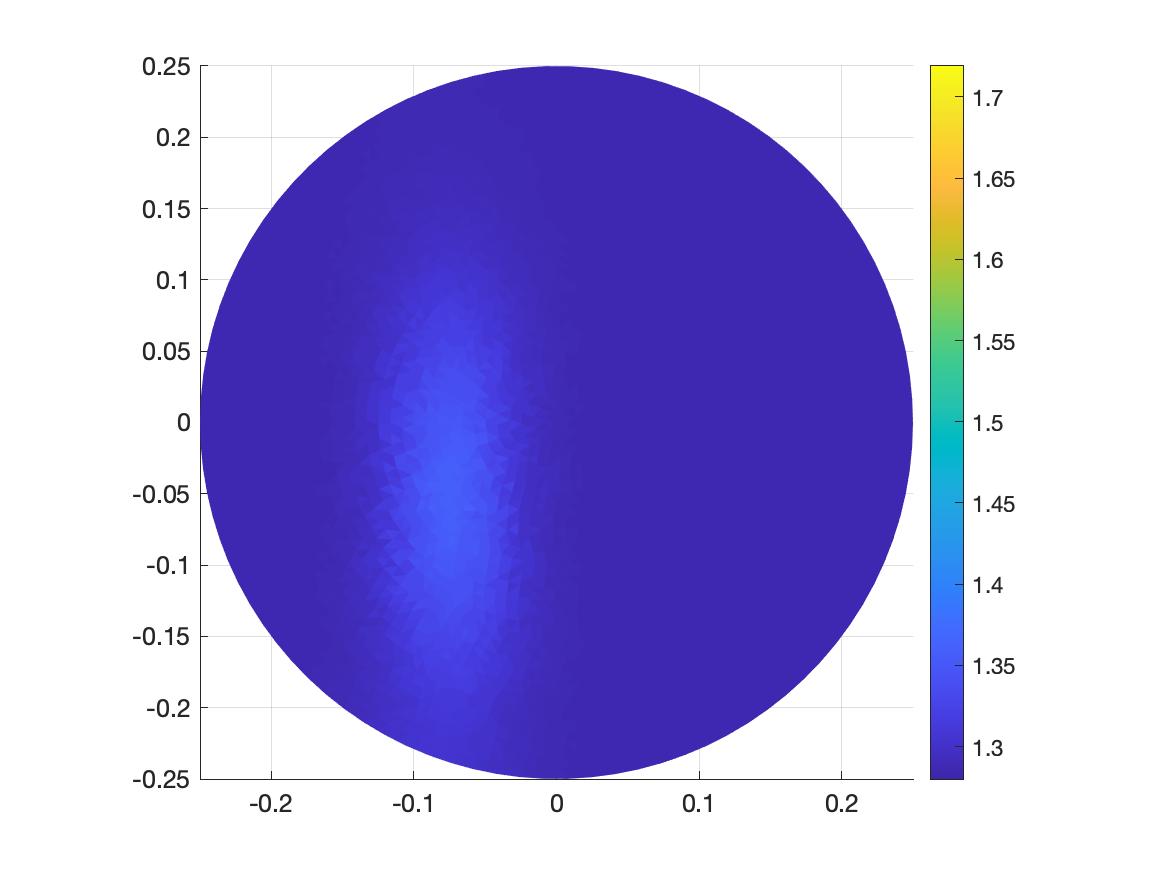}}
		\subfigure[Lac in Astrocytes at $t=2.4$ min]{\includegraphics[width=0.32\textwidth]{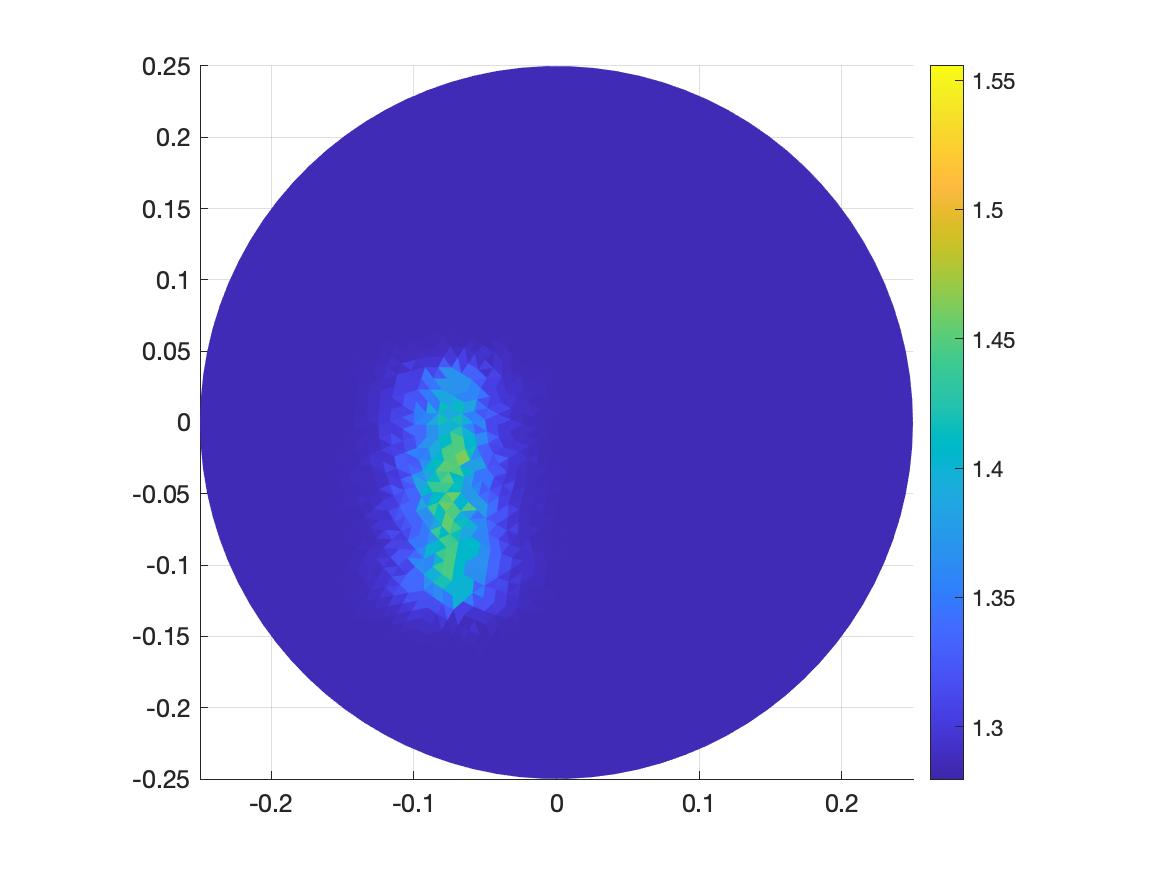}}
		\subfigure[Lac in Astrocytes at $t=4.8$ min]{\includegraphics[width=0.32\textwidth]{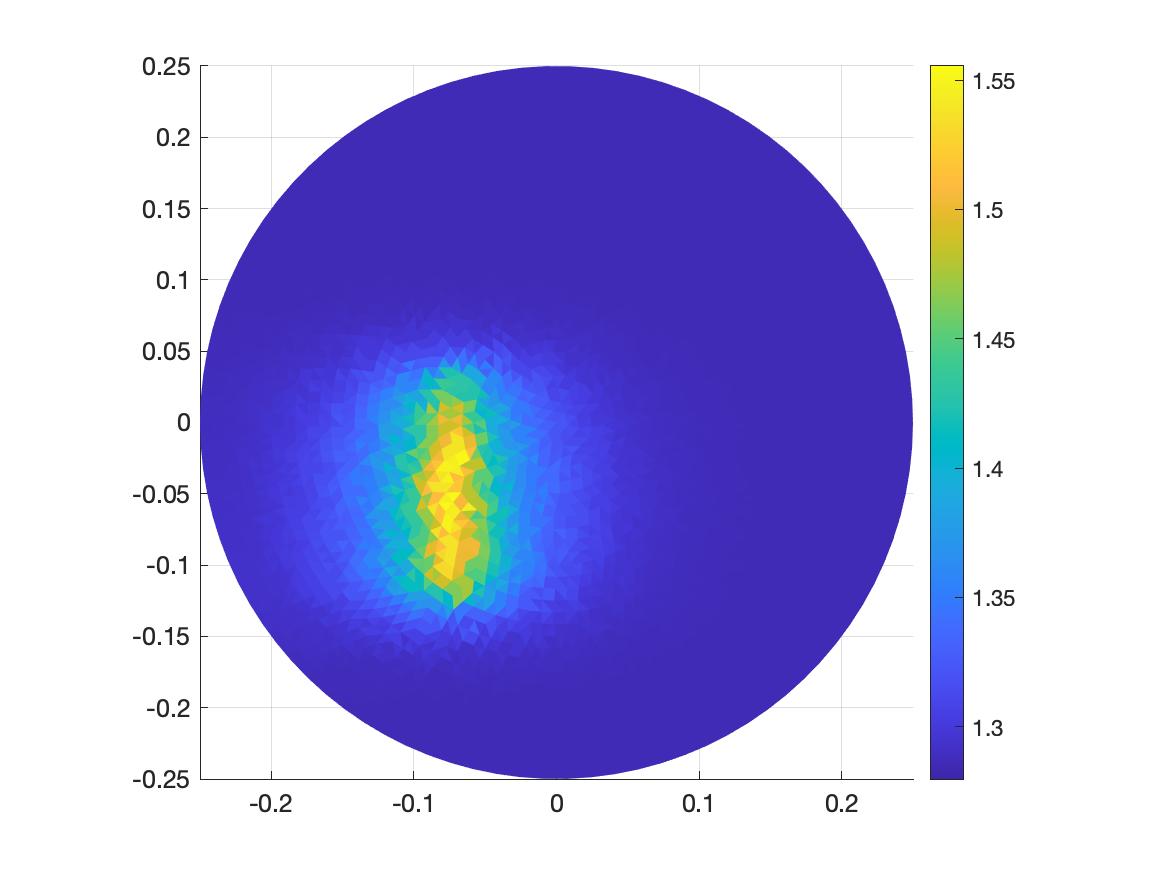}}
		\subfigure[Lac in Astrocytes at $t=6.8$ min]{\includegraphics[width=0.32\textwidth]{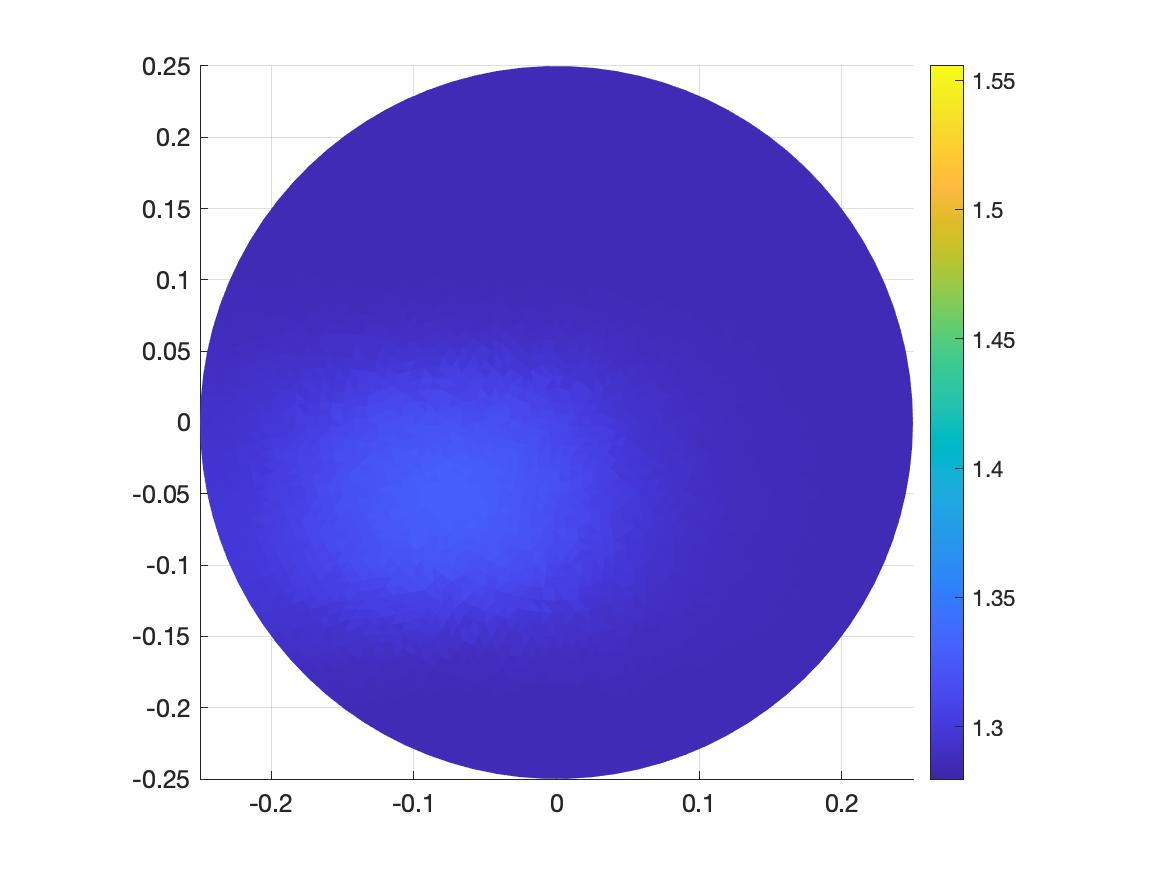}}
		\caption{Time course of lactate concentration in astrocyte with different diffusion tensors in the astrocytic compartment. The top row corresponds to isotropic diffusion, the middle row to diffusion tensor $D_y$ with no horizontal diffusion, and the bottom row to $D_x$ with no vertical l diffusion.}
		\label{fig:anisotropy 1}
	\end{figure*}

	\begin{figure*}[!ht]
		\centering
		\subfigure[Pyr in Astrocytes at $t=2.4$ min]{\includegraphics[width=0.32\textwidth]{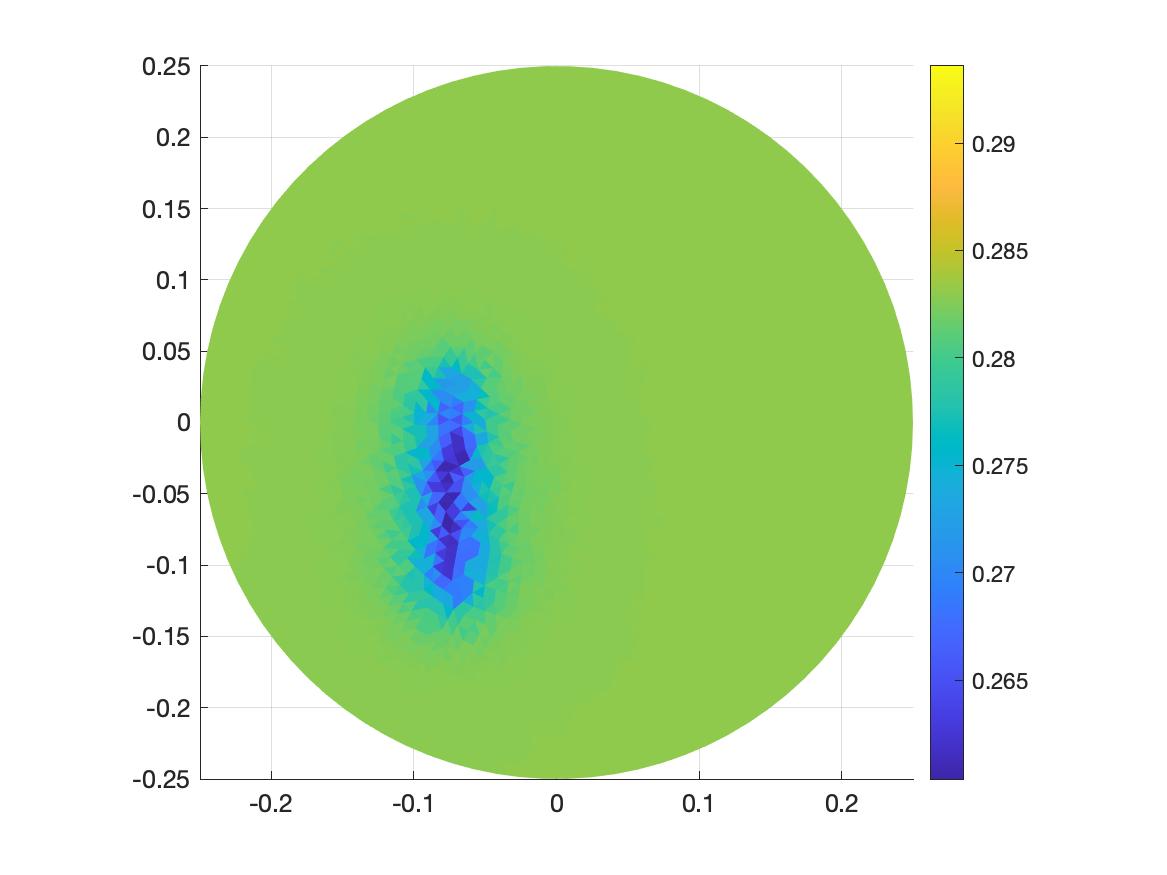}}
		\subfigure[Pyr in Astrocytes at $t=4.8$ min]{\includegraphics[width=0.32\textwidth]{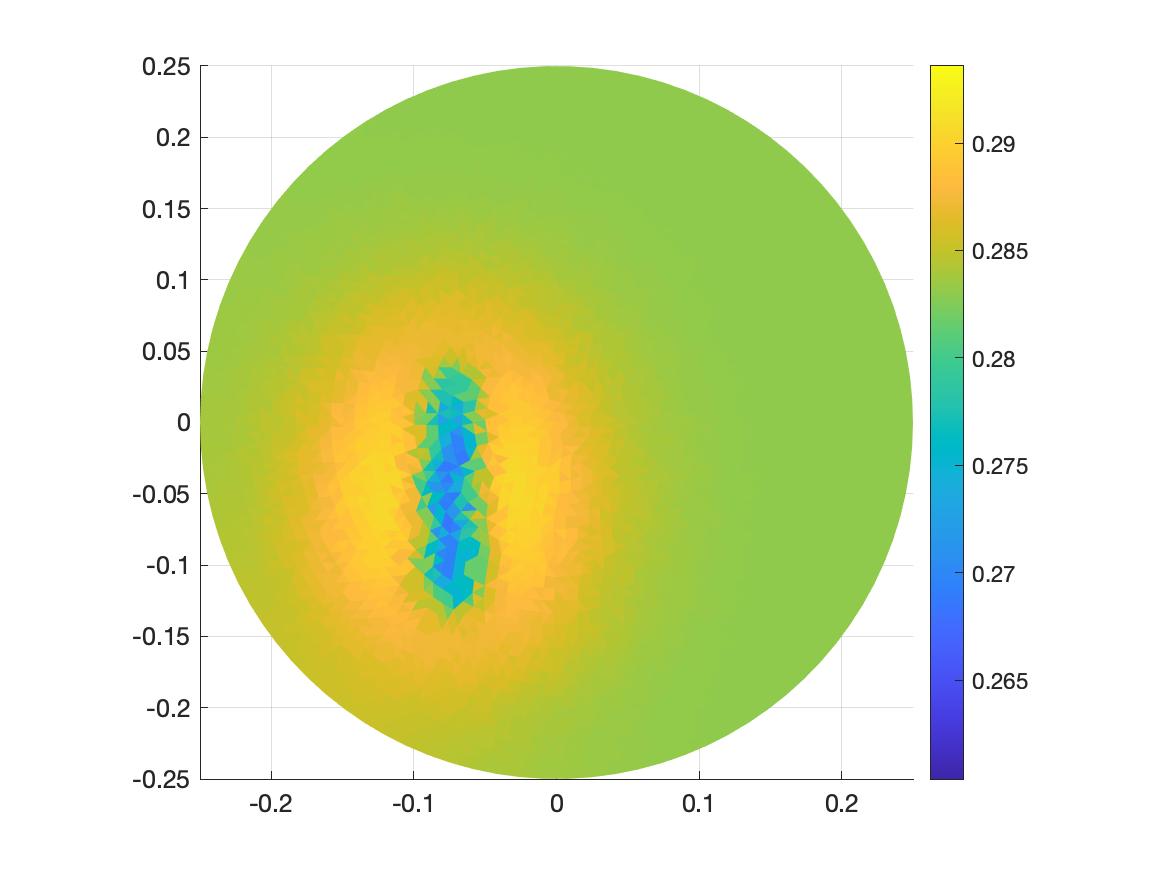}}
		\subfigure[Pyr in Astrocytes at $t=6.8$ min]{\includegraphics[width=0.32\textwidth]{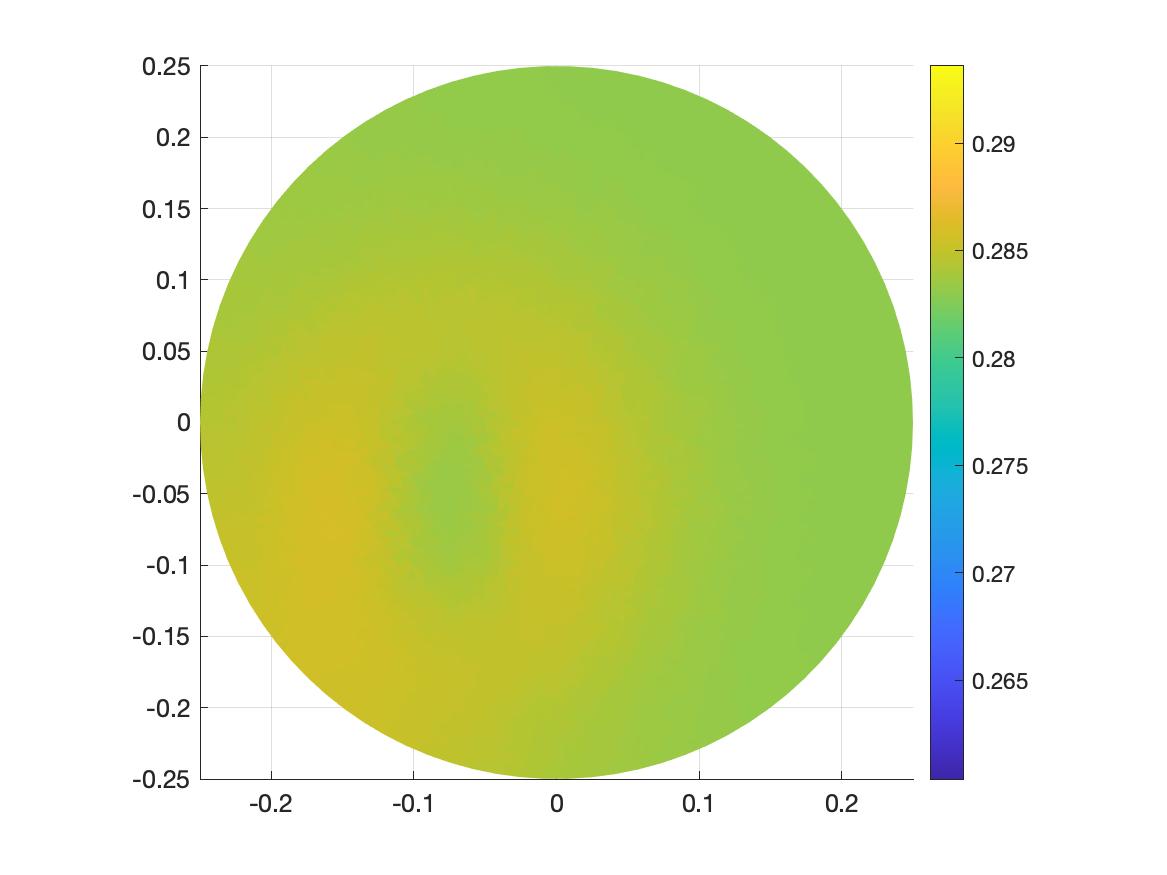}}
		\subfigure[Pyr in Astrocytes at $t=2.4$ min]{\includegraphics[width=0.32\textwidth]{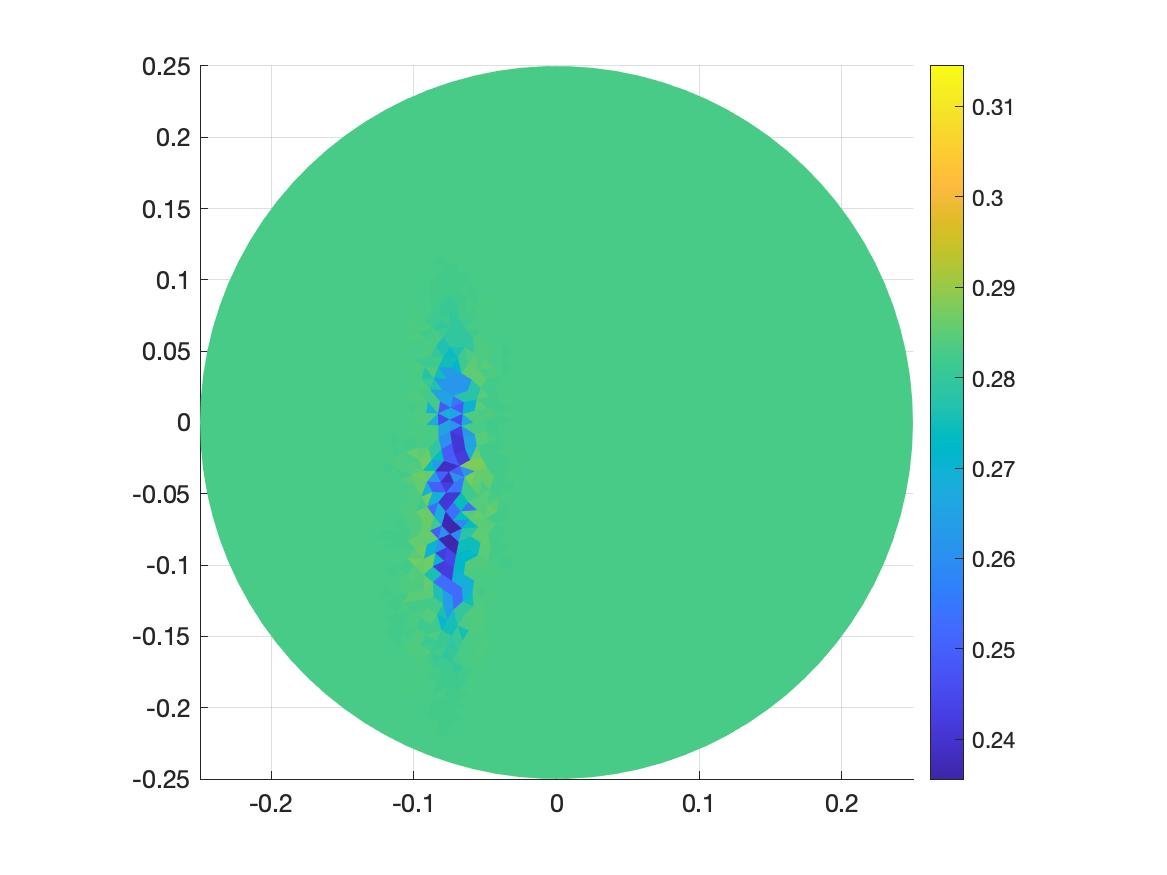}}
		\subfigure[Pyr in Astrocytes at $t=4.8$ min]{\includegraphics[width=0.32\textwidth]{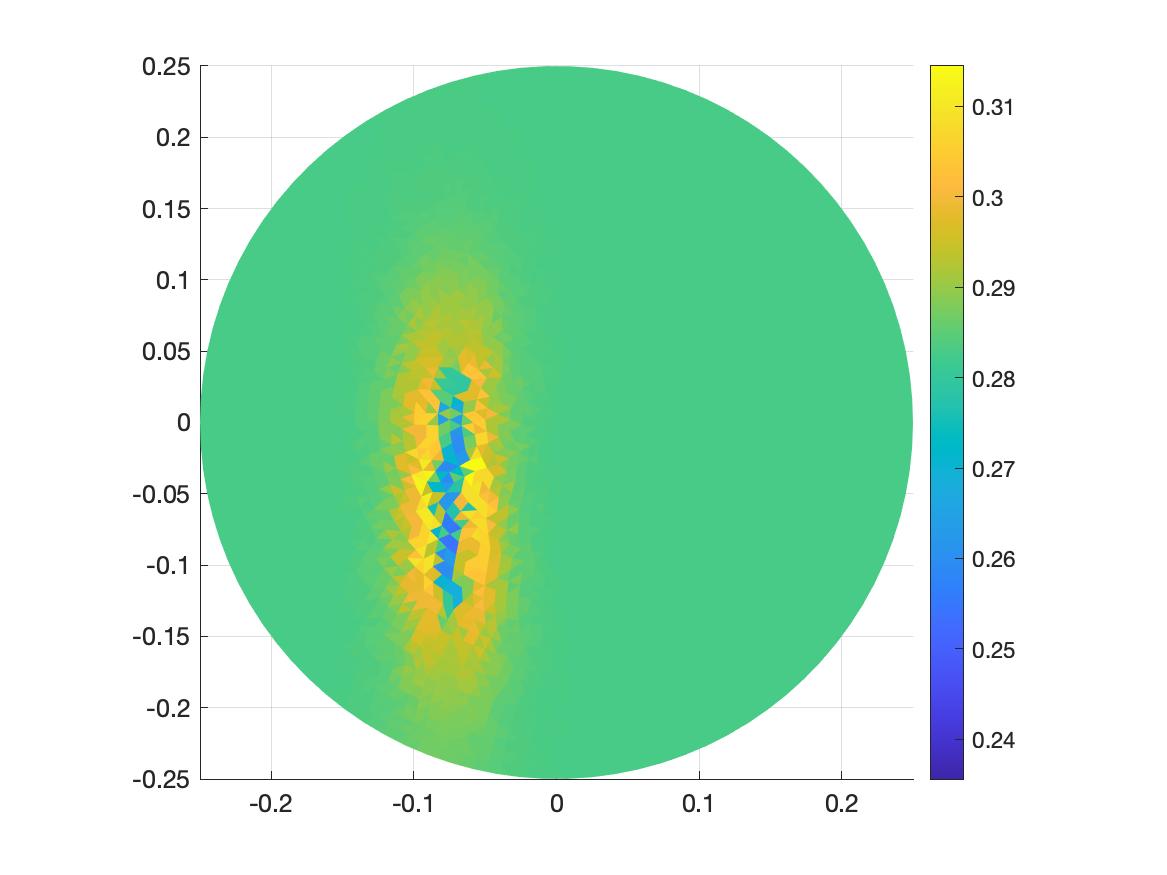}}
		\subfigure[Pyr in Astrocytes at $t=6.8$ min]{\includegraphics[width=0.32\textwidth]{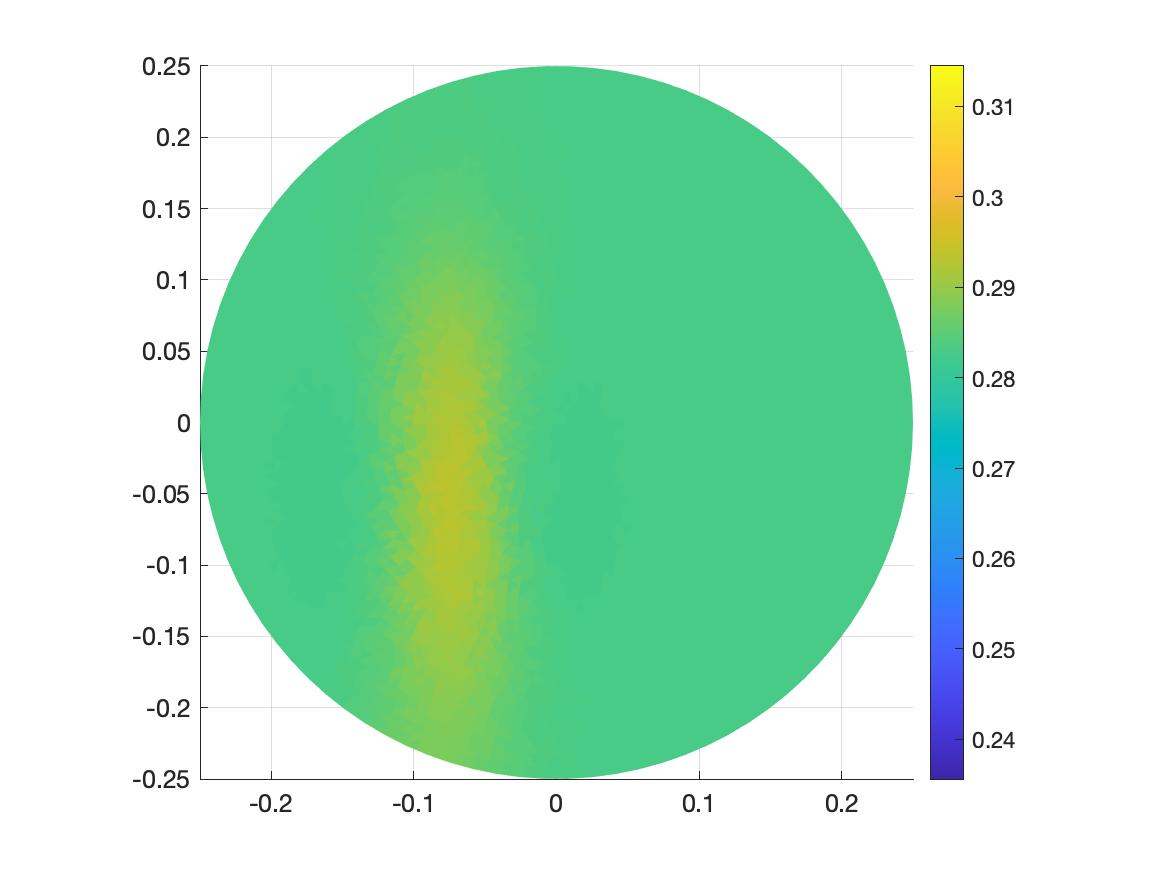}}
		\subfigure[Pyr in Astrocytes at $t=2.4$ min]{\includegraphics[width=0.32\textwidth]{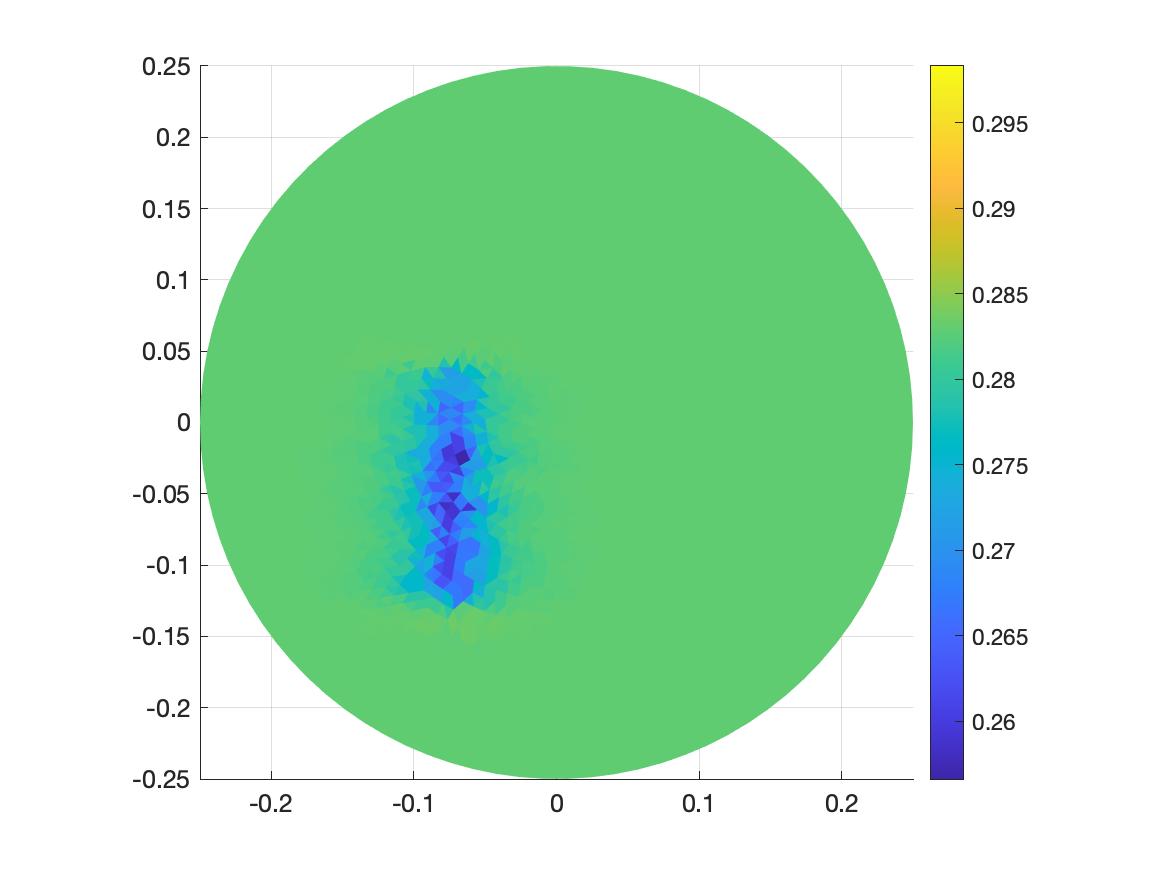}}
		\subfigure[Pyr in Astrocytes at $t=4.8$ min]{\includegraphics[width=0.32\textwidth]{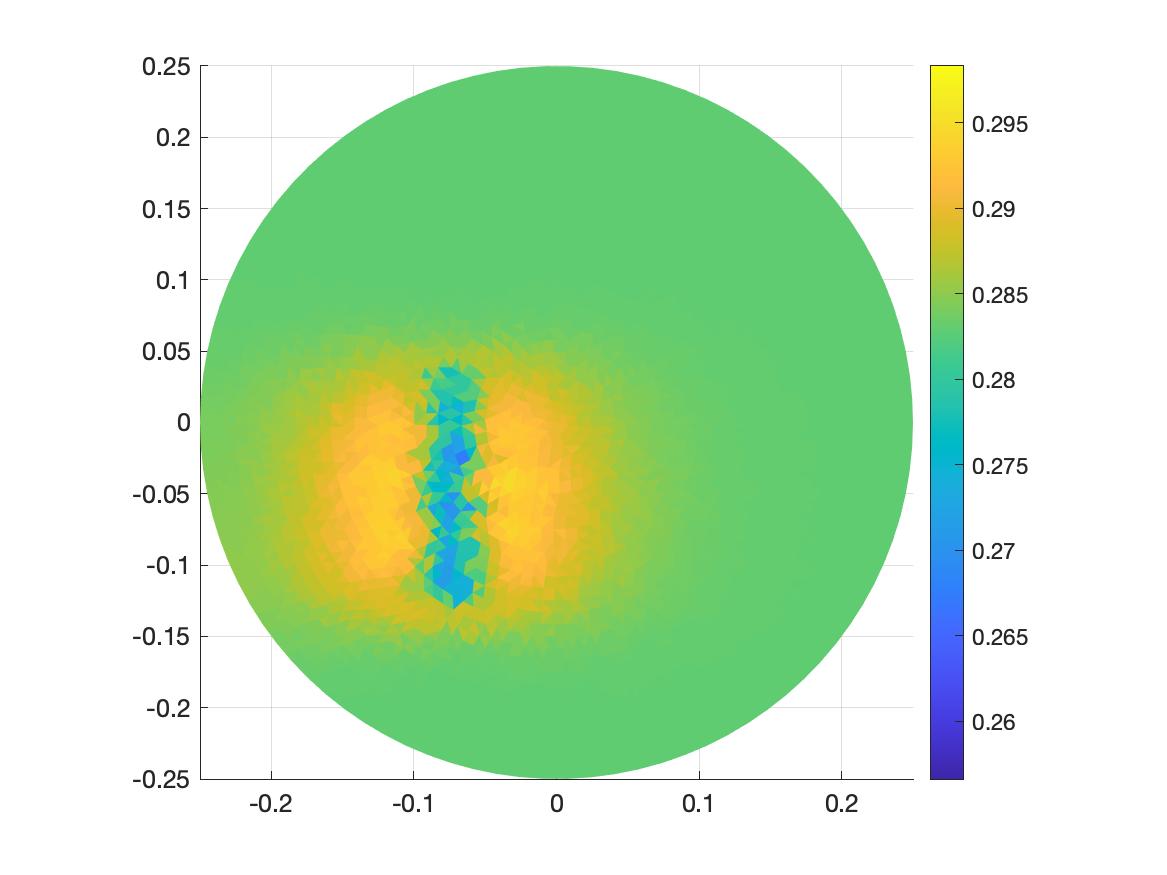}}
		\subfigure[Pyr in Astrocytes at $t=6.8$ min]{\includegraphics[width=0.32\textwidth]{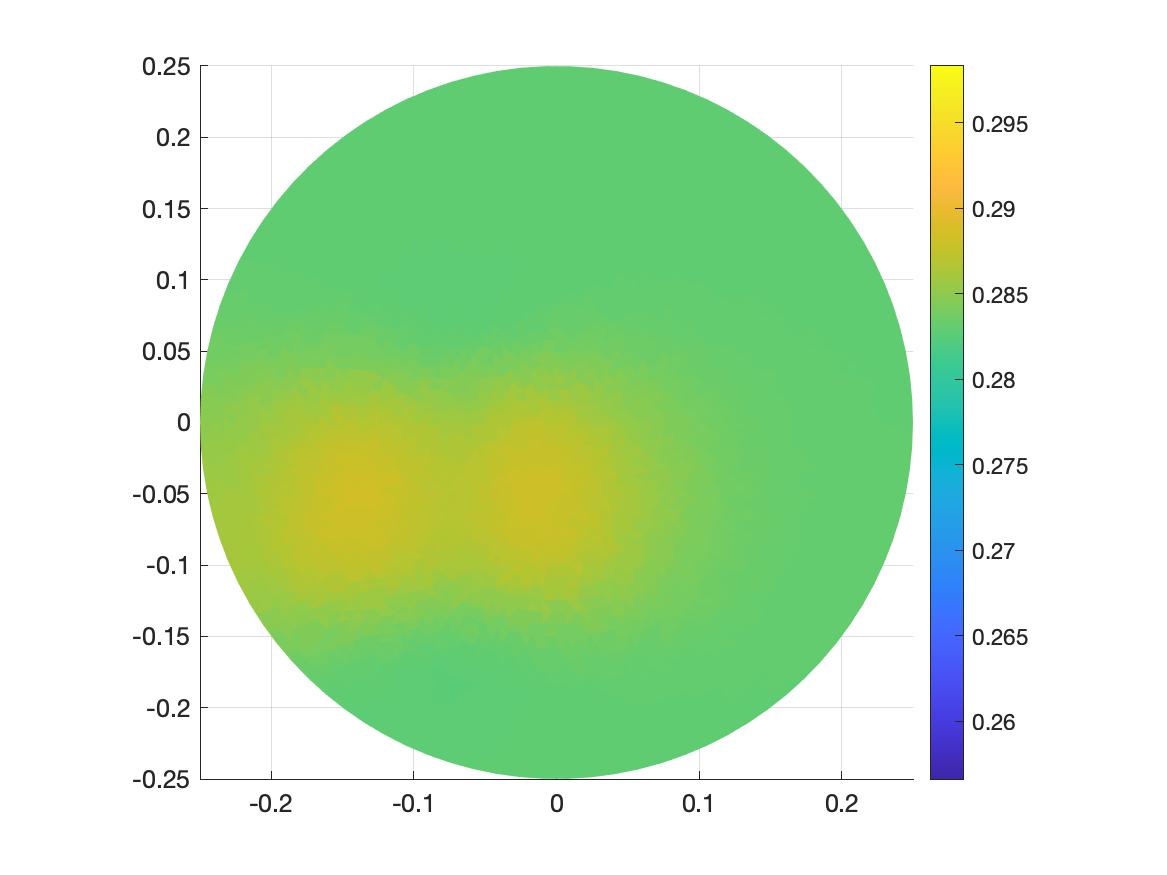}}
		\caption{Time course of pyruvate concentration in astrocyte with different diffusion tensors in the astrocytic compartment. As in Figure~\ref{fig:anisotropy 1}, the top row corresponds to isotropic diffusion, the middle row to diffusion tensor $D_y$ with no horizontal diffusion and the bottom row to $D_x$ with no vertical l diffusion.}
		\label{fig:anisotropy 2}
	\end{figure*} 

	In the same paper it was found that territories of astrocytes were elongated along an axis perpendicular to the pyramidal cell layer, indicating that such morphology could indeed promote diffusion along this axis and result in anisotropic astrocyte coupling. 
	These findings, illustrating that diffusion of molecules may be anisotropic with varying directional preference even within a hippocampal subfield and possibly depending on differences of astrocyte morphology, motivate our third protocol, where we simulate anisotropic diffusion in the current model and test its effect on the diffusion of metabolites. 
	
	To emphasize the effect of the anisotropy, we consider an elliptical activation area, with the major axis ten times the minor axis. The axes of the ellipse are along the Euclidian coordinate axes, the major axis aligned with the vertical direction.
	We begin by simulating the activity with isotropic diffusion tensor given in the baseline simulation, and then modify the diffusion tensor  (\ref{orthotropic}) in the astrocyte to allow diffusion only in the directions parallel or orthogonal to the major axis of the activity ellipse, respectively,
	corresponding to diffusion matrices
	\begin{equation}\label{orthotropic1}
	D_x = \left[\begin{array}{cc} \kappa & 0 \\ 0 & 0\end{array}\right], \quad\mbox{or}  \quad D_y = \left[\begin{array}{cc} 0 & 0 \\ 0 & \kappa \end{array}\right].
	\end{equation}
	
	Figures~\ref{fig:anisotropy 1} and \ref{fig:anisotropy 2} show the effect of the anisotropy on lactate and pyruvate concentration in astrocyte compartment at different time instances. The effect of the anisotropy is as expected, the margin area spreading predominantly in the direction favored by the diffusion tensor. The effect on oxygen and glucose diffusion is less pronounced (data not shown).
	
	\subsection{Conclusions and future work}

	Computational metabolic models are usually assuming well-mixed compartments, and while the importance of diffusion is generally acknowledged, in particular when oxygen is concerned, so far detailed models combining diffusion and metabolism seem to be largely missing. The main aim of this article is to make a substantial contribution toward filling this gap, by proposing a novel spatially distributed computational model of brain energy metabolism. the model accounts for diffusion in ECS and astrocyte network, and allows simulations of differing diffusion patterns. Accounting for diffusion is especially  important in the study of abnormal conditions, e.g., when the metabolite supply is compromised as is the case in ischemia, or when the metabolic demand is elevated such as in seizures. 
	
	The numerical experiments in the present article are mostly concerned with the role of three diffusion-related parameters to the solutions of the distributed model, the gap junction strength, tortuosity of the ECS, and anisotropy of the astrocytic networks. The activity pattern as well as the blood flow are assumed to be within normal parameters. Several extensions will be addresses in forthcoming work. One of the shortcomings of the current model is that the neuronal activity is modeled through an artificial activity function that increases locally the glutamate flux in the clefts. While qualitatively correct, this excitation model does not correctly account for effects of electrolytes to the metabolic needs. A more realistic model should include the double feedback mechanism between metabolism and electrophysiology, where the ATP production is a response to the ion pump action, and vice versa, the electrolyte concentrations changing according to the ATP availability regulating the pump activity.  Another aspect that should be included in the model is the role of astrocyte in the potassium diffusion and extrusion, whose central importance has been recently highlighted. The proposed model will further be adapted to investigate brain energy metabolism under some pathological conditions, including ischemia and cortical spreading depolarization. The model proposed in this paper will be the starting point for all future modifications and extensions. 
	
	\newpage
	\section*{Acknowledgements}
	
	This work made use of the High Performance Computing Resource in the Core Facility for Advanced Research Computing at Case Western Reserve University. The work of D.C was supported in part by NSF Award DMS 1951446. The The work of E.S. was supported in part by NSF Award DMS 2204618.

	\bibliographystyle{unsrt}  
	\clearpage
	\bibliography{references.bib}

\begin{thebibliography}{10}

\bibitem{barros2018current}
L~Felipe Barros, Juan~P Bolanos, Gilles Bonvento, Anne-Karine Bouzier-Sore,
  Angus Brown, Johannes Hirrlinger, Sergey Kasparov, Frank Kirchhoff, Anne~N
  Murphy, Luc Pellerin, et~al.
\newblock Current technical approaches to brain energy metabolism.
\newblock {\em Glia}, 66(6):1138--1159, 2018.

\bibitem{vendel2019need}
Esm{\'e}e Vendel, Vivi Rottsch{\"a}fer, and Elizabeth de~Lange.
\newblock The need for mathematical modelling of spatial drug distribution
  within the brain.
\newblock {\em Fluids and Barriers of the CNS}, 16(1):1--33, 2019.

\bibitem{attwell2001energy}
David Attwell and Simon~B Laughlin.
\newblock An energy budget for signaling in the grey matter of the brain.
\newblock {\em Journal of Cerebral Blood Flow \& Metabolism},
  21(10):1133--1145, 2001.

\bibitem{bonvento2021astrocyte}
Gilles Bonvento and Juan~P Bola{\~n}os.
\newblock Astrocyte-neuron metabolic cooperation shapes brain activity.
\newblock {\em Cell metabolism}, 33(8):1546--1564, 2021.

\bibitem{barros2007enquiry}
L~Felipe Barros and Cristi{\'a}n Mart{\'\i}nez.
\newblock An enquiry into metabolite domains.
\newblock {\em Biophysical journal}, 92(11):3878--3884, 2007.

\bibitem{Tuttle}
A~Tuttle, Diaz~J Riera, and Y.~Mori.
\newblock A computational study on the role of glutamate and nmda receptors on
  cortical spreading depression using a multidomain electrodiffusion model.
\newblock {\em PLoS Comput Biol.}, 15(12), 2019.

\bibitem{sykova2008diffusion}
Eva Sykov{\'a} and Charles Nicholson.
\newblock Diffusion in brain extracellular space.
\newblock {\em Physiological reviews}, 88(4):1277--1340, 2008.

\bibitem{giaume1997metabolic}
Christian Giaume, Arantxa Tabernero, and Jos{\'e}~M Medina.
\newblock Metabolic trafficking through astrocytic gap junctions.
\newblock {\em Glia}, 21(1):114--123, 1997.

\bibitem{terman2019modeling}
David Terman and Min Zhou.
\newblock Modeling the role of the astrocyte syncytium and k+ buffering in
  maintaining neuronal firing patterns.
\newblock {\em Opera Medica et Physiologica}, 5(1):7--16, 2019.

\bibitem{CALVETTI201548}
Daniela Calvetti, Yougan Cheng, and Erkki Somersalo.
\newblock A spatially distributed computational model of brain cellular
  metabolism.
\newblock {\em Journal of Theoretical Biology}, 376:48--65, 2015.

\bibitem{Kuffler}
SW~Kuffler and DD~Potter.
\newblock Glia in the leech central nervous system: physiological properties
  and neuron-glia relationship.
\newblock {\em J Neurophysiol}, 27:290–320, 1964.

\bibitem{Sykova_Nicholson}
Syková Eva and Charles Nicholson.
\newblock Diffusion in brain extracellular space.
\newblock {\em Physiological reviews}, 88(4):1277–1340, 2008.

\bibitem{SYKOVA2004453}
Eva Syková.
\newblock Diffusion properties of the brain in health and disease.
\newblock {\em Neurochemistry International}, 45(4):453--466, 2004.
\newblock Role of Non-synaptic Communication in Information Processing.

\bibitem{2001RPPh...64..815N}
Charles {Nicholson}.
\newblock {Diffusion and related transport mechanisms in brain tissue}.
\newblock {\em Reports on Progress in Physics}, 64(7):815--884, July 2001.

\bibitem{HRABE20041606}
Jan Hrabe, Sabina Hrabĕtová, and Karel Segeth.
\newblock A model of effective diffusion and tortuosity in the extracellular
  space of the brain.
\newblock {\em Biophysical Journal}, 87(3):1606--1617, 2004.

\bibitem{NICHOLSON2000129}
Charles Nicholson, Kevin~C. Chen, Sabina Hrabětová, and Lian Tao.
\newblock Diffusion of molecules in brain extracellular space: theory and
  experiment.
\newblock In {\em Volume Transmission Revisited}, volume 125 of {\em Progress
  in Brain Research}, pages 129--154. Elsevier, 2000.

\bibitem{Nandigam}
Nandigam Ravi and Daniel~M Kroll.
\newblock Three-dimensional modeling of the brain's ecs by minimum
  configurational energy packing of fluid vesicles.
\newblock {\em Biophysical journal}, 92(10):3368--78, 2007.

\bibitem{doi:10.1073/pnas.95.15.8975}
Dmitri~A. Rusakov and Dimitri~M. Kullmann.
\newblock Geometric and viscous components of the tortuosity of the
  extracellular space in the brain.
\newblock {\em Proceedings of the National Academy of Sciences},
  95(15):8975--8980, 1998.

\bibitem{TAO2005525}
A.~Tao, L.~Tao, and C.~Nicholson.
\newblock Cell cavities increase tortuosity in brain extracellular space.
\newblock {\em Journal of Theoretical Biology}, 234(4):525--536, 2005.

\bibitem{JIN20081785}
Songwan Jin, Zsolt Zador, and A.S. Verkman.
\newblock Random-walk model of diffusion in three dimensions in brain
  extracellular space: Comparison with microfiberoptic photobleaching
  measurements.
\newblock {\em Biophysical Journal}, 95(4):1785--1794, 2008.

\bibitem{Verkman}
AS~Verkman.
\newblock Diffusion in the extracellular space in brain and tumors.
\newblock {\em Physical biology}, 10(4):3368--78, 2013.

\bibitem{Shapiro}
B.E. Shapiro.
\newblock Osmotic forces and gap junctions in spreading depression: A
  computational model.
\newblock {\em J Comput Neurosci}, 10:99–120, 2001.

\bibitem{Connell}
R.~O’Connell and Y.~Mori.
\newblock Effects of glia in a triphasic continuum model of cortical spreading
  depression.
\newblock {\em Bull Math Biol}, 78:1943–1967, 2016.

\bibitem{MORI201594}
Yoichiro Mori.
\newblock A multidomain model for ionic electrodiffusion and osmosis with an
  application to cortical spreading depression.
\newblock {\em Physica D: Nonlinear Phenomena}, 308:94--108, 2015.

\bibitem{keener2009mathematical}
James Keener and James Sneyd.
\newblock {\em Mathematical physiology: II: Systems physiology}.
\newblock Springer, 2009.

\bibitem{CALVETTI2018238}
D.~Calvetti, G.~{Capo Rangel}, L.~{Gerardo Giorda}, and E.~Somersalo.
\newblock A computational model integrating brain electrophysiology and
  metabolism highlights the key role of extracellular potassium and oxygen.
\newblock {\em Journal of Theoretical Biology}, 446:238--258, 2018.

\bibitem{overton}
E~Overton.
\newblock \"{U}ber die osmotischen eigenschaften der zelle in ihrer bedeutung
  f\"{u}r die toxikologie und pharmacologie.
\newblock {\em Z Phys Chem}, 22:189--209, 1895.

\bibitem{boron}
Walter~F. Boron.
\newblock Sharpey-schafer lecture: Gas channels.
\newblock {\em Experimental Physiology}, 95(12):1107--1130, 2010.

\bibitem{10.3389/fendo.2013.00137}
Erkki Somersalo and Daniela Calvetti.
\newblock Quantitative in silico analysis of neurotransmitter pathways under
  steady state conditions.
\newblock {\em Frontiers in Endocrinology}, 4, 2013.

\bibitem{doi:10.1146/annurev.micro.50.1.317}
Arthur~L. Koch.
\newblock What size should a bacterium be? a question of scale.
\newblock {\em Annual Review of Microbiology}, 50(1):317--348, 1996.
\newblock PMID: 8905083.

\bibitem{doi:10.1021/jp952903y}
Ping Han and David~M. Bartels.
\newblock Temperature dependence of oxygen diffusion in h2o and d2o.
\newblock {\em The Journal of Physical Chemistry}, 100(13):5597--5602, 1996.

\bibitem{doi:10.1021/ja02220a002}
Tor Carlson.
\newblock The diffusion of oxygen in water.
\newblock {\em Journal of the American Chemical Society}, 33(7):1027--1032,
  1911.

\bibitem{Maxarei}
A.F. Maxarei and Sandall.
\newblock Diffusion coefficients for helium, hydrogen and carbon dioxide in
  water at 25 c.
\newblock {\em AIChE J.}, 26:154--157, 1980.

\bibitem{C2AN36715G}
Bertram~L. Koelsch, Kayvan~R. Keshari, Tom~H. Peeters, Peder E.~Z. Larson,
  David~M. Wilson, and John Kurhanewicz.
\newblock Diffusion mr of hyperpolarized 13c molecules in solution.
\newblock {\em Analyst}, 138:1011--1014, 2013.

\bibitem{Rusakov}
D.~A. Rusakov, L.~P. Savtchenko, K.~Zheng, and J.~M. Henley.
\newblock Shaping the synaptic signal: molecular mobility inside and outside
  the cleft.
\newblock {\em Trends in neurosciences}, 34(7):359–369, 2011.

\bibitem{doi:10.1021/ja01118a065}
L.~G. Longsworth.
\newblock Diffusion measurements, at 25o, of aqueous solutions of amino acids,
  peptides and sugars.
\newblock {\em Journal of the American Chemical Society}, 75(22):5705--5709,
  1953.

\bibitem{BOWEN196430}
William~J. Bowen and Harold~L. Martin.
\newblock The diffusion of adenosine triphosphate through aqueous solutions.
\newblock {\em Archives of Biochemistry and Biophysics}, 107(1):30--36, 1964.

\bibitem{A806004E}
Antonio Doménech, Enrique García-España, José A.~Ramírez, Bernardo Celda,
  Ma~Carmen~Martínez, Daniel Monleón, Roberto Tejero, Andrea Bencini, and
  Antonio Bianchi.
\newblock A thermodynamic{,} electrochemical and molecular dynamics study on
  nad and nadp recognition by
  1{,}4{,}7{,}10{,}13{,}16{,}19-heptaazacyclohenicosane ([21]anen7)†.
\newblock {\em J. Chem. Soc.{,} Perkin Trans. 2}, pages 23--32, 1999.

\bibitem{HASINOFF198753}
Brian~B. Hasinoff, Richard Dreher, and John~P. Davey.
\newblock The association reaction of yeast alcohol dehydrogenase with coenzyme
  is partly diffusion-controlled in solvents of increased viscosity.
\newblock {\em Biochimica et Biophysica Acta (BBA) - Protein Structure and
  Molecular Enzymology}, 911(1):53--58, 1987.

\bibitem{Dzubay5265}
Jeffrey~A. Dzubay and Craig~E. Jahr.
\newblock The concentration of synaptically released glutamate outside of the
  climbing fiber{\textendash}purkinje cell synaptic cleft.
\newblock {\em Journal of Neuroscience}, 19(13):5265--5274, 1999.

\bibitem{neurovascular}
Daniela Calvetti and Erkki Somersalo.
\newblock Dynamic activation model for a glutamatergic neurovascular unit.
\newblock {\em J Theor Biol.}, 274(1):12--29, 2011.

\bibitem{Massucci}
F.~A. Massucci, M.~DiNuzzo, F.~Giove, B.~Maraviglia, I.~P. Castillo,
  E.~Marinari, and A.~De~Martino.
\newblock Energy metabolism and glutamate-glutamine cycle in the brain: a
  stoichiometric modeling perspective.
\newblock {\em BMC systems biology}, 7(103), 2013.

\bibitem{Shulman2001LactateEA}
Robert~G. Shulman, Fahmeed Hyder, and Douglas~L. Rothman.
\newblock Lactate efflux and the neuroenergetic basis of brain function.
\newblock {\em NMR in Biomedicine}, 14, 2001.

\bibitem{doi:10.1097/00004647-199807000-00005}
Peter~L. Madsen, Rasmus Linde, Steen~G. Hasselbalch, Olaf~B. Paulson, and
  Niels~A. Lassen.
\newblock Activation-induced resetting of cerebral oxygen and glucose uptake in
  the rat.
\newblock {\em Journal of Cerebral Blood Flow \& Metabolism}, 18(7):742--748,
  1998.
\newblock PMID: 9663504.

\bibitem{Wallraff}
Anke Wallraff, Benjamin Odermatt, Klaus Willecke, and Christian Steinhäuser.
\newblock Distinct types of astroglial cells in the hippocampus differ in gap
  junction coupling.
\newblock {\em Glia}, 48(1):36--43, 2004.

\bibitem{houades_rouach_ezan_kirchhoff_koulakoff_giaume_2006}
Vanessa Houades, Nathalie Rouach, Pascal Ezan, Frank Kirchhoff, Annette
  Koulakoff, and Christian Giaume.
\newblock Shapes of astrocyte networks in the juvenile brain.
\newblock {\em Neuron Glia Biology}, 2(1):3–14, 2006.

\bibitem{Schools}
Gary~P. Schools, Min Zhou, and Harold~K. Kimelberg.
\newblock Development of gap junctions in hippocampal astrocytes: Evidence that
  whole cell electrophysiological phenotype is an intrinsic property of the
  individual cell.
\newblock {\em Journal of Neurophysiology}, 96(3):1383--1392, 2006.
\newblock PMID: 16775204.

\bibitem{Houades5207}
Vanessa Houades, Annette Koulakoff, Pascal Ezan, Isabelle Seif, and Christian
  Giaume.
\newblock Gap junction-mediated astrocytic networks in the mouse barrel cortex.
\newblock {\em Journal of Neuroscience}, 28(20):5207--5217, 2008.

\bibitem{Anders}
Stefanie Anders, Daniel Minge, Stephanie Griemsmann, Michel~K. Herde, Christian
  Steinhäuser, and Christian Henneberger.
\newblock Spatial properties of astrocyte gap junction coupling in the rat
  hippocampus.
\newblock {\em Philosophical Transactions of the Royal Society B: Biological
  Sciences}, 369(1654):20130600, 2014.

\end{thebibliography}

\end{document}